\documentclass[twocolumn]{aastex701}

\usepackage{enumitem}
\usepackage[nohypertypes={main,acronym}]{glossaries} 
\usepackage{graphicx} 
\usepackage{hyperref}
\hypersetup{colorlinks=true,allcolors=blue}
\usepackage{graphicx}
\usepackage{xcolor}
\usepackage{xspace}
\definecolor{purple}{rgb}{0.54, 0.17, 0.89}
\definecolor{burgundy}{rgb}{0.5, 0.0, 0.13}
\definecolor{royalazure}{rgb}{0.0, 0.22, 0.66}
\definecolor{brickred}{rgb}{0.65, 0.16, 0.16}
\definecolor{aliceblue}{rgb}{0.94, 0.97, 1.0}
\definecolor{honeydew}{rgb}{0.94, 1., 0.94}
\definecolor{lightpurple}{rgb}{0.92, 0.77, 1.}
\definecolor{lightpink}{rgb}{0.96, 0.90, 1.}
\definecolor{lilac}{rgb}{0.73, 0.65, 1.}
\definecolor{cornflower}{RGB}{163, 171, 255}
\definecolor{violet}{RGB}{46, 0, 222}
\definecolor{brightube}{rgb}{0.82, 0.62, 0.91}
\definecolor{languidlavender}{rgb}{0.84, 0.79, 0.87}
\definecolor{lavenderblue}{rgb}{0.8, 0.8, 1.0}
\definecolor{mediumlavendermagenta}{rgb}{0.8, 0.6, 0.8}
\definecolor{palegreen}{RGB}{224, 255, 201}
\definecolor{paleaqua}{RGB}{206, 255, 231}
\definecolor{palebluegreen}{RGB}{190, 255, 223}
\definecolor{darkgreen}{RGB}{0, 92, 7}

\definecolor{DeepTwilight}{RGB}{3, 4, 94}
\definecolor{BrightTealBlue}{RGB}{0, 119, 182}
\definecolor{TurquoiseSurf}{RGB}{0, 180, 216}
\definecolor{FrostedBlue}{RGB}{144, 224, 239}
\definecolor{LightCyan}{RGB}{202, 240, 248}

\definecolor{DeepNavy}{RGB}{11,44,92}
\definecolor{RoyalBlue}{RGB}{32,90,170}
\definecolor{SteelBlue}{RGB}{78,132,199}
\definecolor{SkyTint}{RGB}{220,236,252}

\definecolor{MidnightBlue}{RGB}{15,35,95}
\definecolor{AzureBlue}{RGB}{0,95,184}
\definecolor{CornflowerBlue}{RGB}{84,142,255}
\definecolor{IceBlue}{RGB}{226,239,255}

\definecolor{DeepOcean}{RGB}{0,52,89}
\definecolor{OceanBlue}{RGB}{0,102,153}
\definecolor{CoastalBlue}{RGB}{66,153,203}
\definecolor{SeaMist}{RGB}{223,242,250}

\definecolor{PersianBlue}{RGB}{12, 52, 183}
\definecolor{CornflowerBlue}{RGB}{117, 139, 253}
\definecolor{RosyCopper}{RGB}{205, 83, 52}
\definecolor{LightSeaGreen}{RGB}{12, 164, 165}
\definecolor{DarkCyan}{RGB}{6, 144, 143}

\definecolor{PersianBlue}{RGB}{12, 52, 183}
\definecolor{CornflowerBlue}{RGB}{117, 139, 253}
\definecolor{RosyCopper}{RGB}{205, 83, 52}
\definecolor{LightSeaGreen}{RGB}{12, 164, 165}
\definecolor{DarkCyan}{RGB}{6, 144, 143}

\definecolor{zonared}{HTML}{AB0520}
\definecolor{zonablue}{HTML}{0C234B}

\definecolor{TurquoiseSurfDark}{RGB}{0,139,167}
\definecolor{TurquoiseSurfDeep}{RGB}{0,128,153}
\definecolor{TurquoiseSurfOcean}{RGB}{0,102,122}

\newcommand{\eq}{Equation}
\newcommand{\f}{Figure}
\newcommand{\tab}{Table}

\newcommand{\RMS}{rms\xspace}

\newcommand{\kpc}{kiloparsec\xspace}
\newcommand{\gc}{galactocentric\xspace}
\newcommand{\Gc}{Galactocentric\xspace}

\newacronym{snd}{SNd}{Solar Neighborhood}
\newcommand*{\SNd}{\gls{snd}\xspace}

\newacronym{sn}{SN}{super nova}

\newacronym{ism}{ISM}{interstellar medium}
\newcommand*{\ISM}{\gls{ism}\xspace}

\newacronym{gmc}{GMC}{giant molecular cloud}

\newacronym{mw}{MW}{Milky Way}
\newcommand*{\MW}{\gls{mw}\xspace}

\newacronym{ca}{C\&A}{stellar clusters and associations}

\newacronym{lmc}{LMC}{Large Milky Cloud}

\newacronym{smc}{SMC}{Small Milky Cloud}

\newacronym{bpx}{BPX}{boxy-peanut-X shaped bulge}

\newacronym{bh}{BH}{Black Hole}

\newacronym{ns}{NS}{Neutron Star}

\newacronym{gw}{GW}{gravitational wave}

\newcommand{\kmy}{kinematically\xspace}

\newcommand{\azi}{azimuthal\xspace}
\newcommand{\epi}{epicyclic\xspace}
\newcommand{\axisym}{axisymmetric\xspace}

\newcommand{\perturber}{perturber\xspace}
\newcommand{\pangle}{pitch angle\xspace}

\newacronym{tdyn}{$T_{\rm Dyn}$}{dynamical time}
\newcommand{\Tdyn}{\gls{tdyn}\xspace}
\newacronym{ec}{$e$}{eccentricity}
\newcommand{\ecc}{\gls{ec}\xspace}
\newacronym{ltot}{$L$}{angular momentum}
\newcommand{\Ltot}{\gls{ltot}\xspace}
\newacronym{lz}{$L_z$}{angular momentum in the direction perpendicular to the plane of the disk}
\newcommand{\Lz}{\gls{lz}\xspace}
\newacronym{ilz}{$L_{z,i}$}{initial angular momentum in the direction perpendicular to the plane of the disk}
\newcommand{\iLz}{\gls{ilz}\xspace}
\newacronym{rg}{$R_g$}{guiding center radius\xspace}
\newcommand{\Rg}{\gls{rg}\xspace}

\newacronym{vrd}{$\sigma_R$}{radial velocity dispersion}
\newcommand{\sigmaR}{\gls{vrd}\xspace}

\newcommand{\JR}{$J_R$\xspace}

\newcommand{\DelLz}{$\Delta L_z$\xspace}
\newcommand{\RMSL}{\RMS\!$(\Delta L_z$)\xspace}
\newcommand{\RMSJphi}{\RMS\!$(\Delta J_\phi$)\xspace}
\newcommand{\RMSJR}{\RMS\!$(\Delta J_R$)\xspace}
\newcommand{\RMSEnc}{\RMS\!$(\Delta E_{\rm nc}$)\xspace}

\newcommand{\Fe}{metallicity\xspace}
\newcommand{\Fes}{metallicities\xspace}
\newcommand{\FeH}{$[{\rm Fe}/{\rm H}]$\xspace}
\newcommand{\aFe}{$[\alpha/{\rm Fe}]$\xspace}

\newacronym{ccd}{chrono-chemo-dynamic}{age, chemical, and kinematic}
\newcommand{\CCD}{\gls{ccd}\xspace}
\newacronym{amr}{AMR}{age-\Fe\, relation}
\newcommand{\AMR}{\gls{amr}\xspace}
\newacronym{avr}{AVR}{age-velocity dispersion relation}
\newcommand{\AVR}{\gls{avr}\xspace}

\newacronym{rm}{radial migration}{radial migration}
\newcommand{\rmn}{\acrlong{rm}\xspace}
\newacronym{ct}{cold torquing}{cold torquing}
\newcommand{\cting}{\acrlong{ct}\xspace}
\newacronym{rr}{radial redistribution}{radial redistribution}
\newcommand{\rrn}{\acrlong{rr}\xspace}
\newacronym{orbr}{orbital redistribution}{orbital redistribution}
\newcommand{\orn}{\acrlong{orbr}\xspace}

\newcommand{\TypeCR}{trapped at \CR\ only\xspace}
\newcommand{\TypeOverlap}{trapped with overlap\xspace}
\newcommand{\Lag}{L$_{4/5}$\xspace}

\newacronym{cr}{CR}{corotation}
\newcommand{\CR}{\acrlong{cr}\xspace}
\newacronym{cring}{corotating}{corotating}
\newcommand{\crting}{\acrlong{cring}\xspace}
\newcommand{\crt}{corotate\xspace}

\newacronym{rcr}{$R_{\rm CR}$}{radius of \CR{}}
\newcommand{\RCR}{\gls{rcr}\xspace}

\newacronym{ulr}{ULR}{ultraharmonic resonance}

\newacronym{ilr}{ILR}{inner Lindblad resonance}
\newcommand{\ILR}{\gls{ilr}\xspace}
\newacronym{olr}{OLR}{outer Lindblad resonance}

\newacronym{iolr}{I/OLRs}{inner and outer Lindblad resonances}
\newcommand{\IOLRs}{\gls{iolr}\xspace}

\newacronym{pecc}{PECCARY}{Permutation Entropy and statistiCal Complexity Analysis for astRophYsics}

\newacronym{nsf}{NSF}{National Science Foundation}

\newacronym{hsf}{HSF}{Heising-Simons Foundation}

\newacronym{nasa}{NASA}{National Aeronautics and Space Administration}

\newacronym{ua}{U~Arizona}{University of Arizona}

\newacronym{bmc}{BMC}{Bryn Mawr College}

\newacronym{swat}{SC}{Swarthmore College}

\newacronym{cca}{CCA}{Flatiron Institute's Center for Computational Astrophysics}

\newacronym[first={Spitzer Survey of Stellar Structure in Galaxies \citep[S$^4$G;][]{2010PASP..122.1397S}}]{ssssg}{S$^4$G}{S$^4$G}

\newacronym{jwst}{JWST}{\textit{James Webb Space Telescope}}
\newcommand*{\JWST}{\gls{jwst}\xspace}

\newacronym[first={\textit{Gaia} \citep{2016A&A...595A...1G}}
]{gaia}{\textit{Gaia}}{\textit{Gaia}}

\newacronym[first={\textit{GALAH} \citep{2015MNRAS.449.2604D}}
]{galah}{\textit{GALAH}}{\textit{GALAH}}

\newacronym[first={\textit{LAMOST} \citep{2012RAA....12.1197C}}
]{lamost}{\textit{LAMOST}}{\textit{LAMOST}}

\newacronym[first={\textit{Sloan Digital Sky Survey} \citep{2000AJ....120.1579Y}}
]{sdss}{SDSS}{SDSS}

\newacronym[first={\textit{APOGEE} \citep{2017AJ....154...94M}}]{apogee}{\textit{APOGEE}}{\textit{APOGEE}}

\newacronym[first={Galaxy Zoo \citep{2011MNRAS.410..166L}}
]{gzoo}{Galaxy Zoo}{Galaxy Zoo}

\newacronym[first={Mapping Nearby Galaxies at Apache Point Observatory \citep[MaNGA;][]{2015ApJ...798....7B}}
]{manga}{MaNGA}{MaNGA}

\newacronym[first={\textit{Physics at High Angular resolution in Nearby GalaxieS} survey \citep[PHANGS;][]{phangs_2024}}
]{phangs}{PHANGS}{PHANGS}

\newacronym[first={\textit{Dark Energy Spectroscopic Instrument} (DESI; \citealp{2022AJ....164..207D,Desi}}
]{desi}{DESI}{DESI}

\newacronym{hst}{HST}{\textit{Hubble Space Telescope}\xspace}
\newcommand*{\HST}{\gls{hst}\xspace}

\newcommand{\Python}{\texttt{Python}\xspace}

\newacronym{petar}{\cite{wang_2020}}{\texttt{petar}\xspace}

\DeclareRobustCommand{\galpyfirst}{\texttt{galpy}\footnote{The \texttt{galpy} package can be accessed at \url{https://github.com/jobovy/galpy} \citep{Bovy15}.}}
\newacronym[ first={\protect\galpyfirst}
]{galpy}{\texttt{galpy}}{\texttt{galpy}}

\newacronym[first={Orbital Torus Imaging \citep[OTI;][]{2021ApJ...910...17P}}
]{oti}{OTI}{OTI}

\newacronym[first={Tremaine-Weinberg method \citep[TW;][]{1984ApJ...282L...5T}}
]{tw}{TW}{TW}

\newacronym{bfe}{BFE}{basis function expansion}

\newacronym{mssa}{mSSA}{multivariate Singular Spectrum Analysis}

\newacronym{fof}{FoF}{Friends of Friends}

\newacronym{wdft}{wDFT}{windowed discrete Fourier transform \citep{1986MNRAS.221..195S}}

\newacronym[first={\textit{Latte} \citep{Wetzel16}}
]{latte}{\textit{Latte}}{\textit{Latte}}

\newacronym[first={Feedback In Realistic Environments (FIRE; \citealp{Hopkins18})}
]{fire}{FIRE}{Feedback In Realistic Environments}

\newacronym[first={Auriga Project \citep{2017MNRAS.467..179G}}
]{auriga}{Auriga}{Feedback In Realistic Environments}

\newacronym{loi}{LOI}{Letter of Intent}

\newacronym{fi}{FI}{Future Investigator}

\newacronym{pi}{PI}{Principle Investigator}

\newacronym{coi}{Co-I}{Co-Investigator}

\newacronym{ta}{TA}{Teaching Assistant}

\newacronym{udl}{UDL}{Universal Design for Learning}

\newacronym{sip}{SIP}{Society of Indigenous Physicists}

\newacronym{ymi}{YMI}{The Yellow Mountain Institute}

\newacronym{nn}{NN}{Native Nation}

\newacronym{nsbp}{NSBP}{National Society of Black Physicists}

\newacronym{nshp}{NSHP}{National Society of Hispanic Physicists}

\newacronym{ce}{CE}{Cosmic Explorer}

\newacronym{gis}{GIS}{Geographic Information Systems}

\newacronym{ligo}{LIGO}{Laser Interferometer Gravitational-Wave Observatory}

\newacronym{ipp}{IPP}{Indigenous and Place-Based Partnerships}

\newacronym{norc}{NORC}{National Opinion Research Center}

\received{\today}
\revised{??}
\accepted{??}
\submitjournal{ApJ}

\begin{document}

\shorttitle{Spiral morphology and radial redistribution}
\shortauthors{Daniel}

\title{Spiral Morphology and Radial Migration: Kinematically heating, cooling, and cold}


\correspondingauthor{Kathryne J. Daniel}
\email{kjdaniel@arizona.edu}

\author[0000-0003-2594-8052]{Kathryne J. Daniel}
\affiliation{Department of Astronomy \& Steward Observatory, University of Arizona,
933 North Cherry Avenue,
Tucson, AZ 85721, USA}
\email{kjdaniel@arizona.edu}

\author[0000-0002-6505-9981]{M. E. Wisz}
\affiliation{Department of Physics, Bryn Mawr College,
Bryn Mawr, PA 19010, USA}
\affiliation{Department of Physics, University of California, Merced, 5200 Lake Rd, Merced, CA 95343, USA}
\email{mwisz@ucmerced.edu}

\author[0000-0003-0846-9578]{Karen L. Masters}
\affiliation{Haverford College,
Department of Physics \& Astronomy,
Haverford, PA 19041, USA}
\email{klmasters@haverford.edu}

\author[0000-0002-4013-1799]{Rosemary F.~G.~Wyse}
\affiliation{William H. Miller III Department of Physics \& Astronomy, \\
Johns Hopkins University,
3400 N. Charles St.,
Baltimore, MD 21218, USA}
\email{wyse@jhu.edu}

\author[[0000-0002-3041-7822]{Amy Smock} 
\affiliation{Department of Astronomy \& Steward Observatory, University of Arizona,
933 North Cherry Avenue,
Tucson, AZ 85721, USA}
\email{amysmock@arizona.edu}

\author[0009-0000-9825-9755]{Lipika Chatur}
\affiliation{Department of Astronomy \& Steward Observatory, University of Arizona,
933 North Cherry Avenue,
Tucson, AZ 85721, USA}
\email{lchatur@arizona.edu}



\begin{abstract}
Transient spiral arms are known to drive \rrn of stars and thus could play a central role in shaping disk galaxies, including modifying, over time, the \CCD distributions in the Milky Way.  However, the physical factors governing the efficiency of such processes remain poorly understood. This paper investigates how the morphology of spiral arms -- the number, \pangle, lifetime, and radial dependence of the pattern speed -- influences \orn through \lq \cting ' at the \CR resonance(s). Analytic expressions are derived for the maximum radial excursion of stars trapped at \CR that explicitly account for spiral morphology, predicting that the efficiency of \cting for a density-wave like spiral is greater for more open spiral patterns. Tracer-particle simulations confirm the analytic prediction, in both two- and three-dimensional galactic potentials. In contrast, spirals that have a radially dependent pattern speed such that they \crt with the disk at all radii exhibit the opposite behavior, with \cting becoming more efficient as the spiral winds to smaller \pangle{s} over time. This study further finds that resonant interactions from the same transient spiral causing \cting naturally also produces both kinematic heating and cooling of orbits away from \CR. These results demonstrate that spiral morphology alone cannot predict the efficiency of \cting and suggest that the relationship between spiral \pangle and \rrn provides a potential diagnostic for distinguishing between competing theories of spiral structure.
\end{abstract}



\section{Introduction}

The \rrn, or ``\rmn," of stars is generally accepted to be a key contributor to disk galaxy evolution. 
Multiple physical mechanisms, of both internal and external origin, can change the orbits of disk stars. The combined changes in individual orbital \ecc, \Ltot, and vertical motion
alter the \CCD distributions of stellar populations over the lifetime of a galactic disk.

Almost all mechanisms causing the \rrn of stars change \textit{both} orbital \acrlong{lz} and \acrlong{ec}, and many mechanisms also impact vertical motions.
Orbital scattering off of \glspl{gmc} increases the velocity dispersion of stars in all directions \citep{SS53,Wielen77,Lacey84}.
This picture can be expanded to include direct scattering and resonant effects from spiral arms \citep{BW67,CS85} and bars \citep{2010ApJ...721.1878S,2016AN....337..949F},
where the non-linear response of orbits to overlapping patterns and their resonances \citep{MQ06,Minchev11,Minchev12,Daniel19} can cause strong \rrn.
Even substructure in the \ISM can induce significant \orn \citep{2026arXiv260521579M}.
In addition to internal, secular processes, cosmologically driven environmental effects, such as bombardment by  satellite galaxies \citep{2009MNRAS.397.1599Q,BKW12,Grand16, 2022MNRAS.516.5067C} and accretion of misaligned gas \citep{Khachaturyants22a}, can redistribute disk orbits.

The only physical process that can change orbital \acrlong{lz} (mean orbital size) without simultaneously altering orbital \acrlong{ec} and/or vertical motion \citep{SB02} is `\cting' \citep[this term was coined by][]{Daniel19}.
This happens when a star is trapped in a stable orbit about the \CR resonance of a transient spiral pattern \citep{SB02}. 
\Acrlong{ct} can thus efficiently redistribute orbital angular momenta without  without \kmy heating a stellar population and so there are no significant kinematic signatures of past episodes.

Efficient \cting has been invoked as a contributing mechanism for the formation of global structures in disk galaxies.  
Stars torqued to orbits with mean radius beyond the star forming edge of a disk \citep{Roskar08a,Roskar12} could produce an outer disk of older stars, thus reproducing the observed `U' shaped mean stellar age profile \citep{Roskar08a,RS12,2017MNRAS.467.5022H,DRL17,2026A&A...708A.252F}.
Outward dominated \cting has also been invoked as a genesis mechanism for a thickened disk component \citep{SB09b,Loebman11}. 

Efficient \cting may best be observed using \CCD measures.  
It would flatten any pre-existing radial \Fe gradient \citep{Loebman16,2025ApJ...991..139G}  
and could be responsible for weakening the  stellar \AMR in the \SNd, in agreement with the large scatter in the observed relation  \citep{Edvardsson93, Casagrande11,2024A&A...690A.147H}.
However, the same torques responsible for \cting of stars would also mix metals in the \ISM, thus washing out any (growing) radial gradient in the gas from which the stellar \Fe gradient would have arise in the first place \citep{2021MNRAS.505.4586B,2026arXiv260407076K}.
The presence of stars with solar and super-solar \Fe in the \SNd that are old \citep{1997AJ....114..376C} or have low \acrlong{ec} \citep{Kordopatis15}  may be a smoking gun signal for \cting since these stars are otherwise unexpected \citep{Grenon87,WS89}.
Indeed, the nearly Solar \Fes of young stars and star forming regions in the \SNd, was one of the first indications that the Sun itself likely migrated some \kpc{s} from the inner disk \citep{WFD96,Frankel18,2025arXiv251209987Z}.
\Acrlong{ct} could be responsible for the observed radial dependence of the \aFe bimodality, where stars trapped at \CR with a slowing bar would have traveled outward from the inner disk in this scenario \citep[e.g][]{2025ApJ...983L..10Z_AMR}.

In practice, it is nearly impossible to separate the effects of \cting from other forms of \rrn since investigations, whether simulated \citep[e.g.][]{VC14,VCdON16,2025A&A...701A..88M,2026A&A...706A..31M,2026A&A...707A.147B} or observed \citep[e.g.][]{Hayden15,2022A&A...663A..38K,Kordopatis15,2024MNRAS.535..392L}, necessarily include all mechanisms for \rrn at once.

Based on the assumption that stars are born in nearly circular orbits, one might appeal to the \AVR as a constraint on the degree of \cting versus \kmy heating forms of \rrn.  
However, how the initial stellar velocity dispersion and  subsequent heating of a population depends on redshift is not well constrained.
Old stellar populations were born during an epoch when \MW progenitors were thicker, clumpier, more turbulent, and highly perturbed by satellite bombardment, gas accretion, and feedback \citep[e.g.,][]{2012ApJ...758..106K,2022MNRAS.514..689B}. 
It has been proposed that stars formed under these conditions may have been born on \kmy hotter orbits than in the present day \citep{2021MNRAS.503.1815B,2024MNRAS.527.6926M,2025A&A...700A..89K}. 
To complicate the matter, the \JWST enabled discovery of massive, well-established galaxies at ultra-high redshift \citep[e.g.][]{2022ApJ...940L..14N,2023ApJ...946L..13F,2023Natur.616..266L,2023ApJ...945L..10G,2023Natur.623..499C,2023ApJ...958L..26H,2024MNRAS.535.2068R,2024ApJ...968L..15K}, which is not in agreement with predictions from contemporary theoretical $\Lambda$CDM models.  Followup studies of disk galaxies at cosmic noon show little measurable evolution over these timescales \citep{2025A&A...700A..42E,2026A&A...709A.120J}. 
\cite{2026A&A...709A.120J} appeals to a high cold gas fraction at high redshift as a possible solution.  
A hydrodynamical cosmological simulation, PHOEBOS, that prescribes a weak stellar feedback model closely reproduces galaxy demographics at cosmic dawn \citep{2025MNRAS.543.2760V}.
This line of inquiry is supported by a study of the multi-phase gas kinematics in galaxies from $z{\sim}0.5-8$, showing minimal evolution, where the authors \citep{2025MNRAS.544.2777W} suggest that kinematics based on ionized gas dispersions may overestimate that of the molecular gas phase by a factor of two. 



In a detailed case study, \cite{Wiggins25} followed the individual orbital trajectory of a massive star cluster in a \MW-like galaxy from the Feedback In Realistic Environments (FIRE; \citealp{Hopkins18}) simulation suite. This cluster migrated multiple scale radii across the disk and into a vertically extended orbit.  They showed that the orbit evolved from a combination of mechanisms, primarily \cting from spiral arms and a satellite interaction, causing episodes of kinematic heating, cooling, and cold \orn. 
This study could provide an explanation for the observed population of old, high-\FeH star clusters high Galactic latitudes and large \Gc radii in the \MW \citep{2022AJ....164...85M}.
It also underscores that statistically improbable orbital outcomes should not be treated is impossible.

Any robust model for disk galaxy evolution should be able 
to predict the efficiency of \cting and isolate observable characteristics that can be utilized to test the predictions.  Such models are the foundation from which the effects of \cting (only changes in \Lz) can be distinguished from all other forms of \rrn (changes in both \Lz and non-circular motions). To that end it is useful to consider the physical factors that set the efficiency of \cting{}, and thus its importance to disk evolution.

An episode of \cting is more efficient when a larger fraction of disk stars are trapped at \CR.
A star's \iLz is the primary orbital property governing whether or not it can be trapped at the \CR resonance, with some contribution from its non-circular motions \citep{DW15}, where the range of \iLz that can be trapped is greater for stronger spiral patterns \citep{Contopoulos78,BT87,SB02,DW15}.  Further, the fraction of trapped stars decreases with increasing kinematic temperature \citep[i.e.,~increasing \sigmaR;][]{DW15,DW18,2020MNRAS.495.3295M}.

\Acrlong{ct} is also more efficient when larger changes in \acrlong{lz} \DelLz are possible, where the final effect is often measured as changes in \Rg, since $\Delta L_z \propto \Delta R_g$ in a disk with a flat rotation curve.
The maximum change in \Rg is predicted to scale as the square root of the strength of the perturbing potential \citep{BT87,SB02}, 
\begin{equation}\label{e:maxRgbar}
    \max(\Delta R)~\propto \sqrt{\Phi_1}.
\end{equation}
However, this scaling relation depends on a few key assumptions.
First, the perturbing body is presumed to be in rigid rotation such that its pattern speed is radially independent, $\Omega_p(R)\equiv\Omega_p$.  The derivation for the relation given in \eq~\ref{e:maxRgbar} also relies on the perturbation's shape being radial only, in other words, a bar pattern.  Given that transient spiral patterns, which are clearly \azi{ly} dependent, are presumed to drive \cting it is unclear to what degree \eq~\ref{e:maxRgbar} should be expected to hold in disk galaxies.  

The surface brightness, multiplicity, radial extent, and opening angle of spiral patterns vary widely in nature \citep[see review by][]{2022ARA&A..60...73S}.  
Flocculent patterns, such as those imaged by \JWST in the \textit{Physics at High Angular resolution in Nearby GalaxieS} survey \citep[PHANGS;][]{phangs_2024} are well reproduced by the FIRE-2/\textit{Latte} \citep{Wetzel16} suite of zoom-in simulations.  These patterns are consistent bright star forming regions that may shear over time and not necessarily with high amplitude perturbations \citep[e.g.][]{Elmegreen11}.
On the other end of the spectrum are high amplitude grand design spirals.  These are expected to be long lived \citep{Struck11} where simulated analogs show pattern speeds that change in time \citep{Roskar12,Quinn26}.

The time- and radial-dependence of a spiral's pattern speed is tied to the physics that determines its nature. 
Analytic descriptions of \cting \citep{SB02,DW15} are extensions of theories that assume a bar structure, which is, by construction, time-independent in an appropriately selected rotating frame \citep{LBK72,BT87}.
Density wave theories \citep{LS64} also allow for time-independence in a rotating frame and so offer an starting point for further exploration.
\cite{DW15} found that a spiral pattern speed's radial rate of divergence from the underlying rotation curve is a mathematical analog to the radial range of \iLz that can be trapped at \CR.  This relation predicts that spirals with radially independent pattern speed have the narrowest range of \iLz for \cting while a \crting spiral, such as would be expected from a sheared or swing-amplified spiral \citep{Toomre64,1981seng.proc..111T,GKC12, 2013ApJ...766...34D}, would drive \cting across the radial range of the spiral.
High-resolution, cosmological simulations suggest that there is a wide diversity in the nature of spiral structure even in a given simulation suite \citep{Quinn26,2026MNRAS.548ag774G}.

To date, no study has explored how spiral morphology and time-dependence may impact the efficiency of \cting.  This paper explores these topics.  
In \S\ref{s:Theory} a review of the analytic underpinnings is provided, followed by an intuition-building exercise in \S\ref{s:R1intuition}, together with an analytic treatment illustrating how changes in angular momentum depend on spiral opening angle \S\ref{s:R1spiral}. 
\S\ref{sec:Numerical} presents a numerical exploration of 2D and 3D density-wave models, as well as a 3D model with a radially dependent pattern speed such that is \crt{s}   with the disk, hereafter called a \lq winding' spiral (as used by \cite{2019MNRAS.490.1026H}, but sometimes also called a \lq dynamic' spiral \citep[e.g.,][]{2013A&A...553A..77G,2026MNRAS.548ag774G} in the literature).  
The discussion in \S\ref{s:Discussion} considers  observational constraints (\S\ref{s:Observations}), 
the nature of spiral structure (\S\ref{s:SpiralNature}),
and the resonant kinematic heating and cooling that can be associated with \cting (\S\ref{s:Cooling}).
Conclusions are stated in \S\ref{s:Conclusions}. 


\section{Theoretical Exploration of Radial Excursions}\label{s:Theory}


\Acrlong{ct} is a resonant process.  In general resonances occur when a natural frequency of a system is commensurate with a forcing frequency.  In a spiral galaxy, this occurs when the natural frequency of an orbit, 
\begin{equation}\label{e:frequencies}
    \vec{\Omega} =\omega_R\hat{R} +\omega_\phi \hat{\phi} +\omega_z\hat{z},
\end{equation}
is commensurate with the forcing from a spiral pattern that has pattern speed $\Omega_p$ and $m$-fold azimuthal symmetry.  This occurs when,
\begin{equation}
    \vec{\Omega} \cdot \vec{l} = m\,\Omega_p ,
\end{equation}
where the vector $\vec{l}$ is described by,
\begin{equation}\label{eq:lvec}
    \vec{l} = l_R\hat{R} +l_\phi\hat{\phi} +l_z\hat{z},
\end{equation}
and has integer value coefficients.  Any given spiral pattern will have multiple resonances, but \cting specifically occurs near the \CR resonance,
\begin{equation}
    \omega_\phi = \Omega_p,
\end{equation}
where the orbital frequency of a star  $\omega_\phi$ equals that of the spiral pattern.  
Near the \CR resonance, stars may be trapped in orbits that librate radially about the \RCR and \azi{ly} between spiral arms.  Trapped orbits can librate about \CR indefinitely unless the spiral pattern is transient \citep{SB02}.  In such cases, the orbital \Lz of a trapped orbit can be permanently changed in this process called \cting.

Analytic explorations of \cting are based on a 2D weak bar potential and otherwise circular orbits \citep{BT08,SB02}.  The applicability of these solutions to \cting from spiral arms will be investigated later in this paper. Here, we briefly summarize some important points. 

\subsection{Radial Excursions from a Weak Bar}\label{s:R1bar}

Assume a small perturbation $\Phi_1(R,\phi)$ to an underlying \axisym\ potential $\Phi_0(R)$, such that the perturbed potential has the form
\begin{equation}
    \Phi(R,\phi) = \Phi_0(R) + \Phi_1(R,\phi)
\end{equation} 
and $\left| \Phi_1/\Phi_0 \right| \ll 1$.  This implies that the radial and \azi\ positions of a star in this perturbed potential will a have a $1^{st}$ order corrections such that, 
\begin{equation}\label{eq:Rt}
    R(t) = R_0 + R_1(t),
\end{equation}
and
\begin{equation}\label{eq:phit}
    \phi(t) = \phi_0(t) + \phi_1(t).
\end{equation}

An expression for radial excursions, $R_1$, from the unperturbed radial position, $R_0$, is derived in \cite{BT08} by solving for the equations of motion in a 2D disk perturbed by a weak bar with the form, 
\begin{equation}\label{eq:Phib}
    \Phi_{1,b}(R,\phi)=\Phi_{b}(R)\cos(m \phi).
\end{equation}
Solutions to the equations of motion for this scenario are 
periodic functions of the 
\azi\ position, $\phi$, which is taken to be in a frame that rotates at the bar pattern speed $\Omega_p$, hereafter called the rotating frame.
 
The \glsfirst{rcr} is where the unperturbed orbital frequency, $\Omega_0$, equals that of the perturbing pattern speed, $\Omega_p$. In the linear solution, as $R_0\rightarrow R_{\rm CR}$, the amplitude of radial excursions goes to infinity and linear perturbation theory fails.  A more careful treatment reveals that singularities do not exist at the \CR\ resonance \citep{Hagihara72} and several studies have identified stable orbits trapped near \CR\ \citep[e.g.][]{Contopoulos78,GT82,SB02,DW15}.    
These trapped orbits librate about local maxima in the effective potential $\Phi_{\rm eff}$, which is the potential in the rotating frame.

\begin{figure}
\begin{center}
88\includegraphics[width=\columnwidth]{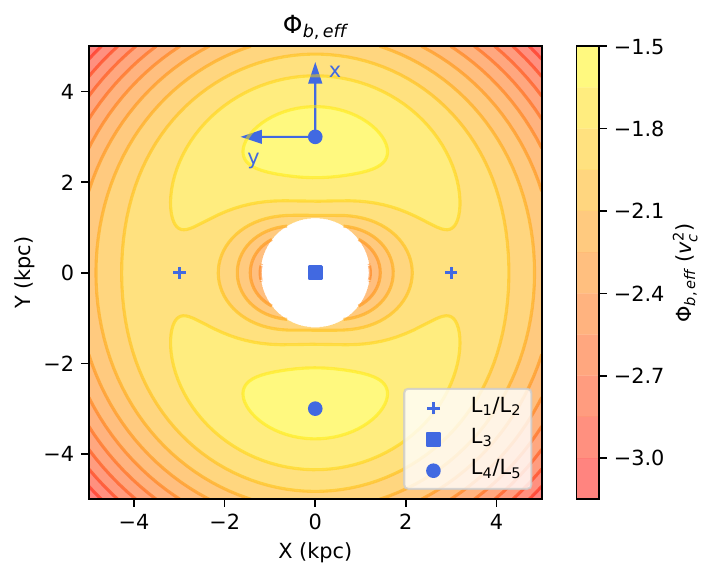}
\caption{Illustration showing the effective potential, $\Phi_{\rm b,eff}$, of a bar.  The bar is horizontally oriented and contours are normalized by the constant circular velocity, $v_c$. 
The stable local maxima (L$_4$ and L$_5$, circles), unstable saddle points (L$_1$ and L$_2$, plus signs), and stable global minimum (L$_3$, square) are labeled.}
\label{fig:PhiEff_Bar}
\end{center}
\end{figure}

Figure~\ref{fig:PhiEff_Bar} illustrates the effective potential for a bar. 
The local maxima, known as the Lagrange points L$_4$ and L$_5$, are labeled.
Assuming a coordinate system where the bar passes through $\phi = 0$ and $\pi$, the
L$_4$ and L$_5$ are located at $R=R_{\rm CR}$ and $\phi=\{\pi/2,3\pi/2\}$, respectively. 
The locations of other Lagrange points are also indicated, where
L$_1$ and L$_2$ are saddle points, and L$_3$ is the minimum at the galactic center.

Equations of motion for orbits that librate about the L$_4$ and L$_5$ can be found by expanding the effective potential around these coordinates.
Let a new frame be defined with origin at the L$_4$, where the $x$-axis points away from the galactic center and the $y$-axis is parallel to the bar.  The orientation of this new coordinate system is shown in Figure~\ref{fig:PhiEff_Bar}.  
Motions about the L$_4$ can be generalized to describe motions about the L$_5$, or any equivalent local maximum from a given weak pattern with $m-$fold \azi\ symmetry.  For the rest of this paper we therefore use the notation \Lag\ to indicate a local maximum in this scenario.

Standard stability analysis reveals that the trajectories of stable orbits about an \Lag\ can be described as a superposition of ellipses \citep[see][p.181-184]{BT08}.  
The smaller of these is associated with \epi\ motion and can be ignored for the purposes of this analysis.  
The more significant ellipse determines the guiding center of an orbit's libration about an \Lag.
The guiding trajectory for this symmetry is nearly symmetric over the $x-$ and $y-$axes and so the axis ratio for a stable orbit can be expressed \citep[\eq~3.126a]{BT08},
\begin{equation}\label{eqn:Y1X1}
\left|\dfrac{Y}{X}\right| = \dfrac{\frac{\partial^2}{\partial x^2}\Phi_{1,b} - \omega^2}{2\Omega_p \omega},
\end{equation}  
where $\omega$ is the frequency of oscillation.

In the limit of a weak bar with perturbing potential amplitude $\Phi_{1,b}\propto\varepsilon$, where $\varepsilon\rightarrow 0$, it can be argued that it is reasonable to assume \citep[][their \eq~3.133]{BT08},
\begin{equation}\label{eqn:omegaapprox}
    \omega^2 = 2\varepsilon \Omega_p^2 = -\frac{\partial^2}{\partial x^2}\Phi_{1,b}.
\end{equation}
As such the guiding ellipse axis ratio gives $Y \sqrt{2\varepsilon} = X$. This relation indicates that it is elongated in the \azi\ direction such that $\phi_1\propto 1$ and $R_1\propto\sqrt{\varepsilon}$.  
The order of time derivatives can be approximated by multiplying by the frequency $\omega\propto \sqrt{\varepsilon}$ (from \eq~\ref{eqn:omegaapprox}).   
These approximations can be used to solve the equations of motion by matching terms of equal order in $\varepsilon$ \citep{GT81,BT08}.

The general solution to the equations of motion obeys,

\begin{equation}\label{eqn:BTR1}
    R_{1,b}^{\rm sym} = \pm\dfrac{\sqrt{2} R_0 \Omega_0}{(\kappa^2 - 4\Omega_0^2)}\sqrt{E_p+ p^2 \cos(2 \phi_1)}
\end{equation}
where $R_{1,b}^{\rm sym}$ and $\phi_1$ are the radial and \azi\ excursions from \Lag, respectively, and the superscript in $R_{1,b}^{\rm sym}$ indicates that this approximation assumes symmetry about the \Lag\ in the $x{-}y$~plane (see arguments around \eq s~\ref{eqn:Y1X1}-\ref{eqn:omegaapprox}. 
The terms $E_p$ and $p$ can be described as the total and potential energy in this transformation.
The amplitude of the potential associated with the oscillatory behavior, $p$, is given by \citep[\eq~3.157]{BT08}, 
\begin{equation}
    p^2 \equiv \frac{4}{R_0^2}|\Phi_b|\frac{4\Omega_0^2-\kappa^2}{\kappa^2},
\end{equation}
and the total energy associated with the oscillatory behavior by,
\begin{equation}
    E_p=\frac{1}{2}(m\dot{\phi_1})^2-p^2\cos(m\phi_1).
\end{equation}

\cite{SB02} use this derivation, in conjunction with horseshoe orbit theory, to describe the radial excursions of stars near \CR.  The great realization of their analysis is that these radial excursions can lead to permanent orbital changes in the presence of \textit{transient} spiral structure.  Assuming stars are in an extreme trap (i.e.,~$E_p \ll p^2$), \cite{SB02} let $E_p\rightarrow0$ in \eq~\ref{eqn:BTR1} and find that the maximum first order radial excursion from an \Lag\ is proportional to the square root of the amplitude of the perturbing potential given by,
\begin{equation}\label{eqn:SBR1}
    \max(R_{1,b}^{\rm sym}) = \sqrt{\dfrac{|\Phi_b|}{-AB}},
\end{equation}
where $A$ and $B$ are Oort's constants for galactic shear ($A \equiv -\frac{1}{2} R \frac{d\Omega}{dR}$) and vorticity ($B \equiv -(\Omega + \frac{1}{2}R\frac{d\Omega}{dR})$). 

The above review of the derivation leading to an expression (\eq~\ref{eqn:SBR1}) for the maximum radial oscillations of orbits trapped at \CR makes two critical assumptions.  First, the radial excursions from \Lag\ are for otherwise circular orbits. In other words, radial excursions due to orbital eccentricity are typically treated as adding a smaller ellipse to represent \epi\ motion.  The role of such non-circular motion in whether or not an orbit can be trapped near \CR\ is examined by \cite{DW15}.

\begin{figure}
\begin{center}
\includegraphics[width=\columnwidth]{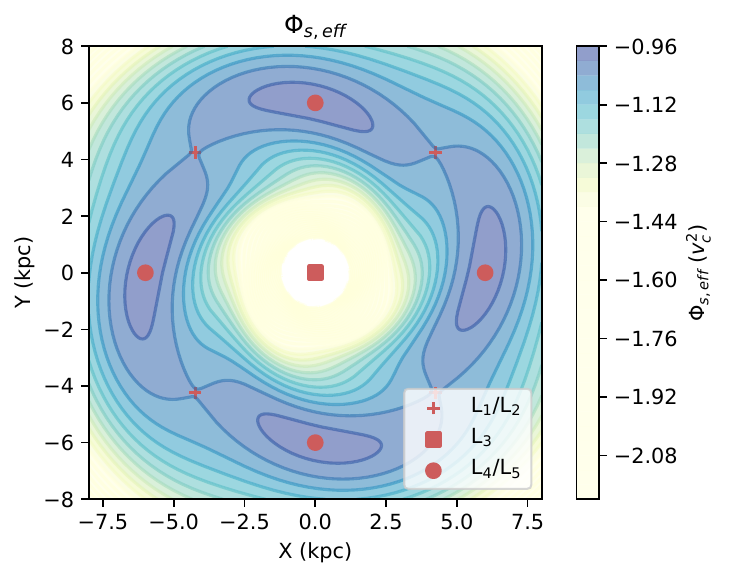}
\caption{The effective potential of a spiral pattern, $\Phi_{\rm s,eff}$, with \pangle $\theta=25^\circ$. A comparison with the effective potential for a bar in \f~\ref{fig:PhiEff_Bar} illustrates how the shape of the contours that close around the \Lag{s} diverges from those of a bar symmetry.}
\label{fig:PhiEff_Spiral}
\end{center}
\end{figure}

The second important assumption, and the assumption that will be examined in this paper, is that the shape of the potential that governs the orientation of the elongation of the primary ellipse oriented around an \Lag\ is symmetric over the $x-$ and $y-$axes.  
A comparison of the effective potential for a spiral pattern, shown in \f~\ref{fig:PhiEff_Spiral} with the effective potential of a bar pattern (\f~\ref{fig:PhiEff_Bar}) illustrates the spiral's divergence from bar symmetry.
The bar symmetry assumption propagates through the equations of motion to the solution found in \eq~\ref{eqn:BTR1} and maximized in \eq~\ref{eqn:SBR1}.  This assumption is appropriate for a bar, where the radial and \azi\ dependencies are separate, but becomes less appropriate as the shape of the potential diverges from this symmetry as is the case for a spiral pattern.

\subsection{Radial Excursions and Morphology}\label{s:R1intuition}

The radial excursions of an orbit that is trapped near \CR\ are caused by the torques exerted on that orbit by the non-\axisym\ perturbation.  Near \CR with a spiral pattern, a star that has instantaneous unperturbed radius inside \CR, $R_0<R_{\rm CR}$, will have an orbital frequency greater than the pattern speed of the \perturber, $\Omega_0>\Omega_p$. As the star encroaches on the lagging side of the spiral arm it experiences a torque that increases it orbital angular momentum. Assuming a flat rotation curve, an increase in angular momentum implies an increase in radial coordinate.  Once the star's radial coordinate is greater than the radius of \CR, $R_0>R_{\rm CR}$, its orbital frequency is less than that of the \perturber, $\Omega_0<\Omega_p$, and the star eventually experiences a torque in the opposite direction causing it to move inward.  This picture provides a physical intuition that the amplitude of the torque is closely related to the amplitude of excursions.

In order to build intuition for how the morphology of a \perturber could influence radial excursions, consider two extreme cases: a bar and a spiral pattern wound so tightly it approaches a ring shape.  
In this study, we define the openness of a spiral pattern by its \pangle, $\theta$, which is the angle between a circle at the radius of \CR and the trailing side of a spiral pattern.
By this definition, a tightly wound spiral has \pangle approaching zero and a bar has $\theta_{\rm bar}=90^\circ$. 

In the case of a bar (summarized in \S\ref{s:R1bar}) the perturbing mass is distributed radially.  Given that torque per unit mass is given by $\vec{\tau}= \vec{r} \times \vec{F}$ it follows that the torque is maximized when the direction of the force from the \perturber is perpendicular to the vector pointing radially to the star being torqued.  In other words, changes in orbital angular momentum, for a given amplitude of the \perturber, are maximized for a bar.

The other extreme case is that of a tightly wound spiral in the limit that it becomes a ring. 
In this limit, for a \perturber\ of a given amplitude, the torque approaches zero since the direction of the force approaches the radial direction.  Thus it might be  expected that lower \pangle spiral patterns would drive smaller radial changes about the \CR\ radius. 

\subsection{Radial Excursions from a Spiral Pattern}\label{s:R1spiral}

\subsubsection{Preliminaries}
In order to explore the relationship between spiral morphology and radial excursions near \CR\  we adopt a Lin-Shu prescription for a spiral perturbation \citep{LS64, LYS69}.
The reasons for this choice are as follows: (a) it provides an analytically accessible calculation for radial excursions as a function of \pangle $\theta$ (see below and Appendix~\ref{s:appendix}); (b) $\Omega_p$ is radially- and time-independent and so can be compared directly to current predictions based on a time-independent bar-like pattern (\S\ref{s:R1bar}); (c) the \pangle $\theta$ is radially-independent; and, (d) most of the torque causing these radial excursions comes from peak spiral amplitude, which is prescribed to be at \CR.  We assert that it is reasonable to assume that any spiral with peak density at \CR\ and radially-independent pattern speed will follow similar scaling laws.

The Lin-Shu prescription for a perturbing spiral potential is,
\begin{equation}\label{eqn:Phis}
    \Phi_{1,s}(R,\phi,t) = \Phi_s\cos \left[\alpha \ln R/R_{\rm CR} + m\Omega_p t - m\phi \right],
\end{equation}
where $\alpha = m \cot \theta$.  
The amplitude of the perturbing spiral potential is given by,
\begin{equation}\label{eqn:PhisAmp}
\Phi_s(R) = \dfrac{2 \pi G \Sigma(R) \epsilon_\Sigma}{k}
\end{equation}
where $\Sigma(R)$ is the surface density of the unperturbed disk, $\epsilon_\Sigma$ is the fractional amplitude of the surface density, and the radial wavenumber is \citep[][eqn. 6.7]{BT08},
\begin{equation}\label{eq:wavenumber}
	k = \dfrac{\alpha}{R} = \dfrac{m \cot \theta}{R}.
\end{equation}

\begin{figure*}
\centering
\includegraphics[width=\textwidth]{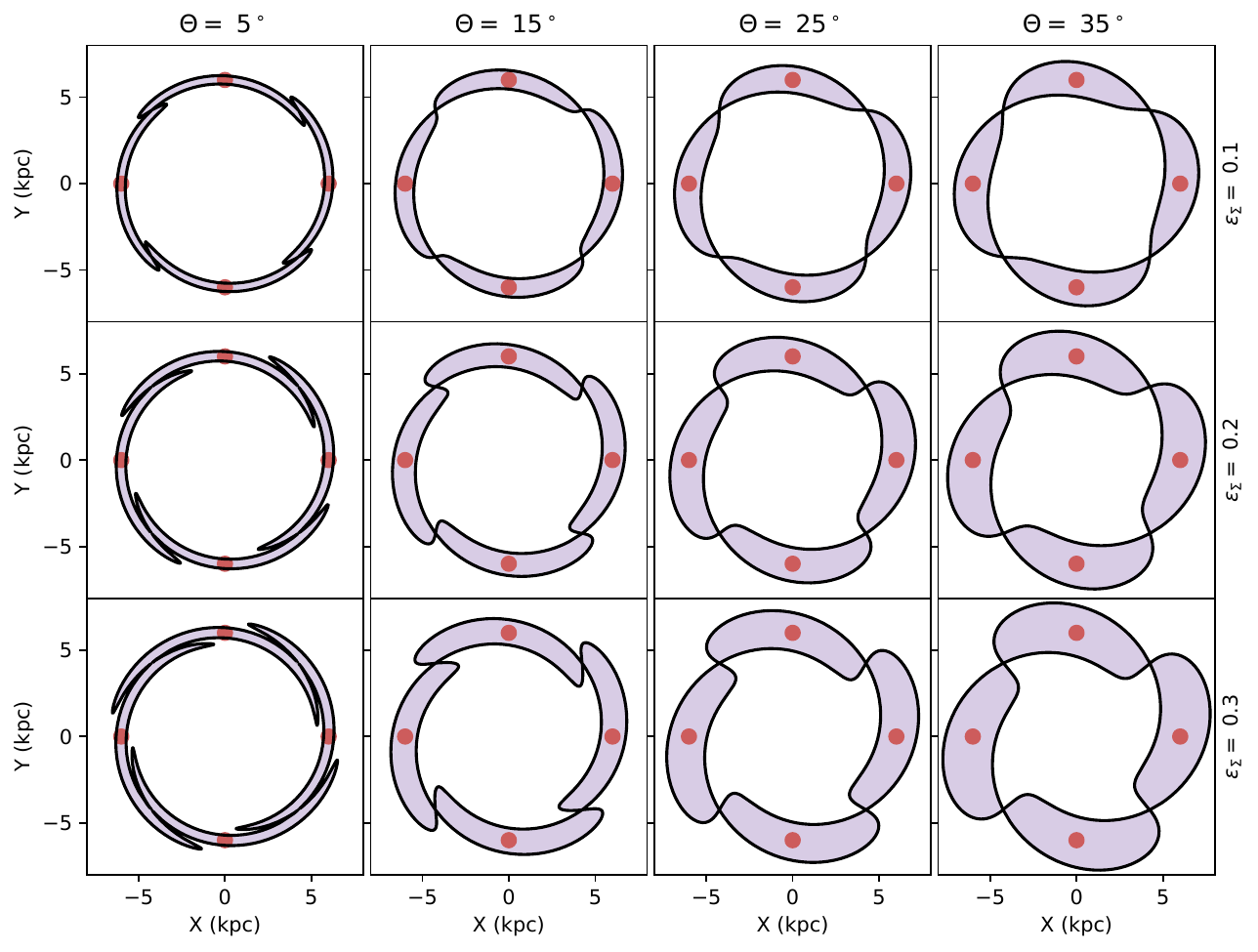}
\caption{Illustration showing the invariant curves $\Phi_{\rm s,eff} = \Phi_{\rm s,eff}({\rm L}_1)$ for an $m=4$ spiral pattern (\eq~\ref{eqn:Phis}) in a Mestel disk.
Positions $\phi=\{\pi/4, 3\pi/4, 5\pi/4, 7\pi/4\}$ correspond to the unstable L$_{1/2}$ saddle points in $\Phi_{\rm s, eff}$ (separatrixes) and are the location of the spiral arms at \RCR$=8$~kpc.
The stable \Lag local maxima (red dot) have azimuthal positions $\phi=\{0, \pi/2, \pi, \pi/2\}$ and are the points around which stars trapped at the \CR resonance librate.  
Panels from left to right show spiral patterns with \pangle{s} $\theta=\{5^\circ,15^\circ,30^\circ,45^\circ\}$ 
and from tip to bottom show spirals with fractional amplitudes $\epsilon_\Sigma=\{0.1,0.3,0.5\}$.  
The size and shape of the region enclosed by the $\Phi_{\rm s,eff} = \Phi_{\rm s,eff}({\rm L}_1)$ curve (shaded purple) roughly corresponds to the largest libration trajectory for trapped stars. Thus the degree of \cting increases for stronger, more open armed spiral patterns.
For all spiral strengths, the degree of \cting approaches zero (\RMSL$\rightarrow 0$) in the limit that the spiral pattern approaches a ring ($\theta\rightarrow0$).}
\label{fig:Equipotential}
\end{figure*}

The equations of motion for stars trapped around the \Lag{s} of a spiral pattern are derived by following the same procedure outlined in \S\ref{s:R1bar}.  The Lagrangian,
\begin{equation}
	\mathcal{L} = \dfrac{1}{2}\dot{R}^2 +\dfrac{1}{2}[R(\dot{\phi} +\Omega_p)]^2 - \Phi(R,\Phi),
\end{equation}
is evaluated in a frame that rotates with the spiral pattern at frequency $\Omega_p$.
As with a bar pattern, a star trapped oscillating about an \Lag\ will have time-dependent radial and \azi\ coordinates given by \eq{s}~\ref{eq:Rt}-\ref{eq:phit}, where unperturbed \azi\ coordinate $\phi_0 = (\Omega_0 - \Omega_p)t$.  
Combining the definitions for the \epi\ frequency evaluated at radius $R_0$,
\begin{equation}\label{eqn:epicyclic}
\begin{array}{cl}
    \kappa^2 & = \left( \dfrac{\partial^2\Phi_{\rm eff}}{dR^2} \right)_{R_0} \\
       & = \left(\dfrac{\partial^2 \Phi}{\partial R^2}\right)_{R_0} + \dfrac{3L_z^2}{R_0^4},
\end{array}
\end{equation}
where $L_z$ is the angular momentum in the $z$-direction, and orbital frequency,
\begin{equation}\label{eqn:Omega}
    \Omega^2 = \dfrac{1}{R} \left( \dfrac{\partial \Phi}{\partial R}\right)_{R_0} = \dfrac{L_z^2}{R_0^4}
\end{equation}
allows the equations of motions to be expressed as, 
\begin{equation}\label{eqn:motionLspiralR1}
	\ddot{R} + (\kappa^2-4\Omega_0^2) R_1 -2 R_0 \Omega_0 \dot{\phi_1} =-\left(\dfrac{\partial \Phi_1}{\partial R}\right)_{R_0}
\end{equation}
and
\begin{equation}\label{eqn:motionLspiralphi1}
	\ddot{\phi_1} + 2 \Omega_0 \dfrac{\dot{R_1}}{R_0} 
=  -\dfrac{1}{R_0^2} \left(\dfrac{\partial \Phi_1}{\partial \phi}\right)_{R_0} .
\end{equation}
The derivatives on the right hand sides of equations~\ref{eqn:motionLspiralR1} \&~\ref{eqn:motionLspiralphi1} for the perturbing spiral potential (\eq~\ref{eqn:Phis}) are give by,
\begin{equation}\label{eqn:dPdR}
\left(\dfrac{\partial \Phi_{1,s}}{\partial R}\right)_{R_0}
 = -\Phi_s k\,\sin[\alpha \ln R/R_0 + -m\phi(t) +m \Omega_p t]_{R_0}
\end{equation}
and
\begin{equation}\label{eqn:dPdp}
\left(\dfrac{\partial \Phi_{1,s}}{\partial \phi}\right)_{R_0} 
 = \Phi_s\, m\sin [\alpha \ln R/R_0 + -m\phi(t) + m \Omega_p t]_{R_0}.
\end{equation}

\subsubsection{Assuming Symmetry About \Lag}

The shape of the potential near a \Lag\ from a low \pangle spiral is, as with the bar, nearly symmetric over the $x-$ and $y-$axes.
In the limit that $\theta \rightarrow 0$ the second derivative of the perturbing potential from \eq~\ref{eqn:Y1X1}, $\frac{\partial^2}{\partial x^2}\Phi_{1,s}\rightarrow \infty$ and thus the ratio $|Y/X| \rightarrow \infty$, indicating a circular orbit for a ring shaped perturbation, as expected from the intuition built in \S\ref{s:R1intuition}.  As $\theta$ increases, the orbit remains greatly elongated in the azimuthal direction.  In words, \textit{as long as the \pangle is small, there can be very little change in the $\hat{R}$ direction}.

In \f~\ref{fig:Equipotential}, invariant zero-velocity contours for $\Phi_{\rm eff} = \Phi_{\rm eff}({\rm L}_1)$ \citep[see][eqns. 3.111-113]{BT08} are shown for four armed ($m=4$) spiral patterns with \pangle{s} $\theta=\{5^\circ,15^\circ,25^\circ,35^\circ\}$ and fractional amplitudes $\epsilon_\Sigma=\{0.1,0.2,0.3\}$ in a Mestel disk.  
The shape of enclosed by these invariant curves roughly corresponds to the trajectory for the largest stable orbits about the \Lag{s}.  Both the spiral amplitude and \pangle independently impact the radial width of the zero-velocity contours closed around the \Lag{s} where these are more narrow for smaller \pangle and weaker spiral amplitude.

By adopting the same order $\varepsilon$ for terms in the equations of motion as for a bar (before \eq~\ref{eqn:BTR1})
we arrive at nearly the same result \eq~\ref{eqn:BTR1},
\begin{equation}
\max(R_{1,s}^{\rm sym}) = \pm\dfrac{\sqrt{2} R_0 \Omega_0}{(\kappa^2 - 4\Omega_0^2)}\sqrt{E_p+ p^2 \cos(2 \phi_1)\tan \theta}
\end{equation}
except that the potential term in the square root is modified by $\tan \theta$.
The maximum radial excursion for a spiral pattern with \pangle $\theta$ can therefore by expressed,
\begin{equation}\label{eqn:SBR1s}
    \max(R_{1,s}^{\rm sym}) = \sqrt{\dfrac{|\Phi_s|}{-AB}\tan \theta},
\end{equation}
where there is both an explicit dependence on spiral \pangle $\theta$ and an implicit dependence in the amplitude of the potential $\Phi_s(\theta)$ (\eq~\ref{eqn:PhisAmp}).  

It can be expected that maximum mean radial excursions for stars trapped at \CR\ can be well approximated by \eq~\ref{eqn:SBR1s} in the limit that the perturbing spiral is extremely tightly wound.  This is because the shape of the perturbed potential is symmetric about the \Lag\ in the $x{-}y$~plane.  As the spiral \pangle opens \eq~\ref{eqn:SBR1s} becomes less appropriate and the equations of motion must be solved explicitly.

\subsubsection{Finding the Radial Equation of Motion}\label{s:EoM}

Symmetry arguments are not valid for orbits trapped near \CR\ of a spiral pattern with a low to moderate \pangle.  We here show that in this case it is possible to solve the equations of motion explicitly.

The $\phi$ directional equation of motion for an orbit in the presence of a spiral pattern (\eq~\ref{eqn:Phis}) is given by \eq~\ref{eqn:motionLspiralphi1}.
The time integral of the left hand side can be expressed,
\begin{equation} \label{eqn:intmotionLspiralphi1}
    \dot{\phi_1} + 2\Omega_0 \dfrac{R_1}{R_0} = - \Phi_s\dfrac{m}{R_0^2} \int dt \sin[-\phi_1(t)].
\end{equation}

In order to solve the integral in \eq~\ref{eqn:intmotionLspiralphi1}, we must assume a form for $\phi_1(t)$.  Stars trapped in stable orbits about a \Lag\ will librate around these local maxima outside the zero-velocity contours in the Jacobi integral (eg., \f~\ref{fig:Equipotential}) in the rotating frame.  Of these stars, those with the greatest radial excursions will most closely approach the local \azi\ minima (eg., between galactocentric coordinates $\phi=\pi/4$ to $\phi=3\pi/4$ in \f~\ref{fig:Equipotential}).  As a first order approximation we here assume \azi\ excursions can be described by,
\begin{equation}\label{eqn:phi1max}
    \phi_1(t) = |\phi_1| \cos(\omega t +\delta)
\end{equation}
where the maximum amplitude for an $m$ armed spiral is $|\phi_1|=2\pi/m$.  
The phase constant can be set to $\delta = 0$ with no loss of generality.    
We note that for spiral patterns that diverge from symmetry about the $x{-}y$~axes so will the shape of the orbit about the \Lag. In these cases the assumed form for $\phi_1(t)$ in \eq~\ref{eqn:phi1max} is less appropriate as trapped orbits will spend more time at larger radii.  

An approximate solution for the radial excursions of an orbit trapped librating about \Lag\ can be found by solving the integral in \eq~\ref{eqn:motionLspiralphi1} and plugging the solution for $\dot{\phi_1}$ into \eq~\ref{eqn:motionLspiralR1} to find $R_1$.  In Appendix~~\ref{s:appendix} a full derivation is given for the case that \azi\ excursions are well described by \eq~\ref{eqn:phi1max}.  
The radial equation of motion for excursions from the \Lag\ can be expressed as,
\begin{equation}\label{eq:R1EoM_pretty}
\begin{array}{cl}
    \ddot{R_1} &+\kappa^2 R_1 \\
    &= \dfrac{\Phi_s}{R_0} \left[m\cot\theta \sin\left(\frac{\cos\tau}{m}\right) +4\pi^2\dfrac{\Omega_0}{\omega} \sin\tau  \right],
\end{array}
\end{equation}
which is simplified from \eq~\ref{eq:R1EoM} to explicitly show model dependencies and where $\omega$ is the frequency of oscillations about the \Lag\ and $\tau = \omega t$.

\begin{figure*}
\centering
\includegraphics[width=\textwidth]{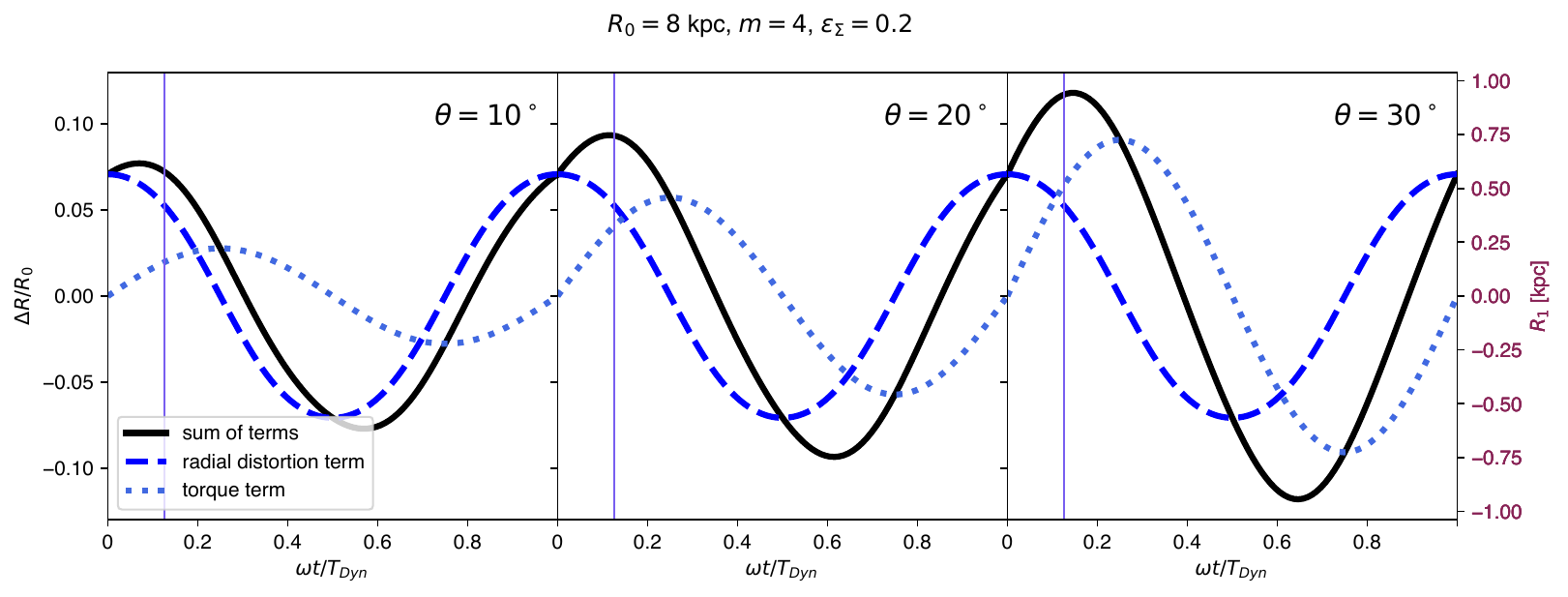}
\caption{Comparison of and total for the contributing terms for radial excursions from \Lag\ from \eq~\ref{eq:R1EoM_pretty}, normalized by the radius of \CR, $R_{CR}=R_0$. These are the term associated with torque from the spiral pattern (light blue, dotted) and the radial distortion term (dark blue, dashed), which are described in \S\ref{s:EoM}.  As \pangle increases the total amplitude of radial excursions (black, solid) increases. A vertical (purple) line marks the amplitude of radial excursions used to approximate radial excursions in \S\ref{s:TSpiralRmax}.
}
\label{fig:R1_terms}
\end{figure*}

The amplitudes of both terms on the right hand side of \eq~\ref{eq:R1EoM_pretty} are time dependent.  This is illustrated in \f~\ref{fig:R1_terms}, where each term is multiplied by the model dependent, time-independent constant $c=\Phi_s/\kappa^2 R_0$ in order to express the values of the curves in appropriate radial units.  The dashed (dark blue) curve shows the contribution from the first term in the right hand side, which arises from the radial derivative of the potential and can be associated with a distortion from an elliptical orbit aligned with the $x{-}y$~axes about the \Lag.  The dotted (light blue) curve shows the contribution to radial excursions originating from the torque and increases with \pangle.  The solid (black) curve shows the sum of these out of phase terms. The maximum radial excursion happens before it reaches the \azi\ coordinate of the \Lag\ as might be inferred from the shape of the zero velocity curves in \f~\ref{fig:Equipotential}.

\subsubsection{Maximum Radial Excursions}\label{s:TSpiralRmax}

This work has considered the three models for the change in mean orbital radius for a star in a trapped orbit about the \Lag\ of a spiral pattern.  These are: 
(1) $\max(R_{1,b}^{\rm sym})$ (\eq~\ref{eqn:SBR1}) is based on the assumption of a bar symmetry and is the most commonly used formula to estimate maximum radial excursions from a transient spiral pattern \cite[e.g.,][]{SB02};
(2) $\max(R_{1,s}^{\rm sym})$ (\eq~\ref{eqn:SBR1s}) is based on a spiral pattern and adopts the same symmetry arguments as for a bar; and,
(3) $R_{1,s}$, which can be found directly from the equation of motion (\eq~\ref{eq:R1EoM_pretty}).

The maximum radial excursion can be obtained from the equation of motion by setting $\ddot{R}=0$. The resulting maximum amplitude of oscillations about the \Lag\ is then expressed as,
\begin{equation}\label{eq:R1longugly}
    |R_{1,s}| = \dfrac{\Phi_s}{\kappa^2 R_0} 
     \max\left[m\cot\theta \sin\left(\frac{\cos\tau}{m}\right)+4\pi^2\dfrac{\Omega_0}{\omega} \sin\tau  \right].
\end{equation}
Numerical methods most accurately solve for the value of $\tau$ when $|R_{1,s}|$ is maximized. However, the maximum amplitude of radial excursions from the \Lag\ for spiral patterns with reasonably small \pangle ($\theta \lesssim 30^\circ$) can be well approximated by adopting the value $\tau=\pi/2m$ for the phase of libration.  The value of radial excursions corresponding to this phase is indicated in \f~\ref{fig:R1_terms} by a vertical (gray, dotted) line. Using this approximation, \eq~\ref{eq:R1longugly} reduces to,
\begin{equation}\label{eq:R1EoMtaumax}
    |R_{1,s}| \approx \dfrac{\Phi_s}{\kappa^2 R_0} 
    \left[ m\cot\theta \sin\left(\frac{1}{\sqrt{2}m}\right) + 2 \sqrt{2}\pi^2\dfrac{\Omega_0}{\omega}  \right].
\end{equation}

\begin{figure*}
\centering
\includegraphics[width=\textwidth]{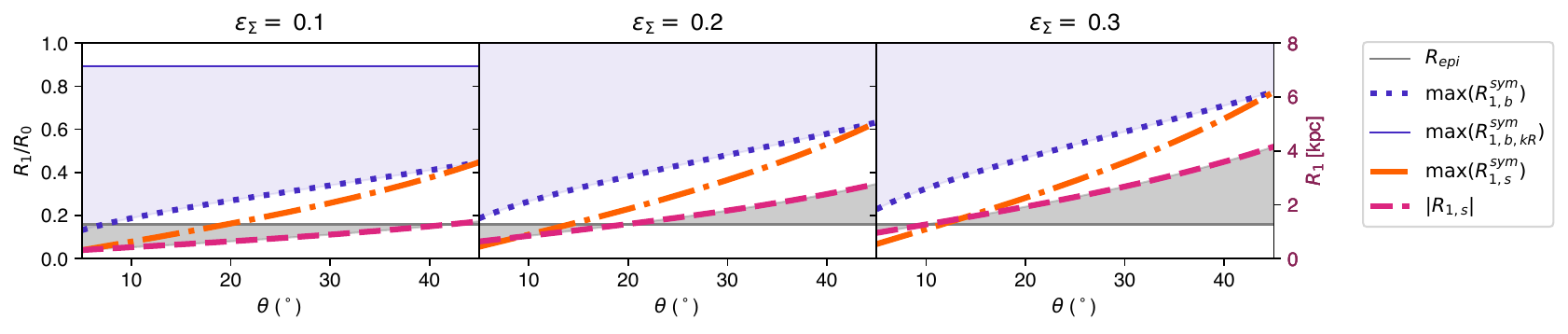}
\caption{Various solutions for the maximum radial excursions of a trapped orbit from the radius of \CR, $R_{CR}=R_0$.  
The model shown is for a Mestel disk with an $m=4$ spiral pattern and fractional amplitude $\epsilon_\Sigma$. 
Maximum radial excursions increase with increasing \pangle $\theta$ in all models.  
The maximum radial excursion using the expression for a bar (purple, dotted; $\max(R_{\rm 1,b}^{\rm sym})$ from \eq~\ref{eqn:SBR1}; \cite{SB02}) exceeds those from the equations of motion (red, dashed; $|R_{\rm 1,s}|$ from \eq~\ref{eq:R1EoMtaumax}) in all cases.
Radial excursions for a spiral, but assuming symmetry arguments as for a bar (orange dot-dashed; $\max(R_{\rm 1,s}^{\rm sym})$ from \eq~\ref{eqn:SBR1s}) show a strong dependence on \pangle going to zero in the tightly wound regime and converging with $\max(R_{\rm 1,b}^{\rm sym})$ as $\theta\rightarrow 45^\circ$.
The amplitude for radial excursions from \epi\ motion is shown (gray, thin, solid; $R_{\rm epi}$) for comparison and exceeds excursions in $|R_{1,s}|$ for models with low to moderate spiral \pangle and strength.
Purple shading shows the difference between the maximum radial excursions using the expression derived under the assumption of bar symmetry ($\max(R_{\rm 1,b}^{\rm sym})$) and the same but modifying the spiral amplitude as $\Phi_{\rm b,kR}=\Phi_s R k$ (purple, thin, solid; $\max(R_{\rm 1,b,kR}^{\rm sym})$) to remove the implicit dependence of $\Phi_{\rm s}$ on \pangle.
}
\label{fig:R1}
\end{figure*}

A comparison of each measure of maximum radial excursions is shown in \f~\ref{fig:R1} for models with \pangle between $5^\circ \leq \theta \leq 45^\circ$ (horizontal axis) and fractional amplitude for the surface density of the perturbation $\epsilon_\Sigma=\{0.1, 0.2, 0.3\}$ (panels left to right).  The vertical axis shows excursions up to the radius of \CR\ since it is un-physical to exceed this range.  Each model uses the same prescription for the spiral potential (\eq~\ref{eqn:Phis}).  For all models, the maximum radial excursions approximated by using $\max(R_{1,b}^{\rm sym})$ from \eq~\ref{eqn:SBR1} (blue, dotted) scale as,
\begin{equation}
    \max(R_{1,b}^{\rm sym}) \propto \sqrt{|\Phi_b|},
\end{equation}
and are almost twice the value when using $|R_{1,s}|$ from \eq~\ref{eq:R1EoMtaumax} (red, dashed), where the maximum radial excursions from \Lag scale as, 
\begin{equation}
    \max|R_{1,s}| \propto |\Phi_s(\theta)| \cot\theta .
\end{equation}
The values estimated when using $\max(R_{1,s}^{\rm sym})$ from \eq~\ref{eqn:SBR1s} (purple, dotted) scale as,
\begin{equation}
    \max(R_{1,s}^{\rm sym}) \propto \sqrt{|\Phi_s(\theta)|\tan \theta}
\end{equation}
and converge to $\max|R_{1,s}|$ at low \pangle and to $\max(R_{1,b}^{\rm sym})$ at high \pangle. 
For comparison, the expected range of \epi\ radial excursions, approximated as \citep[eqn. 3.99]{BT08},
\begin{equation}\label{e:epiX}
    R_{\rm epi}=\sqrt{\dfrac{2 \sigma_R^2}{\kappa^2}}
\end{equation}
where $\kappa^2=2\Omega^2$ in a disk with a flat rotation curve, is shown in gray (solid) with gray shading showing the difference between the amplitude of \epi\ excursions and those from trapping around \Lag\ calculated directly from the equations of motion.  At low to moderate \pangle and spiral amplitude, radial excursions \epi\ motion exceed those from from radial migration at \CR.
Blue shading is included to show the difference between $\max(R_{1,b}^{\rm sym})$ for spiral with amplitude $\Phi_b=\Phi_s$ and $\max(R_{1,b}^{\rm sym})$ with $\Phi_b=\Phi_s Rk$, an approximation that removes the inherent dependence of the amplitude of the potential on \pangle.

For all models considered, radial excursions increase with increasing spiral \pangle.  This is in agreement with the findings of \cite{Papayannopoulos79a,Papayannopoulos79b}, who found that the radial excursions of stars in resonant orbits around local maxima scale with \pangle. 

The maximum radial excursion for an orbit trapped at the \Lag set the upper limit for \cting.  However, most orbits are not in these maximized configurations but rather have first order trajectories that are nested within the trajectory for maximum excursions.  The amplitude of radial excursions for a given star trapped at in libration about the \Lag\ depends on its initial distance from the \Lag.  The greater the radial distance at the azimuth of the \Lag, the larger the overall radial excursion.  
\section{Numerical Exploration of Radial Excursions}\label{sec:Numerical}

This section describes three experiments designed to numerically verify the analytic prediction from \S\ref{s:Theory} that more open spirals should induce larger radial excursions from \cting.  
Each experiment uses tracer particle simulations to evolve the orbits of massless \lq star' particles through a prescribed potential.
Section~\ref{sec:Numerical2D} explores the amplitude of radial changes for a population of stars in a 2D disk perturbed by a Lin-Shu spiral pattern.  This treatment aims to provide a clear verification of the analytic theory.  
Section~\ref{sec:Numerical3D} tests the robustness of the predicted trend using a 3D disk model with an alternate, physically motivated spiral potential.
In contrast to the analytic treatment, which can only approximate the expected \textit{maximum} radial excursion for a single star trapped around the \CR\ resonance, these numerical experiments focus on the \RMS effects of \cting on a population of orbits. 
These population analyses provide insight into how spiral morphology affects the efficiency of \cting in galactic disks.

In \S\ref{sec:3DWinding}, the 3D numerical exploration is modified so that the spiral pattern has radially dependent pattern speed that \crt{s} at all radii $\Omega_p=\Omega_p(R)=\Omega_\phi(R)$.  This experiment provides insight into how the derived scaling relation between spiral \pangle and the amplitude of \cting critically depends on the density wave assumption and underscores the need to better understand the nature of spiral structure.

\subsection{Tracer Particle Simulations}

Tracer particle simulations evolve the orbits of massless \lq star particles' through a prescribed potential.  
The following experiments use an an \axisym underlying potential that is perturbed by a spiral potential.
In each experiment the amplitude of the spiral perturbation is made to be time dependent so that,
\begin{equation}\label{eq:Phist_PhisEpsilont}
    \Phi_s(t) = \Phi_s\, \epsilon(t).
\end{equation}
The adopted prescription for $\epsilon(t)$ has Gaussian form described by,
\begin{equation}\label{eqn:et}
    \epsilon(t) = \epsilon_0\, e^{-(t-2T_{\rm Dyn})^2/2\sigma_t^2},
\end{equation}
so that the peak amplitude occurs two orbital periods, $2T_{\rm Dyn}$, after the simulation begins and the standard deviation is set to $\sigma_t=T_{\rm Dyn}$, where an orbital period is the nominal \Tdyn.
These assumptions ensure slow growth and decay for the perturbation, thus avoiding a non-adiabatic dynamical response.  

Initial phase-space coordinates are selected by sampling an appropriately chosen distribution function for each suite of simulations, 2D (\S\ref{sec:Numerical2D}) or 3D (\S\ref{sec:Numerical3D}).
Sampling for each is done using the \Python based galaxy modeling package \texttt{galpy}\footnote{The \texttt{galpy} package can be accessed at \url{https://github.com/jobovy/galpy} \citep{Bovy15}.}~v1.7. 
Each suite of simulations uses identical initial phase-space positions for each simulation.
The only condition modified for each simulation within the 2D and 3D suites is the assigned spiral pitch angle, which takes a value $\theta= \{10^\circ,20^\circ,30^\circ,40^\circ \}$.  

The winding spiral in \S\ref{sec:3DWinding} is setup to replicate the 3D spiral potential that has $\theta=20^\circ$  (\S\ref{sec:Numerical3D}) at the time when the spiral amplitude peaks, $t_{\rm ref}{=}2T_{\rm Dyn}$.  The shape of the spiral potential is otherwise modified by the \texttt{galpy} wrapper
\texttt{CorotatingRotationWrapperPotential} so that,
\begin{equation}\label{eqn:wrappingwrapper}
    \phi \rightarrow \phi + \frac{v_{\rm c}(R)}{R} \times \left(t - t_{\rm ref}\right) + \phi_{0}
\end{equation}
where $\phi_{0}$ is the position angle at the starting time.

\subsection{2D Model}\label{sec:Numerical2D}

The derivation in \S\ref{s:Theory} predicts that radial excursions around \CR from \cting are greater for more open spiral arms, all other parameters being equal.  It does not assume a particular underlying form for the \axisym potential $\Phi_0$ and the spiral perturbation assumes a Lin-Shu prescription (\eq~\ref{eqn:Phis}).
The analytic prediction is therefore based on a two dimensional model with a specific form for the spiral pattern.  
The numerical experiment presented here aims to closely mirror and validate the analytic model.

The methods for this experiment are the same as those used to build the simulation suite presented in \cite{Daniel19}.  A brief review is given below.

The adopted underlying \axisym\ potential has the form,
\begin{equation}\label{eq:MestelPhi}
    \Phi_0 = v_c^2 \ln(R/R_p),
\end{equation}
where $R_p$ is the scale length for the underlying potential.
The perturbing potential, $\Phi_1(R,\phi,t)$, takes the form described in \eq~\ref{eq:Phist_PhisEpsilont} where the time-independent amplitude for a Lin-Shu spiral potential is given by \eq~\ref{eqn:Phis}.


Initial 4D in-plane phase-space coordinates for $5\times10^4$ star particles are assigned by sampling a distribution function for a warm 2D disk \citep{Dehnen99},
\begin{equation}\label{eqn:fnew}
f_{\rm new}(E,L_z) = \dfrac{\Sigma(R_E)}{\sqrt{2}\pi \sigma_R^2(R_E)} \exp\left\lbrace\dfrac{\Omega(R_E)[L_z-L_c(R_E)]}{\sigma_R^2(R_E)}\right\rbrace,
\end{equation}
where $R_E$ is the orbital radius for a star in a circular orbit with energy $E$, $\Omega(R)$ is the circular frequency at a given radial coordinate, $L_z$ represents orbital angular momentum about the $z$-axis, $L_c(R)$ is the orbital angular momentum for a star in a circular orbit at radius $R$, and $\sigma_R(R)$ is the radial velocity dispersion at radius $R$.  
This 2D distribution function well reproduces many \MW disk properties, including a flat rotation curve \citep[e.g.,][]{Rubin83,Sofue09} and an exponential surface density profile \citep[e.g.,][]{Freeman70,vanderKruit87,Juric08,deJong10}.

The exponential surface density has the form,
\begin{equation}\label{eqn:expR}
    \Sigma(R) = \Sigma_0\, e^{-R/R_\Sigma}, 
\end{equation}
where $R_\Sigma$ is the scale length for the surface density of the disk.
Similarly, the velocity dispersion profile takes exponential form,
\begin{equation}\label{eqn:expsigmaR}
    \sigma_R(R) = \sigma_{R,0}\, e^{-R/R_\sigma}. 
\end{equation}

Scaling constants are set to approximate the \MW disk with the \RCR set to approximate the Solar circle at $R_\odot=8$~kpc.  
Constant values and scalings are given in \tab~\ref{tbl:modelscalefactors}.
Initial positions for the star particles are sampled between $0.7R_{CR}$ ($5.6$~kpc) and $1.4R_{CR}$ ($11.2$~kpc) in order to ensure the region near \CR\ is well sampled.

Orbital trajectories through each potential are integrated using the 2$^{nd}$ order leapfrog orbital integrator described in \citet{DW15}.  

\begin{deluxetable*}{lll}
	\tablecaption{2D Model Scale Constants \label{tbl:modelscalefactors}}
	\tablehead{ 
	    \colhead{Parameter}
	    & \colhead{Value}
	    &  \colhead{Description}}
	\startdata
		$R_\odot$ & 8~kpc & Solar radius\\
		$R_p$ & $1$~kpc & underlying potential scale length\\
		$R_\Sigma$ & {$R_\odot/3=2.7~{\rm kpc}$} & surface density scale length\\
		$R_\sigma$ & $R_\odot=8~{\rm kpc}$ & velocity dispersion scale length\\
		$R_{CR}$ & $R_\odot=8~{\rm kpc}$ & \CR\ radius\\
        $\Sigma_0=\Sigma(R_\odot)$ & 50~M$_\odot$~pc$^{-2}$ & surface density normalization\\
		$v_c$ & $220$~km~s$^{-1}$ & circular velocity\\
        $\sigma_{R,0} = \sigma_R(R_\odot)$ & 0.16$v_c=35.2$~km~s$^{-1}$ & velocity dispersion normalization\\ 
		$\epsilon_0$ & $0.3$ & maximum fractional spiral amplitude\\
		$m$ & $4$ & number of arms\\
	\enddata
	\tablecomments{{The adopted value for $\epsilon_0$ is high \citep[e.g.,][]{RZ95,SJ98,Ma02}, though not unreasonably so. Most important is that it is held constant throughout in order to illustrate the trends in radial excursions around \CR\ for various spiral pitch angles.}}
\end{deluxetable*}

\subsubsection{Orbital Classification Scheme}\label{sec:2DType}

Orbits in the 2D simulations are classified as being trapped in libration about the \CR resonance when they meet the capture criterion proposed by \cite{DW15}.  
That is, a star's orbital energy can be described as,
\begin{equation}\label{eq:Eorbital}
	E = E_{\rm c} + E_{\rm nc}
\end{equation}
where the subscripts indicate that the instantaneous energy is a combination of energy associate with \lq circular' and \lq non-circular' motions.
In the regime where $E=E_{\rm c}$, a star is trapped at \CR when the value \citep[][their equation~17]{Contopoulos78},
\begin{equation}\label{eq:Lambdac}
	\Lambda_c \equiv \left|\dfrac{E_J - h_{\rm CR}}{|\Phi_s|_{\rm CR}} \right| < 1,
\end{equation}
for a spiral potential that has amplitude $|\Phi_s|_{\rm CR}$ evaluated at \RCR, Jacobi energy,
\begin{equation}\label{eq:EJ}
    E_J = E-L_z \Omega_p,
\end{equation}
and where the value Jacobi energy in the unperturbed potential $\Phi_0$  is equal to $E_J=h_{\rm CR}$ when evaluated at \RCR. In this scenario, both of $E_J$ and $h_{\rm CR}$ are conserved.
This formalism can be expanded to include non-circular orbits by evaluating whether or not the expression given by \citep[][their equations 33 \& 34]{DW15},
\begin{equation}\label{eq:Lambdanc}
	\Lambda_{\rm nc,2}(t) \equiv \left| \Lambda_{\rm c} - \left( \dfrac{R_L(t)}{R_{CR}} \right) \left( \dfrac{E_{\rm nc}}{|\Phi_s|_{\rm CR}} \right) \right| < 1,
\end{equation}
is satisfied at any time during the simulation, where the subscript \lq 2' indicates this is an appropriate form for a flat rotation curve as is imposed by our choice for underlying potential (\eq~\ref{eq:MestelPhi}).
The radius
\begin{equation}\label{eq:RL}
	R_L(t) = R(t) \dfrac{v_\phi(t)}{v_c},
\end{equation}
is the time dependent radius that reduces to the guiding center radius in the \epi approximation.  

In this experiment a population of star particles classified as `\TypeCR' indicates that it is composed of orbits with phase-space coordinates that satisfy \eq~\ref{eq:Lambdanc} at some time during the simulation.
In the absence of other influences, it is expected that the \RMS change in a trapped population's orbital angular momentum, \RMSL, will be significant and the \RMS change in non-circular orbital energy, \RMSEnc, will be zero \citep{SB02}.

The \IOLRs and their harmonics are described when the condition,
\begin{equation}\label{eq:Lindblad}
	\kappa = \pm m\,(n+1) \,[\Omega_p-\Omega(R)],
\end{equation}
is satisfied, where $n$ stands for the $n^{th}$ harmonic of the \IOLRs. which have harmonic number $n=0$.
When $n=1$, \eq~\ref{eq:Lindblad} gives the condition for the first harmonics of the \IOLRs, which are the resonances most relevant to the current study.  

\cite{Daniel19} showed that populations composed of orbits that are trapped at \CR and simultaneously meet the first harmonic of the \IOLRs conditions have a significant increases in \RMSEnc.  Such populations may have large values for \RMSL, but their \rrn is not \kmy cold.
The population of orbits trapped at \CR that also meet the first harmonic \IOLRs condition are separately described as `\TypeOverlap.'

\subsubsection{2D Trends with Pitch Angle}\label{sec:2DTrends}

\begin{figure*}[t!]
\centering
\includegraphics[width=18cm]{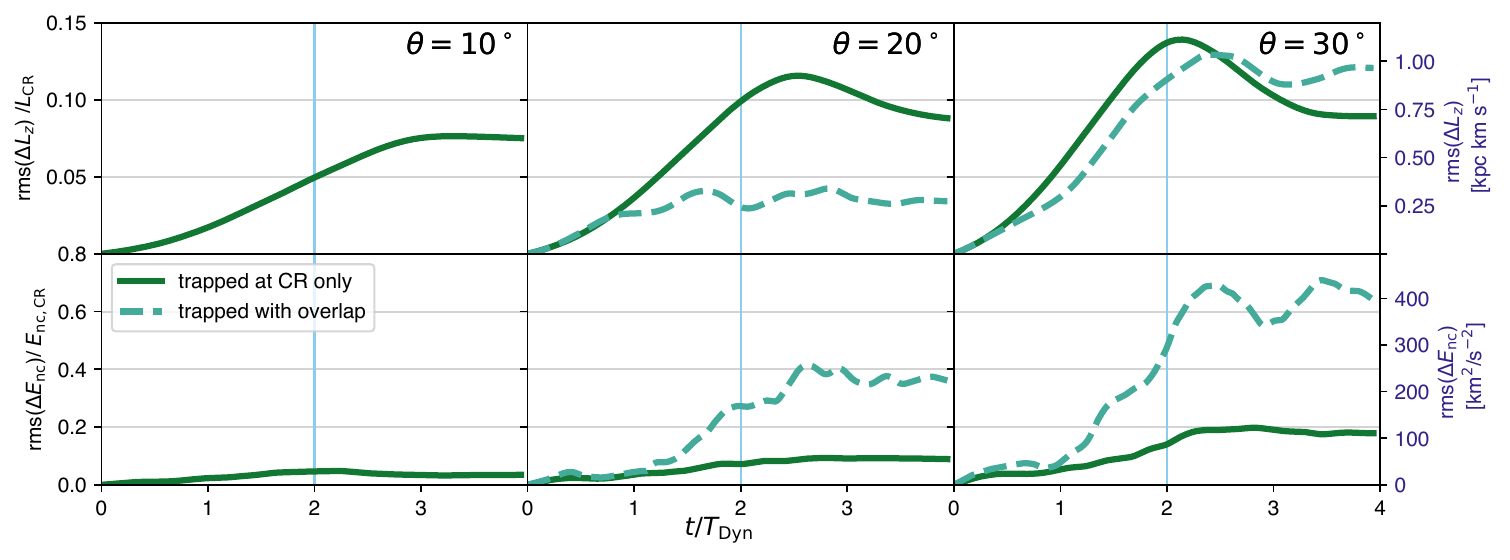}
\caption{Time evolution of the \RMSL (top) and \RMSEnc (bottom) for populations of orbits meeting resonant criteria with spiral patterns that have \pangle $\theta=10^\circ$ (left), $20^\circ$ (middle), and $30^\circ$ (right). 
Units are normalized by models values in the underlying \axisym disk at the \RCR and the vertical (purple) line indicates the time when the spiral reached peak amplitude.
Curves show values for `\TypeCR' (solid, dark green) and `\TypeOverlap' (dashed, teal) populations.
The right side y-axis gives non-normalized model values. 
}
\label{fig:2D_numerical}
\end{figure*}

The 2D simulations demonstrate that a trapped population's \RMS changes in \Lz, \RMSL, is greater for more open spiral patterns.  Figure~\ref{fig:2D_numerical} shows the time evolution of \RMSL (a proxy for changes in \gc orbital size)  and \RMSEnc (proxy for kinematic temperature) for `\TypeCR' (solid, dark green) and `\TypeOverlap' (dashed, teal) populations in the presence of spiral patterns with \pangle $\theta=\{10^\circ$, $20^\circ$, $30^\circ$\}.
Populations `\TypeCR' have more significant \RMSL in systems with more open spiral patterns (increasing \pangle $\theta$) and do not show significant \RMSEnc (kinematic heating). This is the population that experiences \cting.
Comparatively, the populations identified as `\TypeOverlap' have a similar trend in \RMSL with increasing \pangle, but also have a strong increase in \RMSEnc.  No orbits meet the    `\TypeOverlap' criteria for the model with $\theta=10^\circ$.  The \RMSL for `\TypeOverlap' populations is therefore not considered to be from \cting since there is simultaneous kinematic heating.

\begin{figure*}
\centering
\includegraphics[width=\textwidth]{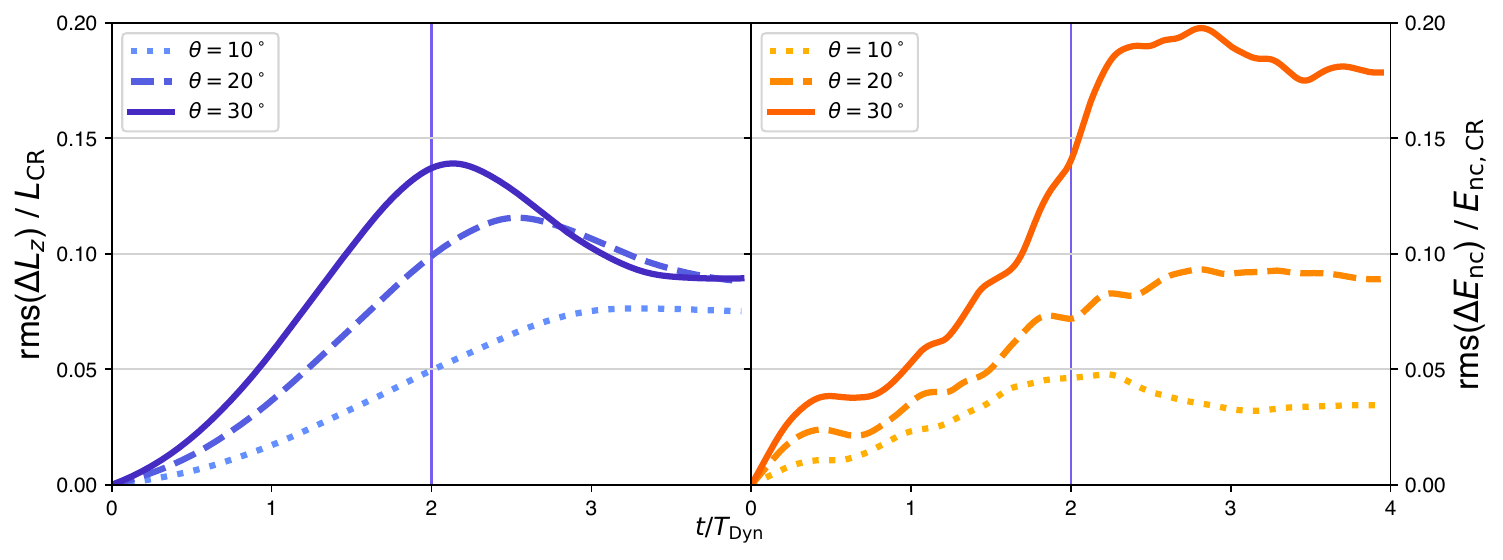}
\caption{Time evolution of \RMSL from \cting and \RMSEnc (fractional kinematic heating) for `\TypeCR' populations in models with various \pangle{s}, $\theta$.
Units are normalized by models values in the underlying \axisym disk at the \RCR and the vertical (purple) line indicates the time when the spiral reached peak amplitude.   
The values for \RMSL in this 2D tracer particle simulation are most consistent with the maximum radial excursions approximated from the equations of motion (\eq~\ref{eq:R1EoMtaumax}).
}
\label{fig:2D_numerical_eq}
\end{figure*}

The time evolution of \RMSL and \RMSEnc for \TypeCR are shown in \f~\ref{fig:2D_numerical_eq}.
The degree of \cting increases with increasing \pangle $\theta$ at the time when the spiral has peak amplitude, but the values for \RMSEnc do not decrease after initial heating. 
The values for \RMSL in this 2D tracer particle simulation are best matched to the maximum radial excursions from \RCR $|R_{\rm 1,s}|$ expected from the equations of motion (\eq~\ref{eq:R1EoMtaumax}) and shown in \f~\ref{fig:R1_terms}~\&~\ref{fig:R1}.  The maximum radial change in guiding center radius for these orbits is $\max(\Delta R_g) = 2|R_{\rm 1,s}|$.
Since the the models so far have assumed a flat rotation curve, $\Delta R_g\propto \Delta L_z$, and therefore the value for maximum radial change in guiding center radius from the equations of motion ($\max(\Delta R_g)/R_0{\sim}0.2$ from \f~\ref{fig:R1_terms}) is slightly larger than the equivalent \RMSL$/L_{\rm CR}$ from \f~\ref{fig:2D_numerical_eq}.

The associated kinematic heating in the right hand panel can be evaluated as an associated contribution to the radial excursions using the \epi approximation from \eq~\ref{e:epiX} and by assuming values from \tab~\ref{tbl:modelscalefactors} for circular velocity, $v_c=220$~km~s$^{-1}$, \CR radius, $R_{\rm CR}=8$~kpc, and the associated value for the \epi frequency, $\kappa=39$~km~s$^{-1}$~kpc$^{-1}$.  The change in radial excursion $\delta R_{\rm epi}$ from an associated fractional change in non-circular energy $\delta E_{\rm nc}=$\RMSEnc$/E_{\rm nc,CR}$ would scale as the square root $\delta R_{\rm epi} \propto \sqrt{\delta E_{\rm nc}}$.
For reference, the peak \RMSL for \TypeCR orbits in the presence of a spiral with $\theta=20^\circ$ corresponds to \RMS$(\Delta R_g) \sim 1$~kpc (smaller than a typical \epi excursion, \RMS$(\Delta R_{\rm epi})\sim 1.3$~kpc), whereas the increase in the radial excursions \textit{around} the guiding center radius ($R_g$) $\Delta R_{\rm epi}\sim 0.2$~kpc.

\subsection{3D Model}\label{sec:Numerical3D}

The 2D numerical study in \S\ref{sec:Numerical2D} is based on the analytic exploration of in-plane orbital changes for stars trapped at the \CR resonance.  We here expand our numerical model to include the vertical dimension.  

The 3D model uses of orbital actions, $J\equiv\{J_R, J_\phi,J_z\}$, which naturally describe orbits in a disk geometry.
The \azi action of an orbit is equal to its orbital angular momentum in the $\hat{z}$ direction, 
\begin{equation}
    J_\phi=L_z,
\end{equation}
and is a proxy for orbital size.
Radial action, $J_R$, is related to orbital \acrlong{ec}.  In the \epi approximation, radial action can be related to the energy associated with non-circular orbits through \citep[][their \eq~3.261]{BT08},
\begin{equation}\label{eq:JREnc}
	J_R \approx \frac{E_{\rm nc}}{\kappa}.
\end{equation}
Similarly, the vertical action is associated with motions perpendicular to the plane and vertical height.  For small excursions, it can be approximated by,
\begin{equation}\label{eq:JzEz}
    J_z \approx \dfrac{E_z}{\nu},
\end{equation}
where $\nu$ is the vertical \epi frequency.
Since the governing physics for \cting is decoupled from vertical motions, $J_z$ is expected to be approximately conserved for small vertical excursions \citep{SSS12}.
Equations~\ref{eq:JREnc} and~\ref{eq:JzEz} break down for orbits that diverge from the \epi approximation or in potentials that are far from smooth \citep{2025MNRAS.537.1620D}. Nonetheless, the fundamental intuition gained from these expressions holds.

The underlying disk in the 3D disk is modeled using the default settings for \texttt{galpy}'s the \texttt{MWPotential14}, which approximates a smooth \MW-like potential.  
This potential includes components for a 
NFW dark matter halo profile \citep{NFW97},
a power law prescription for the bulge potential that has an exponential cutoff, and
and Miyamoto-Nagai disk adjusted to fit \MW measurements \citep{BR13}.
Normalization parameters are given in \tab~\ref{tbl:diskparameters} and are set to approximate \SNd values. 

\begin{deluxetable*}{lll}
	\tablecaption{3D Model Scale Constants \label{tbl:diskparameters}}
	\tablehead{ 
	    \colhead{Parameter}
	    & \colhead{Value}
	    &  \colhead{Description}}
	\startdata
		$r_0=R_\odot$ & $8$~kpc & normalization radius\\
		$R_\Sigma$ & $R_\odot/3=2.7~{\rm kpc}$ & surface density scale length\\
		$R_{\sigma_R}$ & $3R_\odot/4=2.0~{\rm kpc}$ & radial scale length for the radial velocity dispersion\\		
		$R_{\sigma_z}$ & $R_{\sigma_R}=2.0~{\rm kpc}$ & radial scale length for the vertical velocity dispersion\\
		$R_{CR}$ & $R_\odot=8~{\rm kpc}$ & \CR\ radius\\		
		$R_s$ & $2.4$~kpc & radial scale length for spiral density\\
		$z_s$ &  $0.3$~kpc & scale height for spiral pattern\\
		$v_o=v_c(r_0)$ & $220$~km~s$^{-1}$ & normalization value for the circular velocity\\
        $\sigma_{R,0} = \sigma_R(r_0)$ & 0.16$v_o=35.2$~km~s$^{-1}$ & radial velocity dispersion normalization\\ 
        $\sigma_{z,0} = \sigma_z(r_0)$ & 0.08$v_o=17.6$~km~s$^{-1}$ & vertical velocity dispersion normalization\\ 
        $m$ & $4$ & number of arms\\
	\enddata
	\tablecomments{{Each parameter gives the value at $r_0=8~{\rm kpc}$ and is set to approximate the \SNd.}}
\end{deluxetable*}

The form for the 3D spiral arm potential is taken from \cite{CG02}.
This simulation adopts parameters such that the maximum density of the spiral pattern is at $R_{\rm peak}{=}$\RCR and so that the \azi profile is sinusoidal, thus conserving mass.  The form for this spiral potential is given by,
\begin{equation}\label{eq:3DSpiral}
\begin{aligned}
    \Phi_1(R, \phi, z) = -4 \pi G\, z_s\, \rho(R_{\rm CR})\, \exp \left( -\frac{R-R_{\rm CR}}{R_s} \right)\\ 
    \times {\frac{1}{K D} \,\cos(\gamma) \,\mathrm{sech}^{B} \left( \frac{K z}{B} \right)},
\end{aligned}
\end{equation}
where $z_s$ is the scale height for the spiral pattern, $R_s$ is the scale length for the density fall off for the spiral arms, the parameter $\gamma$ is given by, 
\begin{equation}
    \gamma = m \left[\phi - \frac{\ln(R/R_{\rm CR})}{\tan\theta} \right],
\end{equation}
and
\begin{align}
    K &= \frac{m}{R \sin\theta}, \\
    B &= K z_s\, (1 + 0.4 K z_s),\, {\rm and} \\
    D &= \frac{1 + K z_s + 0.3 (K z_s)^2}{1 + 0.3 K z_s}.
\end{align}
This suite of simulations sets the values $z_s=0.3$~kpc and $R_s=2.4$~kpc.
The number of spiral arms ($m=4$) and radius of \CR ($R_{\rm CR}{=}R_0{=}8$~kpc), are set to the same values as in the 2D model (\S\ref{sec:Numerical2D}). 

Initial conditions for disk stars were selected by sampling \texttt{galpy}'s class \texttt{quasiisothermaldf}, which is a 3D action based distribution function modeling an \axisym disk \citep{Binney10,BM11}.  This distribution function has the analytic form,
\begin{equation}\label{eqn:fq}
    f_Q(J_R,J_\phi,J_z) = f_{\sigma_R}(J_R,J_\phi) \times \dfrac{\nu}{2\pi\, \sigma_z^2} \exp\left( -\dfrac{\nu J_z}{\sigma_z^2(R_L)}\right),
\end{equation}
where $\sigma_z$ is the vertical velocity dispersion, and,
\begin{equation}
\begin{aligned}
    f_{\sigma_R}(J_R,J_\phi) = \dfrac{\Omega\, \Sigma(R_L)}{\pi\, \sigma_R(R_L)^2\, \kappa}\Big|_{R_L} \times \left[1+ \tanh{L_z/L_s}\right]\\
    \times \exp\left(-\dfrac{\kappa\, J_R}{\sigma_R^2(R_L)}\right),
\end{aligned}
\end{equation}
where the surface density $\Sigma$ takes the exponential form given by \eq~\ref{eqn:expR} and $L_s$ is a constant of the set to \texttt{galpy}'s default value, which is $L_s=0.00568L_\odot\approx 45$~pc in our model.

\subsubsection{Characterization of Trapped Orbits}\label{sec:3DType}

Initial 6D phase-space positions are selected so that the sampling of orbits that would be trapped around \CR is unbiased by edge effects. 

The methods used to identify trapped orbits in S\ref{sec:2DType} are well suited for the 2D model.  
However, the criterion for orbital trapping given by \eq~\ref{eq:Lambdanc} was developed under the assumption that vertical action is conserved, where this assumption breaks down with increasing disk thickness \citep{SSS12}.
How well \eq~\ref{eq:Lambdanc} holds for a 3D disk of given thickness will be explored in a separate paper.  

In this work we adopt an approximation, that is informed by the findings from \S\ref{sec:2DTrends} and \cite{Daniel19}, in order to isolate orbits that are `\TypeCR' in the 3D model.
The population of `\TypeCR' orbits is characterized as orbits that have initial guiding center radius ($R_L(t=0)$ from \eq~\ref{eq:RL}) to be between the first harmonics of the \IOLRs (see \eq~\ref{eq:Lindblad} and following discussion).  For $n=1$ these radii can be approximated by,
\begin{equation}\label{eqn:LR}
	R_{\rm LR}^{(n+1)=2} = \left(1 \mp \dfrac{\sqrt{2}}{2m} \right) \dfrac{v_c}{\Omega_{p}}.
\end{equation}
Orbits with $R_L(t=0)$ outside this annulus enclosing \RCR are not included in the current analysis.
This characterization will surely have some leakage, but trends in orbital changes that scale with \pangle will be dominated by orbits that are `\TypeCR.'

\subsubsection{3D Trends with Pitch Angle}\label{sec:3DTrends}

\begin{figure*}
\centering
\includegraphics[width=\textwidth]{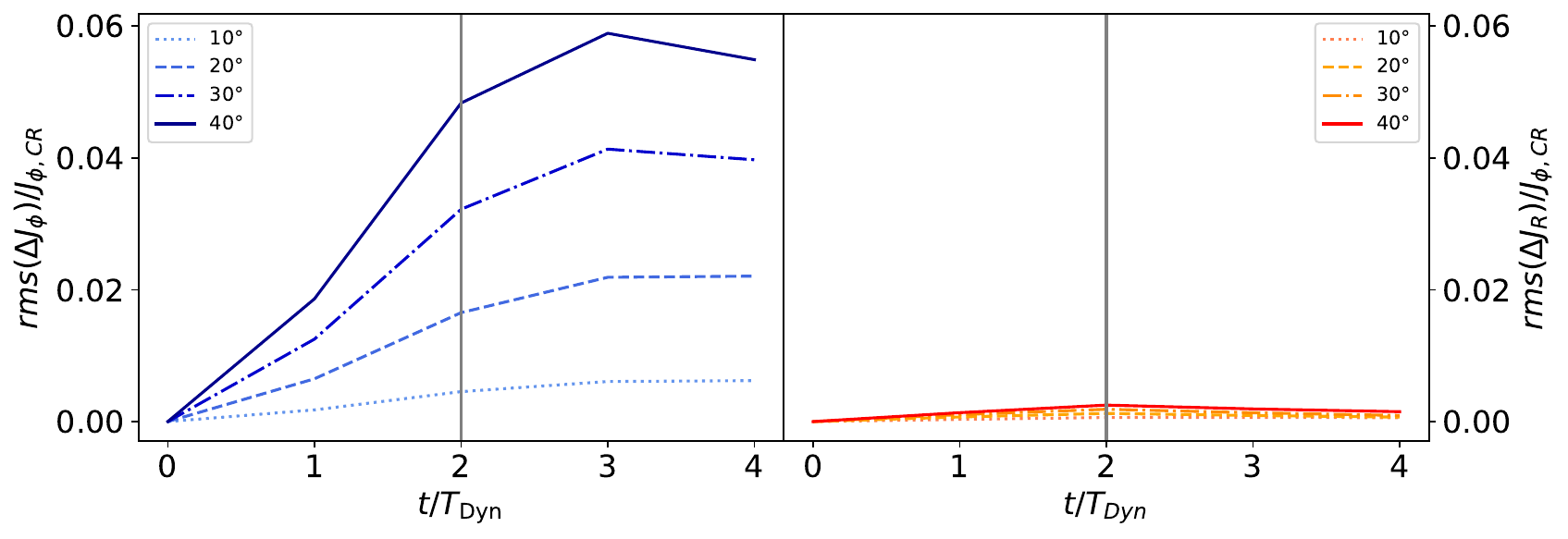}
\caption{Time evolution of \RMSJphi (degree of \cting) and \RMSJR (kinematic heating) for `\TypeCR' populations in models with various \pangle{s}, $\theta$.
Units are normalized by $J_{\phi,{\rm CR}}$, the value of $J_\phi$ at \RCR in the underlying \axisym disk. The vertical gray line indicates the time when the spiral reached peak amplitude.  The degree of \cting increases with increasing \pangle $\theta$.
}
\label{fig:3D_numerical_jphi}
\end{figure*}

\begin{figure}
\centering
\includegraphics[width=\columnwidth]{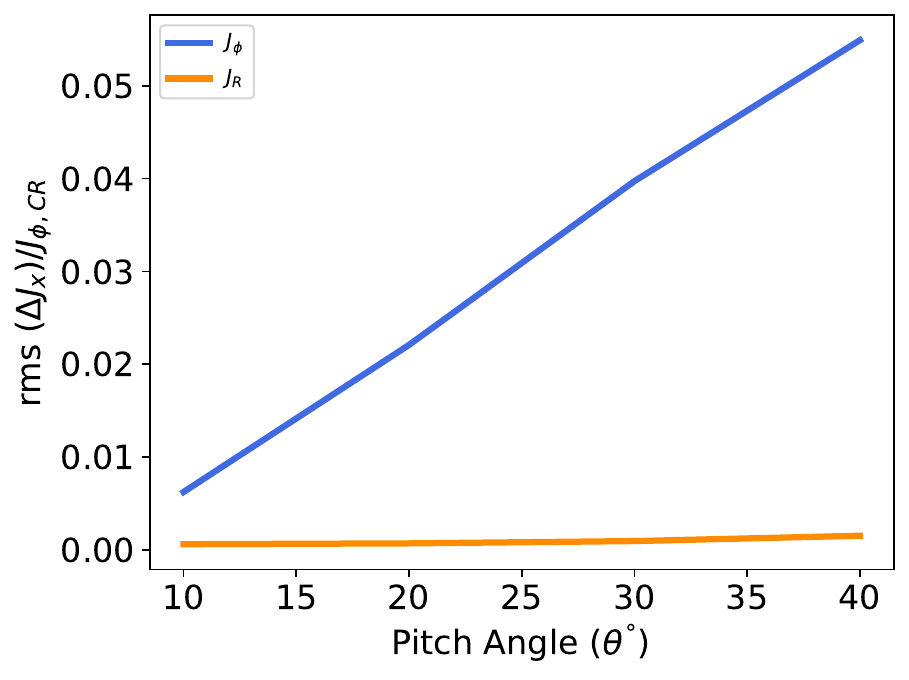}
\caption{Summarizing plot showing the near linear increase in the degree of \cting (increasing value for \RMSJphi with constant \RMSJR$\sim 0$) for `\TypeCR' orbits with increasing \pangle, $\theta$.
}
\label{fig:3D_numerical_jphi_theta}
\end{figure}

The 3D models for \cting have the same trends as the 2D model.  Figure~\ref{fig:3D_numerical_jphi} shows the time evolution in \RMSJphi (proxy for \cting) and \RMSJR (proxy for kinematic heating) for the population of `\TypeCR' orbits in models with spiral \pangle $\theta=\{10^\circ, 20^\circ, 30^\circ, 40^\circ\}$. None of the models indicate significant kinematic heating while the degree of \cting increases for increasing \pangle $\theta$.  Figure~\ref{fig:3D_numerical_jphi_theta} summarizes the trends over both measures, clearly demonstrating a near linear relation, \RMSJphi$\propto\theta$. 



\subsection{Winding Spiral Model}\label{sec:3DWinding}

The previous sections considered spiral patterns with radially independent pattern speed, consistent with the assumptions adopted in the analytic derivation. Here we instead examine a spiral pattern that \crt{s} with the underlying disk at all radii, such that $\Omega_p(R)=\Omega(R)$. This experiment provides a direct comparison between rigidly rotating (e.g.,~density-wave) and winding spiral models.

\begin{figure*}
    \centering
    \includegraphics[width=\textwidth]{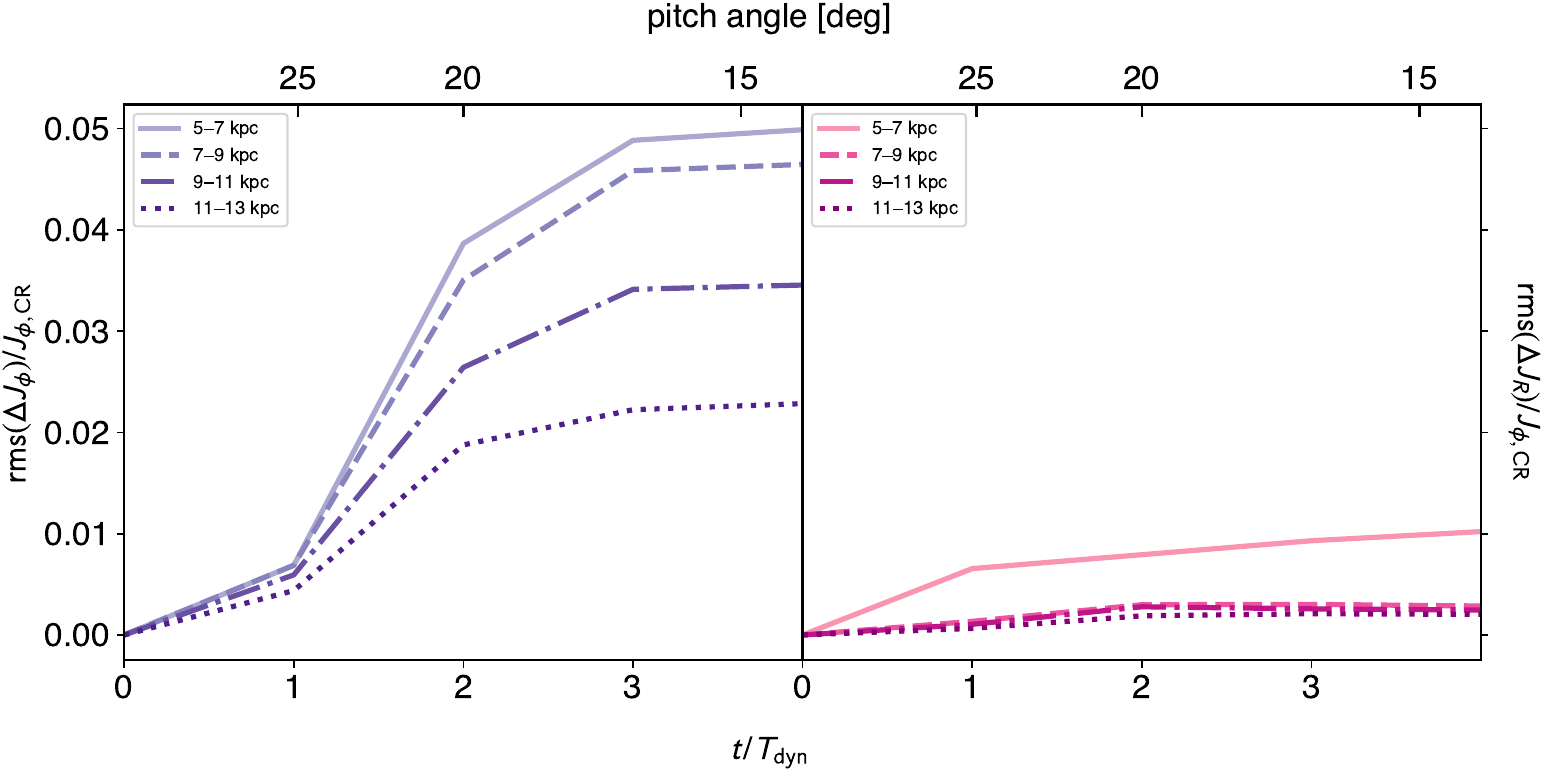}
    \caption{
    Time evolution of \cting from a winding spiral potential. 
    The left panel shows \RMSJphi and  the right shows \RMSJR. The vertical axes are normalized by the value of $J_\phi$ at \RCR in the smooth underlying disk $J_{\phi,{\rm CR}}$ and the x-axis is normalized by the orbital period at $R_0$. 
    Orbits are grouped by initial radial position in 2~kpc wide annuli. All orbits in this simulation are `\TypeCR' since \CR is satisfied at all radii. The time dependent value for the \pangle ($\theta$) is indicated along the top axis.
    In contrast to a rigidly rotating spiral, where \cting is more efficient for more open spirals, the winding spiral has larger \RMSL for smaller \pangle.
}
\label{fig:Winding_numerical_jx}
\end{figure*}

In this experiment no `\TypeCR' population can be defined since \RCR is at all radii.
Figure~\ref{fig:Winding_numerical_jx} shows the time evolution of the \RMS changes in azimuthal action, \RMSJphi, and radial action, \RMSJR, for stars grouped according to their initial guiding-center radius. The spiral pattern has \pangle $\theta=20^\circ$ at $t=t_{\rm ref}=2T_{\rm Dyn}$, corresponding to the same morphology as the fiducial model in \S3.3, but has a more open spiral pattern before this time ($t<t_{\rm ref}$) and winds to progressively smaller \pangle{s} afterwards ($t>t_{\rm ref}$) as differential rotation shears the pattern. The values for $\theta$ over time are given in $5^\circ$ intervals on the top axis.

Unlike the rigidly rotating models, the degree of \cting increases as the spiral winds and the \pangle $\theta$ decreases. Orbits with initial guiding center radii within the annulus inside with peak amplitude (set to be at $R_{\rm peak}=8$~kpc) have the largest \RMS changes in $J_\phi$, closely trending with the annulus that includes $R_{\rm peak}$. Orbits with significantly larger initial guiding center radii exhibit progressively weaker changes. Across the disk, there are only modest increases in \RMSJR and the increases in kinematic temperature remain small compared to the \RMS changes in angular momentum. 

The winding spiral model demonstrates the opposite trend in the dependence of the degree of \cting on spiral morphology. For the rigidly rotating models presented in \S\ref{sec:Numerical2D}-\ref{sec:Numerical3D}, \cting is more efficient for spirals with more open \pangle{s}. In contrast, the winding spiral exhibits progressively larger changes in \RMSL as the \pangle decreases in time. 

This trend reversal reflects the fundamentally different nature of these perturbations. In the rigidly rotating picture, a larger \pangle increases the torque exerted by a spiral of fixed pattern speed. For a winding spiral, however, the \pangle decreases over time as the pattern is continually sheared by differential rotation. Since the spiral pattern \crt{s} everywhere, the \CR condition is met over an extended radial range.  In the winding picture, orbits on the leading side of an arm will continuously be torqued such that they lose \Lz and stream inward while orbits on the trailing side of an arm will stream outward as they are continuously torqued such that they gain \Lz.  Over time, the cumulative effect of continuous torquing will accumulate progressively larger changes in \RMSL even as the spiral winds to smaller \pangle.

This experiment demonstrates that the relation between spiral \pangle and \cting cannot be interpreted from morphology alone. The same observed spiral strength and \pangle can imply qualitatively opposite efficiencies for \cting depending on whether the spiral is, for example, a rigidly rotating density wave or a swing amplified winding spiral. Understanding the dynamical nature of spiral structure is therefore critically important to connect observable characteristics with orbital dynamics.  This claim forms the basis for the discussion in \S\ref{s:Discussion}.

\section{Discussion}\label{s:Discussion}

\subsection{Observational Constraints}\label{s:Observations}

Spiral structure varies considerably between galaxies and is likely to vary over a single galaxy's lifetime.  Thus, the importance of \cting in spiral galaxy evolution is as variable as spiral structure itself.
We can place estimate the degree of \cting in the \MW and in samples of external galaxies by using values for the spiral \pangle ($\theta$), strength (measured in this study using fractional amplitude, $\epsilon_\Sigma$) and pattern speed ($\Omega_p$) that are based on observables.  The current efficiency of \cting can be approximated when combining these measures with estimates for spiral lifetimes and the timescales for strong \cting. 
Observational studies that characterize spiral structure provide meaningful constraints on the range of realistic values for spiral properties and the current role of \cting in both our own Galaxy and external galaxies. 

\subsubsection{Spiral Pitch Angle and Strength}\label{s:ObservationPitchAndStrength}

While the \MW is our closest, and home galaxy, our observation point within the disk makes it challenging to view as a whole system. Based on both star counts, and emission line based gas observations, the spiral arms in our own \MW appear to be of intermediate type between grand design and flocculent \citep{Elmegreen98} with \pangle{s} found to be between 
$10^\circ$ \citep{Reid2019} and $18^\circ$ \citep{Drimmel00}.  This is in agreement with the mean \pangle, $\langle\theta\rangle=15^\circ$, found for a sample of 51 \MW-like (Sbc type) galaxies \citep{Ma99}.  Such values for $\theta$ suggest that the changes in stellar guiding centers around the radius of \CR are well described by the low \pangle regime and thus $R_1$ values from spirals in the \MW's disk are quite small compared to those from a bar.  A low pitch angle lessens the value of $|\Phi_s|$, thus decreasing the range of $E_J$ values allowed for capture and decreasing the probability of \cting. 

The amplitude of the perturbing potential ($\epsilon_\Sigma$) is also extremely important in evaluating $R_1$. \citet{Eilers2020} recently used \textit{Gaia} \citep{2016A&A...595A...1G} data to estimate a 10\% arm amplitude in the \MW. Another observational method is to measure amplitudes in external galaxies similar to the \MW. \citet{Elmegreen11} found amplitudes of 0.1-0.24 for galaxies in the Spitzer Survey of Stellar Structure in Galaxies \citep[S$^4$G;][]{2010PASP..122.1397S} with similar morphology to the \MW. These values for spiral strength can be used as constraints on the efficiency of \cting in the \MW.

We can place some limits on this, using observations of external galaxies, noting however that pitch angles are notoriously difficult to constrain observationally. Various studies in the nearby Universe suggest most galaxies today are in the low \pangle regime. For example \citet{Kennicutt1981,Ma99,BP99,SJ98} found a range in \pangle{s} between $0^\circ$ (although many might class spiral arms with zero \pangle{s} as rings rather than spirals) and $49^\circ$, with median $\theta_{\rm med}=8^\circ-22^\circ$. In a Bayesian analysis of around 100 nearby galaxies with stellar masses $9.5 < \log(M_\star/M_\odot) < 10.0$,  \citet{Lingard2021}, found most had \pangle{s} in the range $10^\circ$ and $20^\circ$. Using \HST and/or \JWST images, several recent studies suggest pitch angles are typically larger in high redshift galaxies \citep[e.g.][]{Reshetnikov2023,Chugunov2025}, however observationally there is some ambiguity between spiral arms with more open \pangle{s} and tidal features, which may complicate the interpretation of this result, and in a similar study, \citet{Kuhn2026} concluded there was no detectable redshift evolution of pitch angles. 

\subsubsection{Spiral Pattern Speed}\label{s:ObservationPatternSpeed}

The radial location of \CR is fundamentally tied to the pattern speed of a spiral since, by definition, the lowest order expression for the \CR resonance is $\Omega_p=\Omega_0$.

Methods for measuring pattern speeds in external galaxies are primarily aimed at studying bar patterns since they these high amplitude structures are typically the dominant perturbation in a disk galaxy.  
The Tremaine-Weinberg method \citep[TW;][]{1984ApJ...282L...5T} combines surface brightness and radial velocities to find bar pattern speeds and is the most widely used for this purpose.  However, TW has been used to measure whether or not spiral pattern speeds are equal to the bar pattern speed \citep{2025RNAAS...9..236P}.  
The lowest surface brightness regions that are \azi{ly} located between the arms of a bar have been associated with position of the \Lag.  Under this assumption these ``dark gaps" have been used to estimate the radial coordinate for a bar's \RCR and thus estimate its pattern speed \citep{2017MNRAS.470.3819B}.
In recent years the Font-Beckman method \citep{2011ApJ...741L..14F}, which uses the switching of streaming motions at \CR, has gained traction.  This method is particularly pertinent for the study of spiral pattern speeds since it has a demonstrated ability to find more than one \RCR, where the majority are presumably associated with spiral perturbations \citep{2014ApJS..210....2F,2018ApJ...854..182B}. 
Studies to find the pattern speed of the \MW{'s} bar have found a wide range of values \citep[review in][their figure~10]{2025NewAR.10001721H} which are potentially converging to a value near ${\sim}35-40$~km~s$^{-1}$~kpc$^{-1}$.

The \azi color of stellar populations can be used under the assumption that star formation is triggered by a dense pattern moving through the \ISM  \citep{2019NatAs...3..178P, 2022MNRAS.512..366A}.  In this model, the direction of the \azi color gradient, or other measures of stellar population age, should swap at the \RCR of a rigidly rotating spiral since the pattern moves faster/slower than the underlying stellar population outside/inside \RCR.
Efforts to use this method have produced a wide range of results, including evidence for density waves, multiple offset modes, and material or winding spirals \citep{2025A&A...701A.183Q}.
Some studies did not find sufficient evidence to support a density wave theory at all, but rather determined their results were consistent with material or winding spirals 
\citep{2019MNRAS.487.1808M,2025ApJ...981..115S},
a conclusion that is supported by the observed wide range of \pangle{s}  \citep{2019MNRAS.490.1470P, Lingard2021}.
Finding the pattern speeds for spirals in the \MW{} has proven to be challenging. 
Nonetheless, there is a long history of attempts to determine $\Omega_p$ for \MW spirals with patterns speeds ranging from $\Omega_p=18-30$~km~s$^{-1}$~kpc$^{-1}$ \citep{DL05,FFT01,LMD01}, corresponding to $R_{\rm CR}=7.3-12.2$~kpc, assuming $v_c=220$~km~s$^{-1}$.  

Measuring arm amplitudes in external galaxies has been done in several ways. Assuming density change in the galactic disk goes as NIR surface brightness, \cite{RZ95} find $m=2$ Fourier amplitudes between $\sim0.15$ and $\sim0.6$ (however this can include bar component as well as arms). \cite{Elmegreen99,Elmegreen11} extend this analysis to define the fractional strength ($\epsilon_\Sigma$) of the spiral arms by assuming a sinusoidal variance in the azimuthal direction.  They find, $\epsilon_\Sigma=(I_{max}/I_{min}-1)/(I_{max}/I_{min}+1)$ where $I_{max}/I_{min}=10^{A_r/2.5}$ and $A_r$ is twice the Fourier amplitude.  As discussed below, optically identified flocculent spirals have low density NIR perturbations (if any), with $\epsilon_\Sigma$ between 0.15 and 0.26 \citep{Elmegreen99,Elmegreen11}.  \cite{Elmegreen11} used 46 galaxies from the S$^4$G to illustrate a trend in spiral amplitude with type.  They found that grand design, intermediate and flocculent galaxies have an average $\langle\epsilon_\Sigma\rangle=0.25\pm0.09$, $0.16\pm0.06$ and $0.15\pm0.07$ respectively, with $\epsilon_\Sigma$  increasing with Hubble type.  

The range of observationally derived values for spiral \pangle, strength, and pattern speed allow an informative exercise in assessing the proposition that the Sun may have arrived in the Solar circle via \cting
\citep[e.g.][]{WFD96,Frankel18,2025ApJ...983L..10Z_AMR,2025arXiv251209987Z}.
A relatively strong spiral ($\epsilon_\Sigma=0.2$) with an average \pangle ($\theta=15^\circ$) could have caused the Sun to migrate outward $\max(\Delta R_g){\sim}2$~kpc in one transient spiral episode if it were a rigidly rotating spiral when using the estimated maximum excursion for either \eq~\ref{eqn:SBR1s} or \eq~\ref{eq:R1EoMtaumax}. This scenario would require a spiral pattern speed of $\Omega_s=25-30$~km~s$^{-1}$~kpc$^{-1}$, corresponding to $R_{CR}=7.2-8.8$~kpc, which is within the upper measured limits for current \MW spirals.  In this scenario, the \cting of the Sun would not have cause the Sun to migrate more than the radial excursions from \epi motion, where the key difference would be the associated kinematic heating.  However, should the radial change from the same spiral pattern be estimated using the expression based on bar symmetry (\eq~\ref{eqn:SBR1}), one would conclude that the Sun could have migrated $\max(\Delta R_g)\gtrsim 6$~kpc.  
The expected radial change  (aka. \RMS value) near \CR from a spiral with the same parameters in the 3D scenario (see \f~\ref{fig:3D_numerical_jphi}) is slightly less than in the maximum value in the 2D case, where \RMS$(\Delta R_g)\lesssim 2$~kpc.  These two scenarios are in agreement since the \RMS value is expected to be less than the maximum value.  
However, a winding spiral would have nearly double the expected radial change for the same measured values (see \f~\ref{fig:3D_numerical_jphi_theta}), where \RMS$(\Delta R_g)\lesssim 4$~kpc, 
though the amplitude chosen for this exercise is high for flocculent and material spiral patterns.
The maximum and expected radial migration of the Sun from \cting using observed parameters sets a foundation for understanding the realistic efficiency of \cting from a single transient spiral episode in the \MW.  

\subsubsection{Spiral Lifetime}\label{s:ObservationLifetime}

For a star's angular momentum to be maximally and permanently changed by a rigidly rotating spiral, the perturbing potential must exist for more than half the lifetime of the radial oscillation. We have shown in \f{s}~\ref{fig:2D_numerical}-\ref{fig:3D_numerical_jphi} that the value of \RMSL maximizes after the spiral reaches peak amplitude at 2~\Tdyn in our models.  \cite{EE84} suggested that flocculent spirals have the shortest lifetimes, on the order of $\sim 10^8$~yrs, while grand design spirals that are stimulated by external interactions and/or a bar are likely stable modes and have estimated lifetimes on the order of several Gyr \citep[eg.][]{EE83,Struck11}.  Spiral lifetime and (re)generation rate are important for the efficiency of \cting since multiple transient spiral patterns with a large range in pattern speeds are necessary for radial migration to effectively mix the galactic disk at all radii.  

Rigidly rotating spiral patterns with the range of currently measured values for \MW-like spirals cannot effectively mix the disk from \cting unless there is a large range of spiral pattern speeds with short lifetimes.  This efficiency is increased if each generation of spiral pattern has incrementally larger/smaller $\Omega_p$. 
Alternately, a time dependent pattern speed could cause strong \cting from a given spiral pattern should the pattern slowly slow down or speed up.

\subsubsection{Chemical Trends}\label{s:ChemicalVariations}

A strong perturbation that \crt{s} with the disk is expected to drive coherent radial streaming motions.  This happens as orbits trapped near \CR are torqued by the leading/trailing side of the \perturber and lose/gain angular momentum, thus causing their guiding center radii to move toward/away from the galactic center. 
As a consequence, azimuthal variations in the mean \Fe of stars is predicted for bar patterns \citep{2013A&A...553A.102D,Filion23} as well as around spiral arms near \CR  \citep{2016MNRAS.460L..94G,2023MNRAS.525.3318H, 2025MNRAS.537.1620D}.
These variations may even possible in the absence of a radial \Fe gradient so long as there is a radially changing vertical \Fe gradient \citep{2018A&A...611L...2K}.
Since the radial range of \cting is greatest for a spiral pattern that \crt{s} \citep{DW15}, it follows that the radial range of azimuthal variations would be greatest for \crting spirals and minimized for spirals that rotate rigidly.
In other words, the radial extent of azimuthal variations could provide a measurable insight into the nature of spiral structure. 


Another potential result of \rrn is a flattening of the \azi{ly} averaged radial \Fe gradient.   
Models that include \cting \citep{Loebman16} are able to reproduce the skew in the \FeH distribution of stars within various annuli near the plane ($|z|<0.5$~kpc) of the \MW's disk \citep{Hayden15}.
\citet{Wisz25} found that the gas-phase radial \Fe gradients from Mapping Nearby Galaxies at Apache Point Observatory \citep[MaNGA;][]{2015ApJ...798....7B} data were stronger for galaxies that were identified as having spiral structure in Galaxy Zoo \citep{2011MNRAS.410..166L}, but did not find any correlation in  with spiral pitch angle. 
Morphologically based assessments of the radial \Fe gradient have mixed results \citep[e.g.][]{2016A&A...595A..62P, 2016RMxAA..52..171S, 2018A&A...609A.119S,2023RMxAA..59..213B} but none of these studies look specifically at spiral \pangle.
It is difficult to draw any direct conclusions since the observed \Fe gradient is shaped by the net \rrn of stars from many overlapping spirals over the history of a disk \citep{2026arXiv260700077S,Quinn26}.


\subsection{The nature of spiral structure}\label{s:SpiralNature}

It is unclear whether spiral arms in nature have a radially independent pattern speed, are \crting, both, or neither.  In this study, the efficiency of \cting is explored for the two extreme cases.

Density wave spiral patterns \citep{LS64, LS66,1996ssgd.book.....B} and spiral modes \citep{SC14} have radially independent pattern speeds.  However, multiple overlapping modes could mimic winding spirals and thus confuse the physical interpretation \citep{SC14}.
Even so, \acrlong{cring} spirals occur in galaxy simulations \citep{WBS11,GKC12,2013A&A...553A..77G,BSW13}.
Spirals that \crt could be low amplitude, sheared material arms or strong, swing amplified perturbations \citep{GL65,Toomre64,1981seng.proc..111T,2013ApJ...766...34D}. 
However, a recent analysis of the Auriga Project \citep{2017MNRAS.467..179G} simulations \cite{2026MNRAS.548ag774G} found that spiral structure is diverse and context dependent, finding evidence for multiple formation mechanisms, including manifold spirals \citep{2012MNRAS.426L..46A} associated with the ends of bars.

Attempts to characterize the spiral pattern speeds in simulated galaxies, both isolated \citep{Roskar12} and in the cosmological context \citep{Quinn26}, using a windowed discrete Fourier transform \citep{1986MNRAS.221..195S} (wDFT) found that high amplitude patterns were not winding but are time-dependent.  Patterns in both studies were transient with lifetimes lasting from a few dynamical times to many Gyr and steadily sped up or slowed down, but are rarely steady.

Kinematic structures identified using \textit{Gaia} have been used to support past density-wave like activity in the \SNd \citep{2018MNRAS.479L.108K,2019MNRAS.484.3291T,2020MNRAS.493.2111G}, where \cite{2019MNRAS.484.3291T} identified up to seventeen features that are consistent with post-resonant signatures of transient spirals \cite{2019MNRAS.484.3154S,Smock26} each with a different pattern speed.  
\cite{2021RAA....21....9H} used cluster motions and age distributions around local arms to suggest that \CR of the closest density-wave like spiral is near Solar circle.
Clustering of stellar velocities in \SNd{}, called moving groups, have been attributed to various resonances with the bar and spiral arms \citep{2000AJ....119..800D,2008A&A...490..135A,2020ApJ...890..117D} though disentangling the specific structures that contribute to the formation of such moving groups, much less their pattern speeds, is an intractable question when using kinematics alone \citep{2019MNRAS.490.1026H}.

Flocculent spirals are not associated with large amplitude density disturbances as are grand design spirals \citep{EE84,EEL03,Elmegreen11}, but have colors and \pangle{s} consistent with star formation regions that have been sheared to create short spiral-like structures \citep{EE84}.  The mechanism for \cting via such sheared spirals is not true \cting in the sense described by \cite{SB02} where the perturbation has a single pattern speed.  Rather, sheared perturbations follow $\Omega_p(R)\propto\Omega$.  In agreement with our argument from \S\ref{sec:3DWinding}~\&~\ref{s:SpiralNature}, \cite{GKC12} find that a strong spiral perturbation where $\Omega_p(R)\propto\Omega$ produces large amplitude excursions.  However, as flocculent spirals do not likely represent strong perturbations in the potential, it is probable that the captured fraction of stars for flocculent spirals is small.

It may be that \cting should be associated with grand design spirals, as these have the largest pitch angles and the strongest perturbing potentials.  However, assuming these are stable modes with a constant pattern speed, grand design spirals must have a high turnover rate in order to induce \cting around multiple radii.  Historically, numerical simulations have exhibited high amplitude, short-lived spirals that spontaneously arise without necessitating external interaction \citep[eg.][]{MPQ70,Roskar08a}.  \cite{Sellwood11} presents an argument in favor of multiple short-lived, high density spirals based on these simulation results and moving groups.  Whether or not \MW spirals are best described by the long lived type which are stimulated by the bar or tidal interactions \citep{EE83} or the short lived type that naturally arise in galaxy simulations \citep{Sellwood11} is critical to understanding the efficiency of \cting in disk galaxies.  Interestingly, \cite{MF10} and \cite{Minchev11} found that the combination of two long-lived perturbations (eg. a bar and a spiral) with differing pattern speeds can give rise to \cting across much of the disk without greatly increasing non-circular motions.


\subsection{Associated Kinematic Heating and Cooling}\label{s:Cooling}
Even in cases of strong \cting, the overall effect of any (not winding) perturbation will be to \kmy heat.
This is because the same perturbation that traps stars at \CR will necessarily also induce kinematic heating from its \IOLRs \citep[e.g.,][]{SB02,2019MNRAS.484.3154S,Smock26} along with their associated harmonic \citep{Daniel19} and vertical resonances \citep[e.g.,][]{1996MNRAS.279.1263S,2002MNRAS.337..578P,2000MNRAS.319....1T,2014MNRAS.437.1284Q,2019A&A...629A..52L,2020MNRAS.495.3175S,BeS23,McClure25}.  In a 3D~disk, moderate vertical kinematic heating and cooling may be directly associated with \cting  \citep{Loebman11,Minchev12, VCdON16}.
The net kinematic effect of transient spiral patterns on disk orbits is to \kmy heat, even as \cting induces \kmy cold changes in \Lz.  Indeed, how the high degree of inferred \rrn of \MW disk stars \citep[e.g.][]{Frankel18,2020ApJ...896...15F} can be resolved with the observed vertical and radial velocity dispersions of the disk \citep[e.g.,][]{2021MNRAS.506.1761S} presents an outstanding question.  In other words, how can the \MW's disk so \kmy cool (\citealt{2024arXiv241108944H}; see also a rebuttal \citep{2025arXiv250117907S} that also cannot account for additional forms of kinematic heating).  

While the net effect from the passage of a transient spiral, or any other perturbation, is kinematic heating, \textit{a non-negligible population of stars will be resonantly circularized}.
Resonant kinematic cooling from spirals can be understood from a toy model of star in a circular disk in a disk with a flat rotation curve, as in \S\ref{sec:Numerical2D}. The star's orbit will have kinetic energy that is independent of radius and associated only with circular motion.  Its potential energy depends logarithmically on radius ($\Phi_0 \propto\ln (R/R_p)$). A circular orbit has the minimum orbital energy allowed for given angular momentum and $E\rightarrow E_{\rm c}$ as $E_{\rm nc}\rightarrow 0$  (see~\eq~\ref{eq:Eorbital}).  

Now consider the disk is perturbed by a weak spiral pattern that has a constant pattern speed, $\Omega_p$.  The resulting potential is time-independent in a frame rotating with the spiral pattern and so the Jacobi energy (\eq~\ref{eq:EJ}) is conserved.  Therefore a small change in angular momentum, $\delta L_z$, will scale linearly with a small change orbital energy, $\delta E$, through the Jacobi relation so that
\begin{equation}
    \delta Lz = \delta E.
\end{equation}    
Should that star be at the \ILR it would have a small loss in angular momentum, $\delta L_z$, that would have an associated small decrease in guiding center radius, $\delta R_g$,
\begin{equation}
    \delta Lz=v_c \delta R_g.
\end{equation}
However, in order to conserve $E_J$, there must be a gain in non-circular energy that grows in amplitude with $\delta L_z$ as
\begin{equation}\label{eq:dEnc}
    \delta E_{\rm nc}= \Omega_p\, \delta L_z - v_c^2 \ln\left(\dfrac{\delta L_z}{R_p v_c}\right).
\end{equation}
Similarly, a star in a non-circular orbit that is resonant with the \ILR could gain angular momentum, but would circularize as its potential energy increased and the energy associated with non-circular motion decrease as governed by \eq~\ref{eq:dEnc}.

A clear illustration of resonant kinematic cooling from transient spirals is shown in figures throughout \cite{Smock26}. 
In this study, \citeauthor{Smock26} used tracer particle simulations to characterize possible chrono-kinematic signatures for the post-resonant corrugations/wrinkles that were first identified in the \SNd using \textit{Gaia} DR2 data \citep{2019MNRAS.484.3291T}.  
Wrinkles are well reproduced by the orbital responses to the \IOLRs of a transient spiral, where changes in \Lz and \JR follow lines of conserved $E_J$ \citep{2019MNRAS.484.3154S,Smock26}.

Since any given distribution function that approximates the \MW's disk has more stars in \kmy cooler orbits than \kmy hotter orbits, the fraction of stars that \kmy heat at the resonances is greater than the number that cools.  
Again, the overall effect from the resonances is kinematic heating \citep[e.g.][]{LBK72,CS85}, but a non-negligible number of orbits are also resonantly circularized.

Kinematic cooling in association with orbital \rrn in \MW-like galaxies from the FIRE/\textit{Latte} suite of simulations was investigated by \cite{2026arXiv260700077S}. 
Though they did not identify the mechanism for kinematic cooling, it is known that these simulated galaxies host ongoing spiral structure \citep{Quinn26}, each presumably hosting a series of resonances by which stars could be \kmy cooled.

Other potential mechanisms for kinematic cooling have been identified as well.
A process for kinematic cooling was identified by \cite{2020A&A...638A.144K} in a high-resolution N--Body study.  \citeauthor{2020A&A...638A.144K} found that inner disk stars can circularize as they are moved outward by resonant trapping with a slowing bar. They further offer observational \CCD signatures, such as old, metal rich stars in nearly circular orbits in the \SNd.  \cite{2026ApJ..1000..198S} showed that kinematic cooling can also result from orbital scattering off of massive clouds.

A population of very metal-poor giant stars (\FeH$<-1.7$) in the \MW with thin disk-like orbits  \citep{GonzalezRiveradeLaVernhe2024} was recently discovered in a combined sample drawing from the \textit{Gaia}, \textit{APOGEE} \citep{2017AJ....154...94M}, \textit{GALAH} \citep{2015MNRAS.449.2604D}, and \textit{LAMOST} \citep{2012RAA....12.1197C} datasets.  These stars could be remnants from an in-plane merger \citep{2024ApJ...977..278R,2026MNRAS.548ag563S}, but resonant cooling of thick-disk stars is a plausible scenario.
\section{Conclusions}\label{s:Conclusions}

This paper explores how the morphology and nature of spiral structure in galaxies may impact the \orn of disk stars.  A stepwise approach is taken, beginning with an intuition building exercise, to motivate an examination of the \rrn from \cting. This is followed by an analytic treatment to identify scaling relations for how changes in orbital angular momentum from \cting depend on spiral \pangle (opening angle) from a rigidly rotating (e.g., a density-wave) spiral pattern. This results in the following results: 

\begin{itemize}
\item Analytic expressions for the maximum radial excursions of stars trapped at the \CR resonance with a spiral pattern under various sets of assumptions.  Unlike previous derivations that are based on bar symmetry, these expressions account for the non-symmetry of spiral patterns about the radial axis.

\item \textit{More open spiral patterns drive larger changes in orbital angular momentum} than tightly wound spirals of the same strength.

\item The maximum changes in orbital size from a single episode of \textit{\cting would be overestimated by many times} with an expression derived under the assumption of a bar symmetry (\eq~\ref{eqn:SBR1}; as assumed in derivations from \citealt{SB02} and \citealt{BT08}).

\item For low- to moderate-pitch-angle, the maximum radial change in an orbit's guiding center from \cting is best approximated by maximizing the equations of motion (\eq~\ref{eq:R1EoM_pretty}).  This change scales approximately as
\begin{equation}
    \max(\Delta R_g) = 2\max(|R_{\rm 1,s}|) \propto |\Phi_s(\theta)|\cot\theta,
\end{equation}
which is significantly smaller (${\sim}50\%$) than would be expected from approximating this change using \eq~\ref{eqn:SBR1} and the \pangle dependent amplitude of the potential. 

\item For most observed spiral amplitudes and \pangle{s}, radial excursions from \cting are comparable to, or smaller than, those expected from ordinary \epi motion. Previous estimates therefore overestimate the efficiency of \cting driven by spiral structure.
\end{itemize}

The analytical treatment is followed by a numerical exploration. These experiments include 2D and 3D verifications of the scaling relation from \eq~\ref{eq:R1EoM_pretty} as well as a 3D study of a winding spiral. These experiments provide the following results:  

\begin{itemize}
\item Tracer-particle simulations in both 2D and 3D potentials with a rigidly rotating spiral pattern confirm the analytic prediction. Populations `\TypeCR' exhibit progressively larger changes in \RMSL with increasing \pangle and have minimal change in kinematic temperature.

\item When the spiral pattern \crt{s} with the underlying disk ($\Omega_p(R)=\Omega(R)$) and therefore winds from differential rotation, the trend reverses. As a spiral winds to lower \pangle{s}, stars at all radii experience \cting where their changes in angular momentum progressively increase over the lifetime of the spiral. This demonstrates that \textit{the relationship between spiral morphology and \cting  fundamentally depends on the physical nature of spiral structure.}
\end{itemize}

Discussions on the net kinematic temperature change of the disk population are included throughout.  In particular, \cting is here associated with only \kmy cold (no change in kinematic temperature) changes in angular momentum, while separate processes that are \kmy heating and cooling are often necessarily coeval with the trapping at \CR that leads to \cting. This results in the following conclusions:

\begin{itemize}
\item Trapping at \CR leads to \cting, which is \kmy cold.  However, interactions with other resonances away from \CR from the same (non-winding) spiral \textit{simultaneously \kmy heat and \kmy cool orbits in the disk}. The net effect is to increase non-circular motions in the disk (\kmy heat), but \textit{resonant circularization (kinematic cooling) naturally accompanies resonant \rrn, especially at the \ILR}.  
\item Resonant cooling provides a physical explanation for recently identified populations of unexpectedly cold stellar populations in the \MW \citep[e.g.,][]{GonzalezRiveradeLaVernhe2024} and kinematic cooling in simulations \citep[e.g.,][]{2026arXiv260700077S}.

\item The 3D winding spiral experiment showed nearly zero kinematic heating across the disk.

\item Spiral morphology alone is insufficient to determine the efficiency of \cting and its associated kinematic heating and cooling. Observable quantities such as \pangle, spiral amplitude, and pattern speed can only be interpreted in the context of the underlying nature of spiral structure, offering a potential avenue for distinguishing between rigidly rotating density waves and winding spiral models.
\end{itemize}

The dependence of \cting on spiral morphology provides a new theoretical diagnostic for the nature of spiral structure. Density-wave and winding spiral models predict opposite trends with \pangle, suggesting that observational constraints on \cting, together with measurements of spiral morphology, may distinguish between competing theories for the nature of spiral structure.

\section*{Acknowledgements}
KJD, AS, and LC acknowledge support from the Heising Simons Foundation grants \#2022-3927 \& \#2026-6820. 
They also respectfully acknowledge the University of Arizona is on the land and territories of Indigenous peoples. Today, Arizona is home to 22 federally recognized tribes, with Tucson being home to the O’odham and the Yaqui. The University strives to build sustainable relationships with sovereign Native Nations and Indigenous communities through education offerings, partnerships, and community service. 
RFGW is grateful for support from Schmidt Sciences.
\bibliographystyle{aasjournal}
\bibliography{
mypublications,
mybibliography
} 

@ARTICLE{2025MNRAS.544.2777W,
       author = {{Wisnioski}, E. and {Mendel}, J.~T. and {Leaman}, R. and {Tsukui}, T. and {{\"U}bler}, H. and {F{\"o}rster Schreiber}, N.~M.},
        title = "{Evolution of gas velocity dispersion in discs from z {\ensuremath{\sim}} 8 to z {\ensuremath{\sim}} 0.5}",
      journal = {\mnras},
         year = 2025,
        month = dec,
       volume = {544},
       number = {3},
        pages = {2777-2794},
          doi = {10.1093/mnras/staf1606},
archivePrefix = {arXiv},
       eprint = {2505.24129},
 primaryClass = {astro-ph.GA},
       adsurl = {https://ui.adsabs.harvard.edu/abs/2025MNRAS.544.2777W}
}

@ARTICLE{2014ApJS..210....2F,
       author = {{Font}, J. and {Beckman}, J.~E. and {Querejeta}, M. and {Epinat}, B. and {James}, P.~A. and {Blasco-herrera}, J. and {Erroz-Ferrer}, S. and {P{\'e}rez}, I.},
        title = "{Interlocking Resonance Patterns in Galaxy Disks}",
      journal = {\apjs},
         year = 2014,
        month = jan,
       volume = {210},
       number = {1},
          eid = {2},
        pages = {2},
          doi = {10.1088/0067-0049/210/1/2},
archivePrefix = {arXiv},
       eprint = {1310.3415},
 primaryClass = {astro-ph.GA},
       adsurl = {https://ui.adsabs.harvard.edu/abs/2014ApJS..210....2F}
}

@ARTICLE{2025MNRAS.543.2760V,
       author = {{van Donkelaar}, Floor and {Capelo}, Pedro R. and {Mayer}, Lucio and {Reed}, Darren S. and {Quinn}, Thomas R.},
        title = "{Introducing the PHOEBOS simulation: galaxy properties at the dawn of galaxy formation}",
      journal = {\mnras},
         year = 2025,
        month = nov,
       volume = {543},
       number = {3},
        pages = {2760-2780},
          doi = {10.1093/mnras/staf1638},
archivePrefix = {arXiv},
       eprint = {2507.04927},
 primaryClass = {astro-ph.GA},
       adsurl = {https://ui.adsabs.harvard.edu/abs/2025MNRAS.543.2760V}
}

@ARTICLE{2026A&A...709A.120J,
       author = {{Jeanneau}, Alexandre and {Richard}, Johan and {Bouch{\'e}}, Nicolas F. and {Krajnovi{\'c}}, Davor and {Ciocan}, Bianca-Iulia and {Freundlich}, Jonathan and {Epinat}, Beno{\^\i}t and {Contini}, Thierry},
        title = "{MUSE-DARK: II. 3D morpho-kinematic modelling of lensed galaxies: Tully-Fisher relation of z {\ensuremath{\sim}} 1 star-forming galaxies}",
      journal = {\aap},
         year = 2026,
        month = may,
       volume = {709},
          eid = {A120},
        pages = {A120},
          doi = {10.1051/0004-6361/202659953},
archivePrefix = {arXiv},
       eprint = {2603.28856},
 primaryClass = {astro-ph.GA},
       adsurl = {https://ui.adsabs.harvard.edu/abs/2026A&A...709A.120J}
}

@ARTICLE{2025A&A...700A..42E,
       author = {{Espejo Salcedo}, J.~M. and {Pastras}, S. and {V{\'a}cha}, J. and {Pulsoni}, C. and {Genzel}, R. and {F{\"o}rster Schreiber}, N.~M. and {Jolly}, J.-B. and {Barfety}, C. and {Chen}, J. and {Tozzi}, G. and et al.},
        title = "{Galaxy morphologies at cosmic noon with JWST: A foundation for exploring gas transport with bars and spiral arms}",
      journal = {\aap},
         year = 2025,
        month = aug,
       volume = {700},
          eid = {A42},
        pages = {A42},
          doi = {10.1051/0004-6361/202554725},
archivePrefix = {arXiv},
       eprint = {2503.21738},
 primaryClass = {astro-ph.GA},
       adsurl = {https://ui.adsabs.harvard.edu/abs/2025A&A...700A..42E}
}

@ARTICLE{2020ApJ...896...15F,
       author = {{Frankel}, Neige and {Sanders}, Jason and {Ting}, Yuan-Sen and {Rix}, Hans-Walter},
        title = "{Keeping It Cool: Much Orbit Migration, yet Little Heating, in the Galactic Disk}",
      journal = {\apj},
         year = 2020,
        month = jun,
       volume = {896},
       number = {1},
          eid = {15},
        pages = {15},
          doi = {10.3847/1538-4357/ab910c},
archivePrefix = {arXiv},
       eprint = {2002.04622},
 primaryClass = {astro-ph.GA},
       adsurl = {https://ui.adsabs.harvard.edu/abs/2020ApJ...896...15F}
}

@ARTICLE{2021MNRAS.506.1761S,
       author = {{Sharma}, Sanjib and {Hayden}, Michael R. and {Bland-Hawthorn}, Joss and {Stello}, Dennis and {Buder}, Sven and {Zinn}, Joel C. and {Kallinger}, Thomas and {Asplund}, Martin and {De Silva}, Gayandhi M. and {D'Orazi}, Valentina and et al.},
        title = "{Fundamental relations for the velocity dispersion of stars in the Milky Way}",
      journal = {\mnras},
         year = 2021,
        month = sep,
       volume = {506},
       number = {2},
        pages = {1761-1776},
          doi = {10.1093/mnras/stab1086},
archivePrefix = {arXiv},
       eprint = {2004.06556},
 primaryClass = {astro-ph.GA},
       adsurl = {https://ui.adsabs.harvard.edu/abs/2021MNRAS.506.1761S}
}

@ARTICLE{1996MNRAS.279.1263S,
       author = {{Sridhar}, S. and {Touma}, J.},
        title = "{Adiabatic evolution and capture into resonance: vertical heating of a growing stellar disc}",
      journal = {\mnras},
         year = 1996,
        month = apr,
       volume = {279},
        pages = {1263},
          doi = {10.1093/mnras/279.4.1263},
       adsurl = {https://ui.adsabs.harvard.edu/abs/1996MNRAS.279.1263S}
}

@ARTICLE{2000MNRAS.319....1T,
       author = {{Tremaine}, Scott and {Yu}, Qingjuan},
        title = "{Resonant capture, counter-rotating discs, and polar rings}",
      journal = {\mnras},
         year = 2000,
        month = nov,
       volume = {319},
       number = {1},
        pages = {1-7},
          doi = {10.1046/j.1365-8711.2000.03653.x},
archivePrefix = {arXiv},
       eprint = {astro-ph/0001025},
 primaryClass = {astro-ph},
       adsurl = {https://ui.adsabs.harvard.edu/abs/2000MNRAS.319....1T}
}

@ARTICLE{2014MNRAS.437.1284Q,
       author = {{Quillen}, Alice C. and {Minchev}, Ivan and {Sharma}, Sanjib and {Qin}, Yu-Jing and {Di Matteo}, Paola},
        title = "{A vertical resonance heating model for X- or peanut-shaped galactic bulges}",
      journal = {\mnras},
         year = 2014,
        month = jan,
       volume = {437},
       number = {2},
        pages = {1284-1307},
          doi = {10.1093/mnras/stt1972},
archivePrefix = {arXiv},
       eprint = {1307.8441},
 primaryClass = {astro-ph.GA},
       adsurl = {https://ui.adsabs.harvard.edu/abs/2014MNRAS.437.1284Q}
}

@ARTICLE{2002MNRAS.337..578P,
       author = {{Patsis}, P.~A. and {Skokos}, Ch. and {Athanassoula}, E.},
        title = "{Orbital dynamics of three-dimensional bars - III. Boxy/peanut edge-on profiles}",
      journal = {\mnras},
         year = 2002,
        month = dec,
       volume = {337},
       number = {2},
        pages = {578-596},
          doi = {10.1046/j.1365-8711.2002.05943.x},
       adsurl = {https://ui.adsabs.harvard.edu/abs/2002MNRAS.337..578P}
}

@ARTICLE{2025arXiv250117907S,
       author = {{Sellwood}, J A and {Binney}, J},
        title = "{A comment on ``Why is the Galactic disk so cool?'', by Hamilton et al}",
      journal = {arXiv e-prints},
         year = 2025,
        month = jan,
          eid = {arXiv:2501.17907},
        pages = {arXiv:2501.17907},
          doi = {10.48550/arXiv.2501.17907},
archivePrefix = {arXiv},
       eprint = {2501.17907},
 primaryClass = {astro-ph.GA},
       adsurl = {https://ui.adsabs.harvard.edu/abs/2025arXiv250117907S}
}

@ARTICLE{2020MNRAS.495.3175S,
       author = {{Sellwood}, J.~A. and {Gerhard}, Ortwin},
        title = "{Three mechanisms for bar thickening}",
      journal = {\mnras},
         year = 2020,
        month = jul,
       volume = {495},
       number = {3},
        pages = {3175-3191},
          doi = {10.1093/mnras/staa1336},
archivePrefix = {arXiv},
       eprint = {2005.05184},
 primaryClass = {astro-ph.GA},
       adsurl = {https://ui.adsabs.harvard.edu/abs/2020MNRAS.495.3175S}
}

@ARTICLE{2019A&A...629A..52L,
       author = {{{\L}okas}, Ewa L.},
        title = "{Anatomy of a buckling galactic bar}",
      journal = {\aap},
         year = 2019,
        month = sep,
       volume = {629},
          eid = {A52},
        pages = {A52},
          doi = {10.1051/0004-6361/201936056},
archivePrefix = {arXiv},
       eprint = {1906.03916},
 primaryClass = {astro-ph.GA},
       adsurl = {https://ui.adsabs.harvard.edu/abs/2019A&A...629A..52L}
}

@ARTICLE{2025RNAAS...9..236P,
       author = {{Pearlstein}, Tessa and {Masters}, Karen and {G{\'e}ron}, Tobias},
        title = "{Measuring Bar and Spiral Arm Pattern Speeds in MaNGA Barred Spiral Galaxies}",
      journal = {Research Notes of the American Astronomical Society},
         year = 2025,
        month = sep,
       volume = {9},
       number = {9},
          eid = {236},
        pages = {236},
          doi = {10.3847/2515-5172/ae0305},
       adsurl = {https://ui.adsabs.harvard.edu/abs/2025RNAAS...9..236P}
}

@ARTICLE{2024ApJ...968L..15K,
       author = {{Kuhn}, Vicki and {Guo}, Yicheng and {Martin}, Alec and {Bayless}, Julianna and {Gates}, Ellie and {Puleo}, AJ},
        title = "{JWST Reveals a Surprisingly High Fraction of Galaxies Being Spiral-like at 0.5 {\ensuremath{\leq}} z {\ensuremath{\leq}} 4}",
      journal = {\apjl},
         year = 2024,
        month = jun,
       volume = {968},
       number = {2},
          eid = {L15},
        pages = {L15},
          doi = {10.3847/2041-8213/ad43eb},
archivePrefix = {arXiv},
       eprint = {2312.12389},
 primaryClass = {astro-ph.GA},
       adsurl = {https://ui.adsabs.harvard.edu/abs/2024ApJ...968L..15K}
}

@ARTICLE{2024MNRAS.535.2068R,
       author = {{Rowland}, Lucie E. and {Hodge}, Jacqueline and {Bouwens}, Rychard and {Mancera Pi{\~n}a}, Pavel E. and {Hygate}, Alexander and {Algera}, Hiddo and {Aravena}, Manuel and {Bowler}, Rebecca and {da Cunha}, Elisabete and {Dayal}, Pratika and et al.},
        title = "{REBELS-25: discovery of a dynamically cold disc galaxy at z = 7.31}",
      journal = {\mnras},
         year = 2024,
        month = dec,
       volume = {535},
       number = {3},
        pages = {2068-2091},
          doi = {10.1093/mnras/stae2217},
archivePrefix = {arXiv},
       eprint = {2405.06025},
 primaryClass = {astro-ph.GA},
       adsurl = {https://ui.adsabs.harvard.edu/abs/2024MNRAS.535.2068R}
}

@ARTICLE{2023ApJ...958L..26H,
       author = {{Huang}, Shuo and {Kawabe}, Ryohei and {Kohno}, Kotaro and {Saito}, Toshiki and {Mizukoshi}, Shoichiro and {Iono}, Daisuke and {Michiyama}, Tomonari and {Tamura}, Yoichi and {Hayward}, Christopher C. and {Umehata}, Hideki},
        title = "{J0107a: A Barred Spiral Dusty Star-forming Galaxy at z = 2.467}",
      journal = {\apjl},
         year = 2023,
        month = dec,
       volume = {958},
       number = {2},
          eid = {L26},
        pages = {L26},
          doi = {10.3847/2041-8213/acff63},
archivePrefix = {arXiv},
       eprint = {2310.01782},
 primaryClass = {astro-ph.GA},
       adsurl = {https://ui.adsabs.harvard.edu/abs/2023ApJ...958L..26H}
}

@ARTICLE{2023Natur.623..499C,
       author = {{Costantin}, Luca and {P{\'e}rez-Gonz{\'a}lez}, Pablo G. and {Guo}, Yuchen and {Buttitta}, Chiara and {Jogee}, Shardha and {Bagley}, Micaela B. and {Barro}, Guillermo and {Kartaltepe}, Jeyhan S. and {Koekemoer}, Anton M. and {Cabello}, Cristina and et al.},
        title = "{A Milky Way-like barred spiral galaxy at a redshift of 3}",
      journal = {\nat},
         year = 2023,
        month = nov,
       volume = {623},
       number = {7987},
        pages = {499-501},
          doi = {10.1038/s41586-023-06636-x},
archivePrefix = {arXiv},
       eprint = {2311.04283},
 primaryClass = {astro-ph.GA},
       adsurl = {https://ui.adsabs.harvard.edu/abs/2023Natur.623..499C}
}

@ARTICLE{2023ApJ...945L..10G,
       author = {{Guo}, Yuchen and {Jogee}, Shardha and {Finkelstein}, Steven L. and {Chen}, Zilei and {Wise}, Eden and {Bagley}, Micaela B. and {Barro}, Guillermo and {Wuyts}, Stijn and {Kocevski}, Dale D. and {Kartaltepe}, Jeyhan S. and et al.},
        title = "{First Look at z > 1 Bars in the Rest-frame Near-infrared with JWST Early CEERS Imaging}",
      journal = {\apjl},
         year = 2023,
        month = mar,
       volume = {945},
       number = {1},
          eid = {L10},
        pages = {L10},
          doi = {10.3847/2041-8213/acacfb},
archivePrefix = {arXiv},
       eprint = {2210.08658},
 primaryClass = {astro-ph.GA},
       adsurl = {https://ui.adsabs.harvard.edu/abs/2023ApJ...945L..10G}
}

@ARTICLE{2026arXiv260521579M,
       author = {{Modak}, Shaunak and {Hamilton}, Chris and {Ostriker}, Eve C. and {Tremaine}, Scott},
        title = "{Interstellar Medium-Driven Orbital Transport -- I. Radial Heating and Migration}",
      journal = {arXiv e-prints},
         year = 2026,
        month = may,
          eid = {arXiv:2605.21579},
        pages = {arXiv:2605.21579},
          doi = {10.48550/arXiv.2605.21579},
archivePrefix = {arXiv},
       eprint = {2605.21579},
 primaryClass = {astro-ph.GA},
       adsurl = {https://ui.adsabs.harvard.edu/abs/2026arXiv260521579M}
}

@ARTICLE{GonzalezRiveradeLaVernhe2024,
       author = {{Gonz{\'a}lez Rivera de La Vernhe}, Isaure and {Hill}, Vanessa and {Kordopatis}, Georges and {Gran}, Felipe and {Fern{\'a}ndez-Alvar}, Emma and {Ardern-Arentsen}, Anke and {Thomas}, Guillaume F. and {Sestito}, Federico and {Navarrete}, Camila and {Martin}, Nicolas F. and {Starkenburg}, Else and {Viswanathan}, Akshara and {Battaglia}, Giuseppina and {Venn}, Kim A. and {Vitali}, Sara},
        title = "{The Pristine survey: XXIV. The Galactic underdogs: Dynamic tales of a Milky Way metal-poor population}",
      journal = {\aap},
         year = 2024,
        month = dec,
       volume = {692},
          eid = {A131},
        pages = {A131},
          doi = {10.1051/0004-6361/202450513},
archivePrefix = {arXiv},
       eprint = {2406.05728},
 primaryClass = {astro-ph.GA},
       adsurl = {https://ui.adsabs.harvard.edu/abs/2024A&A...692A.131G}
}

@ARTICLE{2010PASP..122.1397S,
       author = {{Sheth}, Kartik and {Regan}, Michael and {Hinz}, Joannah L. and {Gil de Paz}, Armando and {Men{\'e}ndez-Delmestre}, Kar{\'\i}n and {Mu{\~n}oz-Mateos}, Juan-Carlos and {Seibert}, Mark and {Kim}, Taehyun and {Laurikainen}, Eija and {Salo}, Heikki and et al.},
        title = "{The Spitzer Survey of Stellar Structure in Galaxies (S4G)}",
      journal = {\pasp},
         year = 2010,
        month = dec,
       volume = {122},
       number = {898},
        pages = {1397-1414},
          doi = {10.1086/657638},
archivePrefix = {arXiv},
       eprint = {1010.1592},
 primaryClass = {astro-ph.CO},
       adsurl = {https://ui.adsabs.harvard.edu/abs/2010PASP..122.1397S}
}

@ARTICLE{2023RMxAA..59..213B,
    author = {{Barrera-Ballesteros}, J.~K. and {S{\'a}nchez}, S.~F. and {Espinosa-Ponce}, C. and {L{\'o}pez-Cob{\'a}}, C. and {Carigi}, L. and {Lugo-Aranda}, A.~Z. and {Lacerda}, E. and {Bruzual}, G. and {Hernandez-Toledo}, H. and {Boardman}, N. and et al.},
    title = "{SDSS-IV MaNGA: The Radial Distribution of Physical Properties within Galaxies in the Nearby Universe}",
    journal = {\rmxaa},
         year = 2023,
        month = oct,
       volume = {59},
        pages = {213-258},
          doi = {10.22201/ia.01851101p.2023.59.02.06},
archivePrefix = {arXiv},
       eprint = {2206.07058},
 primaryClass = {astro-ph.GA},
       adsurl = {https://ui.adsabs.harvard.edu/abs/2023RMxAA..59..213B}
}

@ARTICLE{2018A&A...609A.119S,
    author = {{S{\'a}nchez-Menguiano}, L. and {S{\'a}nchez}, S.~F. and {P{\'e}rez}, I. and {Ruiz-Lara}, T. and {Galbany}, L. and {Anderson}, J.~P. and {Kr{\"u}hler}, T. and {Kuncarayakti}, H. and {Lyman}, J.~D.},
    title = "{The shape of oxygen abundance profiles explored with MUSE: evidence for widespread deviations from single gradients}",
    journal = {\aap},
    year = 2018,
    month = feb,
    volume = {609},
    eid = {A119},
    pages = {A119},
    doi = {10.1051/0004-6361/201731486},
    archivePrefix = {arXiv},
    eprint = {1710.01188},
    primaryClass = {astro-ph.GA},
    adsurl = {https://ui.adsabs.harvard.edu/abs/2018A&A...609A.119S}
}

@ARTICLE{2016RMxAA..52..171S,
       author = {{S{\'a}nchez}, S.~F. and {P{\'e}rez}, E. and {S{\'a}nchez-Bl{\'a}zquez}, P. and {Garc{\'\i}a-Benito}, R. and {Ibarra-Mede}, H.~J. and {Gonz{\'a}lez}, J.~J. and {Rosales-Ortega}, F.~F. and {S{\'a}nchez-Menguiano}, L. and {Ascasibar}, Y. and {Bitsakis}, T. and et al.},
        title = "{Pipe3D, a pipeline to analyze Integral Field Spectroscopy Data: II. Analysis sequence and CALIFA dataproducts}",
      journal = {\rmxaa},
         year = 2016,
        month = apr,
       volume = {52},
        pages = {171-220},
          doi = {10.48550/arXiv.1602.01830},
archivePrefix = {arXiv},
       eprint = {1602.01830},
 primaryClass = {astro-ph.IM},
       adsurl = {https://ui.adsabs.harvard.edu/abs/2016RMxAA..52..171S}
}

@ARTICLE{2016A&A...595A..62P,
       author = {{P{\'e}rez-Montero}, E. and {Garc{\'\i}a-Benito}, R. and {V{\'\i}lchez}, J.~M. and {S{\'a}nchez}, S.~F. and {Kehrig}, C. and {Husemann}, B. and {Duarte Puertas}, S. and {Iglesias-P{\'a}ramo}, J. and {Galbany}, L. and {Moll{\'a}}, M. and et al.},
        title = "{The dependence of oxygen and nitrogen abundances on stellar mass from the CALIFA survey}",
      journal = {\aap},
         year = 2016,
        month = oct,
       volume = {595},
          eid = {A62},
        pages = {A62},
          doi = {10.1051/0004-6361/201628601},
archivePrefix = {arXiv},
       eprint = {1608.04677},
 primaryClass = {astro-ph.GA},
       adsurl = {https://ui.adsabs.harvard.edu/abs/2016A&A...595A..62P}
}

@ARTICLE{2015ApJ...798....7B,
       author = {{Bundy}, Kevin and {Bershady}, Matthew A. and {Law}, David R. and {Yan}, Renbin and {Drory}, Niv and {MacDonald}, Nicholas and {Wake}, David A. and {Cherinka}, Brian and {S{\'a}nchez-Gallego}, Jos{\'e} R. and {Weijmans}, Anne-Marie and et al.},
        title = "{Overview of the SDSS-IV MaNGA Survey: Mapping nearby Galaxies at Apache Point Observatory}",
      journal = {\apj},
         year = 2015,
        month = jan,
       volume = {798},
       number = {1},
          eid = {7},
        pages = {7},
          doi = {10.1088/0004-637X/798/1/7},
archivePrefix = {arXiv},
       eprint = {1412.1482},
 primaryClass = {astro-ph.GA},
       adsurl = {https://ui.adsabs.harvard.edu/abs/2015ApJ...798....7B}
}

@ARTICLE{1986MNRAS.221..195S,
       author = {{Sellwood}, J.~A. and {Athanassoula}, E.},
        title = "{Unstable modes from galaxy simulations}",
      journal = {\mnras},
         year = 1986,
        month = jul,
       volume = {221},
        pages = {195-212},
          doi = {10.1093/mnras/221.2.195},
       adsurl = {https://ui.adsabs.harvard.edu/abs/1986MNRAS.221..195S}
}

@ARTICLE{Chugunov2025,
       author = {{Chugunov}, Ilia V. and {Marchuk}, Alexander A. and {Mosenkov}, Aleksandr V.},
        title = "{Less wound and more asymmetric: JWST confirms the evolution of spiral structure in galaxies at z {\ensuremath{\lesssim}} 3}",
      journal = {\pasa},
         year = 2025,
        month = jan,
       volume = {42},
          eid = {e029},
        pages = {e029},
          doi = {10.1017/pasa.2025.6},
archivePrefix = {arXiv},
       eprint = {2501.11670},
 primaryClass = {astro-ph.GA},
       adsurl = {https://ui.adsabs.harvard.edu/abs/2025PASA...42...29C}
}

@ARTICLE{Reshetnikov2023,
       author = {{Reshetnikov}, V.~P. and {Marchuk}, A.~A. and {Chugunov}, I.~V. and {Usachev}, P.~A. and {Mosenkov}, A.~V.},
        title = "{The possible evolution of pitch angles of spiral galaxies}",
      journal = {\aap},
         year = 2023,
        month = dec,
       volume = {680},
          eid = {L14},
        pages = {L14},
          doi = {10.1051/0004-6361/202348449},
archivePrefix = {arXiv},
       eprint = {2311.16915},
 primaryClass = {astro-ph.GA},
       adsurl = {https://ui.adsabs.harvard.edu/abs/2023A&A...680L..14R}
}

@ARTICLE{Kuhn2026,
       author = {{Kuhn}, Vicki and {Guo}, Yicheng and {Rentschler}, Sophie and {Castillo}, Maxmillian and {Nandi}, Gourab and {Dugdale}, Ellie and {Mitiku}, Tsinat},
        title = "{Spiral Arms across Cosmic Time: JWST Measurements of the Pitch Angles of Spiral Galaxies at z < 3.5}",
      journal = {\apj},
         year = 2026,
        month = aug,
       volume = {1006},
       number = {2},
          eid = {137},
        pages = {137},
          doi = {10.3847/1538-4357/ae77fe},
archivePrefix = {arXiv},
       eprint = {2606.11315},
 primaryClass = {astro-ph.GA},
       adsurl = {https://ui.adsabs.harvard.edu/abs/2026ApJ..1006..137K}
}

@ARTICLE{2020A&A...638A.144K,
       author = {{Khoperskov}, S. and {Di Matteo}, P. and {Haywood}, M. and {G{\'o}mez}, A. and {Snaith}, O.~N.},
        title = "{Escapees from the bar resonances. Presence of low-eccentricity metal-rich stars at the solar vicinity}",
      journal = {\aap},
         year = 2020,
        month = jun,
       volume = {638},
          eid = {A144},
        pages = {A144},
          doi = {10.1051/0004-6361/201937188},
archivePrefix = {arXiv},
       eprint = {1911.12424},
 primaryClass = {astro-ph.GA},
       adsurl = {https://ui.adsabs.harvard.edu/abs/2020A&A...638A.144K}
}

@ARTICLE{2026MNRAS.548ag563S,
       author = {{Sestito}, Federico and {Fern{\'a}ndez-Alvar}, Emma and {Brooks}, Rebecca and {Olson}, Emma and {Carigi}, Leticia and {Jofr{\'e}}, Paula and {Silva}, Danielle de Brito and {Eldridge}, Camilla J.~L. and {Vitali}, Sara and {Venn}, Kim A. and et al.},
        title = "{An ancient system hidden in the Galactic plane?}",
      journal = {\mnras},
         year = 2026,
        month = may,
       volume = {548},
       number = {2},
          eid = {stag563},
        pages = {stag563},
          doi = {10.1093/mnras/stag563},
archivePrefix = {arXiv},
       eprint = {2409.13813},
 primaryClass = {astro-ph.GA},
       adsurl = {https://ui.adsabs.harvard.edu/abs/2026MNRAS.548ag563S}
}

@ARTICLE{2024ApJ...977..278R,
       author = {{Re Fiorentin}, Paola and {Spagna}, Alessandro and {Lattanzi}, Mario G. and {Cignoni}, Michele and {Vitali}, Sara},
        title = "{Icarus Revisited: An Ancient, Metal-poor Accreted Stellar Stream in the Disk of the Milky Way}",
      journal = {\apj},
         year = 2024,
        month = dec,
       volume = {977},
       number = {2},
          eid = {278},
        pages = {278},
          doi = {10.3847/1538-4357/ad8887},
archivePrefix = {arXiv},
       eprint = {2410.12581},
 primaryClass = {astro-ph.GA},
       adsurl = {https://ui.adsabs.harvard.edu/abs/2024ApJ...977..278R}
}

@ARTICLE{2017AJ....154...94M,
       author = {{Majewski}, Steven R. and {Schiavon}, Ricardo P. and {Frinchaboy}, Peter M. and {Allende Prieto}, Carlos and {Barkhouser}, Robert and {Bizyaev}, Dmitry and {Blank}, Basil and {Brunner}, Sophia and {Burton}, Adam and {Carrera}, Ricardo and et al.},
        title = "{The Apache Point Observatory Galactic Evolution Experiment (APOGEE)}",
      journal = {\aj},
         year = 2017,
        month = sep,
       volume = {154},
       number = {3},
          eid = {94},
        pages = {94},
          doi = {10.3847/1538-3881/aa784d},
archivePrefix = {arXiv},
       eprint = {1509.05420},
 primaryClass = {astro-ph.IM},
       adsurl = {https://ui.adsabs.harvard.edu/abs/2017AJ....154...94M}
}

@ARTICLE{2015MNRAS.449.2604D,
       author = {{De Silva}, G.~M. and {Freeman}, K.~C. and {Bland-Hawthorn}, J. and {Martell}, S. and {de Boer}, E. Wylie and {Asplund}, M. and {Keller}, S. and {Sharma}, S. and {Zucker}, D.~B. and {Zwitter}, T. and et al.},
        title = "{The GALAH survey: scientific motivation}",
      journal = {\mnras},
         year = 2015,
        month = may,
       volume = {449},
       number = {3},
        pages = {2604-2617},
          doi = {10.1093/mnras/stv327},
archivePrefix = {arXiv},
       eprint = {1502.04767},
 primaryClass = {astro-ph.GA},
       adsurl = {https://ui.adsabs.harvard.edu/abs/2015MNRAS.449.2604D}
}

@ARTICLE{2012RAA....12.1197C,
       author = {{Cui}, Xiang-Qun and {Zhao}, Yong-Heng and {Chu}, Yao-Quan and {Li}, Guo-Ping and {Li}, Qi and {Zhang}, Li-Ping and {Su}, Hong-Jun and {Yao}, Zheng-Qiu and {Wang}, Ya-Nan and {Xing}, Xiao-Zheng and et al.},
        title = "{The Large Sky Area Multi-Object Fiber Spectroscopic Telescope (LAMOST)}",
      journal = {Research in Astronomy and Astrophysics},
         year = 2012,
        month = sep,
       volume = {12},
       number = {9},
        pages = {1197-1242},
          doi = {10.1088/1674-4527/12/9/003},
       adsurl = {https://ui.adsabs.harvard.edu/abs/2012RAA....12.1197C}
}

@ARTICLE{2018ApJ...854..182B,
       author = {{Beckman}, John E. and {Font}, Joan and {Borlaff}, Alejandro and {Garc{\'\i}a-Lorenzo}, Bego{\~n}a},
        title = "{Precision Determination of Corotation Radii in Galaxy Disks: Tremaine-Weinberg versus Font-Beckman for NGC 3433}",
      journal = {\apj},
         year = 2018,
        month = feb,
       volume = {854},
       number = {2},
          eid = {182},
        pages = {182},
          doi = {10.3847/1538-4357/aaa965},
archivePrefix = {arXiv},
       eprint = {1801.07476},
 primaryClass = {astro-ph.GA},
       adsurl = {https://ui.adsabs.harvard.edu/abs/2018ApJ...854..182B}
}

@ARTICLE{1984ApJ...282L...5T,
       author = {{Tremaine}, S. and {Weinberg}, M.~D.},
        title = "{A kinematic method for measuring the pattern speed of barred galaxies.}",
      journal = {\apjl},
         year = 1984,
        month = jul,
       volume = {282},
        pages = {L5-L7},
          doi = {10.1086/184292},
       adsurl = {https://ui.adsabs.harvard.edu/abs/1984ApJ...282L...5T}
}

@ARTICLE{2011ApJ...741L..14F,
       author = {{Font}, Joan and {Beckman}, John E. and {Epinat}, Beno{\^\i}t and {Fathi}, Kambiz and {Guti{\'e}rrez}, Leonel and {Hernandez}, Olivier},
        title = "{Resonant Structure in the Disks of Spiral Galaxies, Using Phase Reversals in Streaming Motions from Two-dimensional H{\ensuremath{\alpha}} Fabry-Perot Spectroscopy}",
      journal = {\apjl},
         year = 2011,
        month = nov,
       volume = {741},
       number = {1},
          eid = {L14},
        pages = {L14},
          doi = {10.1088/2041-8205/741/1/L14},
archivePrefix = {arXiv},
       eprint = {1109.5574},
 primaryClass = {astro-ph.GA},
       adsurl = {https://ui.adsabs.harvard.edu/abs/2011ApJ...741L..14F}
}

@ARTICLE{2026A&A...706A..31M,
       author = {{Minchev}, I. and {Attard}, K. and {Ratcliffe}, B. and {Martig}, M. and {Walcher}, J. and {Khoperskov}, S. and {Bernaldez}, J.~P. and {Marques}, L. and {Sysoliatina}, K. and {Chiappini}, C. and et al.},
        title = "{The impact of radial migration on disk galaxy star formation histories: I. Biases in spatially resolved estimates}",
      journal = {\aap},
         year = 2026,
        month = jan,
       volume = {706},
          eid = {A31},
        pages = {A31},
          doi = {10.1051/0004-6361/202556988},
archivePrefix = {arXiv},
       eprint = {2508.18367},
 primaryClass = {astro-ph.GA},
       adsurl = {https://ui.adsabs.harvard.edu/abs/2026A&A...706A..31M}
}

@ARTICLE{2026A&A...707A.147B,
       author = {{Bernaldez}, J.~P. and {Minchev}, I. and {Ratcliffe}, B. and {Marques}, L. and {Sysoliatina}, K. and {Walcher}, J. and {Khoperskov}, S. and {Martig}, M. and {de Jong}, R.~S. and {Steinmetz}, M.},
        title = "{The impact of radial migration on disk galaxy star formation histories: II. Role of bar strength, disk thickness, and merger history}",
      journal = {\aap},
         year = 2026,
        month = mar,
       volume = {707},
          eid = {A147},
        pages = {A147},
          doi = {10.1051/0004-6361/202557002},
archivePrefix = {arXiv},
       eprint = {2508.19340},
 primaryClass = {astro-ph.GA},
       adsurl = {https://ui.adsabs.harvard.edu/abs/2026A&A...707A.147B}
}

@ARTICLE{2025A&A...701A..88M,
       author = {{Marques}, L. and {Minchev}, I. and {Ratcliffe}, B. and {Khoperskov}, S. and {Steinmetz}, M. and {Wenger}, T.~V. and {Buck}, T. and {Martig}, M. and {Kordopatis}, G. and {Schultheis}, M. and et al.},
        title = "{Bar-spiral interaction induces radial migration and star formation bursts}",
      journal = {\aap},
         year = 2025,
        month = sep,
       volume = {701},
          eid = {A88},
        pages = {A88},
          doi = {10.1051/0004-6361/202554020},
archivePrefix = {arXiv},
       eprint = {2502.02651},
 primaryClass = {astro-ph.GA},
       adsurl = {https://ui.adsabs.harvard.edu/abs/2025A&A...701A..88M}
}

@ARTICLE{2022A&A...663A..38K,
       author = {{Khoperskov}, Sergey and {Gerhard}, Ortwin},
        title = "{Chemo-kinematics of the Milky Way spiral arms and bar resonances: Connection to ridges and moving groups in the solar vicinity}",
      journal = {\aap},
         year = 2022,
        month = jul,
       volume = {663},
          eid = {A38},
        pages = {A38},
          doi = {10.1051/0004-6361/202141836},
archivePrefix = {arXiv},
       eprint = {2111.15211},
 primaryClass = {astro-ph.GA},
       adsurl = {https://ui.adsabs.harvard.edu/abs/2022A&A...663A..38K}
}

@ARTICLE{2026ApJ..1000..198S,
       author = {{Struck}, Curtis and {Elmegreen}, Bruce G. and {D'Onghia}, Elena},
        title = "{Scattering, Migration, Recircularization and Relaxation to Build out Galaxy Disks with Exponential Profiles}",
      journal = {\apj},
         year = 2026,
        month = apr,
       volume = {1000},
       number = {2},
          eid = {198},
        pages = {198},
          doi = {10.3847/1538-4357/ae4d1b},
archivePrefix = {arXiv},
       eprint = {2602.21171},
 primaryClass = {astro-ph.GA},
       adsurl = {https://ui.adsabs.harvard.edu/abs/2026ApJ..1000..198S}
}

@ARTICLE{2023ApJ...946L..13F,
       author = {{Finkelstein}, Steven L. and {Bagley}, Micaela B. and {Ferguson}, Henry C. and {Wilkins}, Stephen M. and {Kartaltepe}, Jeyhan S. and {Papovich}, Casey and {Yung}, L.~Y. Aaron and {Arrabal Haro}, Pablo and {Behroozi}, Peter and {Dickinson}, Mark and et al.},
        title = "{CEERS Key Paper. I. An Early Look into the First 500 Myr of Galaxy Formation with JWST}",
      journal = {\apjl},
         year = 2023,
        month = mar,
       volume = {946},
       number = {1},
          eid = {L13},
        pages = {L13},
          doi = {10.3847/2041-8213/acade4},
archivePrefix = {arXiv},
       eprint = {2211.05792},
 primaryClass = {astro-ph.GA},
       adsurl = {https://ui.adsabs.harvard.edu/abs/2023ApJ...946L..13F}
}

@ARTICLE{2023Natur.616..266L,
       author = {{Labb{\'e}}, Ivo and {van Dokkum}, Pieter and {Nelson}, Erica and {Bezanson}, Rachel and {Suess}, Katherine A. and {Leja}, Joel and {Brammer}, Gabriel and {Whitaker}, Katherine and {Mathews}, Elijah and {Stefanon}, Mauro and et al.},
        title = "{A population of red candidate massive galaxies  600 Myr after the Big Bang}",
      journal = {\nat},
         year = 2023,
        month = apr,
       volume = {616},
       number = {7956},
        pages = {266-269},
          doi = {10.1038/s41586-023-05786-2},
archivePrefix = {arXiv},
       eprint = {2207.12446},
 primaryClass = {astro-ph.GA},
       adsurl = {https://ui.adsabs.harvard.edu/abs/2023Natur.616..266L}
}

@ARTICLE{2022ApJ...940L..14N,
       author = {{Naidu}, Rohan P. and {Oesch}, Pascal A. and {van Dokkum}, Pieter and {Nelson}, Erica J. and {Suess}, Katherine A. and {Brammer}, Gabriel and {Whitaker}, Katherine E. and {Illingworth}, Garth and {Bouwens}, Rychard and {Tacchella}, Sandro and et al.},
        title = "{Two Remarkably Luminous Galaxy Candidates at z {\ensuremath{\approx}} 10-12 Revealed by JWST}",
      journal = {\apjl},
         year = 2022,
        month = nov,
       volume = {940},
       number = {1},
          eid = {L14},
        pages = {L14},
          doi = {10.3847/2041-8213/ac9b22},
archivePrefix = {arXiv},
       eprint = {2207.09434},
 primaryClass = {astro-ph.GA},
       adsurl = {https://ui.adsabs.harvard.edu/abs/2022ApJ...940L..14N}
}

@ARTICLE{2024MNRAS.527.6926M,
       author = {{McCluskey}, Fiona and {Wetzel}, Andrew and {Loebman}, Sarah R. and {Moreno}, Jorge and {Faucher-Gigu{\`e}re}, Claude-Andr{\'e} and {Hopkins}, Philip F.},
        title = "{Disc settling and dynamical heating: histories of Milky Way-mass stellar discs across cosmic time in the FIRE simulations}",
      journal = {\mnras},
         year = 2024,
        month = jan,
       volume = {527},
       number = {3},
        pages = {6926-6949},
          doi = {10.1093/mnras/stad3547},
archivePrefix = {arXiv},
       eprint = {2303.14210},
 primaryClass = {astro-ph.GA},
       adsurl = {https://ui.adsabs.harvard.edu/abs/2024MNRAS.527.6926M}
}

@ARTICLE{2021MNRAS.503.1815B,
       author = {{Bird}, Jonathan C. and {Loebman}, Sarah R. and {Weinberg}, David H. and {Brooks}, Alyson M. and {Quinn}, Thomas R. and {Christensen}, Charlotte R.},
        title = "{Inside out and upside-down: The roles of gas cooling and dynamical heating in shaping the stellar age-velocity relation}",
      journal = {\mnras},
         year = 2021,
        month = may,
       volume = {503},
       number = {2},
        pages = {1815-1827},
          doi = {10.1093/mnras/stab289},
archivePrefix = {arXiv},
       eprint = {2005.12948},
 primaryClass = {astro-ph.GA},
       adsurl = {https://ui.adsabs.harvard.edu/abs/2021MNRAS.503.1815B}
}

@ARTICLE{2022MNRAS.514..689B,
       author = {{Belokurov}, Vasily and {Kravtsov}, Andrey},
        title = "{From dawn till disc: Milky Way's turbulent youth revealed by the APOGEE+Gaia data}",
      journal = {\mnras},
         year = 2022,
        month = jul,
       volume = {514},
       number = {1},
        pages = {689-714},
          doi = {10.1093/mnras/stac1267},
archivePrefix = {arXiv},
       eprint = {2203.04980},
 primaryClass = {astro-ph.GA},
       adsurl = {https://ui.adsabs.harvard.edu/abs/2022MNRAS.514..689B}
}

@ARTICLE{2025A&A...700A..89K,
       author = {{Khoperskov}, Sergey and {Steinmetz}, Matthias and {Haywood}, Misha and {van de Ven}, Glenn and {Krajnovi{\'c}}, Davor and {Ratcliffe}, Bridget and {Minchev}, Ivan and {Di Matteo}, Paola and {Kacharov}, Nikolay and {Marques}, L{\'e}a and et al.},
        title = "{Rediscovering the Milky Way with an orbit superposition approach and APOGEE data: II. Chrono-chemo-kinematics of the disc}",
      journal = {\aap},
         year = 2025,
        month = aug,
       volume = {700},
          eid = {A89},
        pages = {A89},
          doi = {10.1051/0004-6361/202453305},
archivePrefix = {arXiv},
       eprint = {2411.16866},
 primaryClass = {astro-ph.GA},
       adsurl = {https://ui.adsabs.harvard.edu/abs/2025A&A...700A..89K}
}

@ARTICLE{2012ApJ...758..106K,
       author = {{Kassin}, Susan A. and {Weiner}, Benjamin J. and {Faber}, S.~M. and {Gardner}, Jonathan P. and {Willmer}, C.~N.~A. and {Coil}, Alison L. and {Cooper}, Michael C. and {Devriendt}, Julien and {Dutton}, Aaron A. and {Guhathakurta}, Puragra and et al.},
        title = "{The Epoch of Disk Settling: z \raisebox{-0.5ex}\textasciitilde 1 to Now}",
      journal = {\apj},
         year = 2012,
        month = oct,
       volume = {758},
       number = {2},
          eid = {106},
        pages = {106},
          doi = {10.1088/0004-637X/758/2/106},
archivePrefix = {arXiv},
       eprint = {1207.7072},
 primaryClass = {astro-ph.CO},
       adsurl = {https://ui.adsabs.harvard.edu/abs/2012ApJ...758..106K}
}

@ARTICLE{2020MNRAS.495.3295M,
       author = {{Mikkola}, Daniel and {McMillan}, Paul J. and {Hobbs}, David},
        title = "{Radial migration and vertical action in N-body simulations}",
      journal = {\mnras},
         year = 2020,
        month = jul,
       volume = {495},
       number = {3},
        pages = {3295-3306},
          doi = {10.1093/mnras/staa1223},
archivePrefix = {arXiv},
       eprint = {2004.13646},
 primaryClass = {astro-ph.GA},
       adsurl = {https://ui.adsabs.harvard.edu/abs/2020MNRAS.495.3295M}
}

@ARTICLE{phangs_2024,
       author = {{Williams}, Thomas G. and {Lee}, Janice C. and {Larson}, Kirsten L. and {Leroy}, Adam K. and {Sandstrom}, Karin and {Schinnerer}, Eva and {Thilker}, David A. and {Belfiore}, Francesco and {Egorov}, Oleg V. and {Rosolowsky}, Erik and {Sutter}, Jessica and {DePasquale}, Joseph and {Pagan}, Alyssa and {Berger}, Travis A. and {Anand}, Gagandeep S. and {Barnes}, Ashley T. and {Bigiel}, Frank and {Boquien}, M{\'e}d{\'e}ric and {Cao}, Yixian and {Chastenet}, J{\'e}r{\'e}my and {Chevance}, M{\'e}lanie and {Chown}, Ryan and {Dale}, Daniel A. and {Deger}, Sinan and {Eibensteiner}, Cosima and {Emsellem}, Eric and {Faesi}, Christopher M. and {Glover}, Simon C.~O. and {Grasha}, Kathryn and {Hannon}, Stephen and {Hassani}, Hamid and {Henshaw}, Jonathan D. and {Jim{\'e}nez-Donaire}, Mar{\'\i}a J. and {Kim}, Jaeyeon and {Klessen}, Ralf S. and {Koch}, Eric W. and {Li}, Jing and {Liu}, Daizhong and {Meidt}, Sharon E. and {M{\'e}ndez-Delgado}, J. Eduardo and {Murphy}, Eric J. and {Neumann}, Justus and {Neumann}, Lukas and {Neumayer}, Nadine and {Oakes}, Elias K. and {Pathak}, Debosmita and {Pety}, J{\'e}r{\^o}me and {Pinna}, Francesca and {Querejeta}, Miguel and {Ramambason}, Lise and {Romanelli}, Andrea and {Sormani}, Mattia C. and {Stuber}, Sophia K. and {Sun}, Jiayi and {Teng}, Yu-Hsuan and {Usero}, Antonio and {Watkins}, Elizabeth J. and {Weinbeck}, Tony D.},
        title = "{PHANGS-JWST: Data-processing Pipeline and First Full Public Data Release}",
      journal = {\apjs},
         year = 2024,
        month = jul,
       volume = {273},
       number = {1},
          eid = {13},
        pages = {13},
          doi = {10.3847/1538-4365/ad4be5},
archivePrefix = {arXiv},
       eprint = {2401.15142},
 primaryClass = {astro-ph.GA},
       adsurl = {https://ui.adsabs.harvard.edu/abs/2024ApJS..273...13W}
}

@ARTICLE{2022ARA&A..60...73S,
       author = {{Sellwood}, J.~A. and {Masters}, Karen L.},
        title = "{Spirals in Galaxies}",
      journal = {\araa},
         year = 2022,
        month = aug,
       volume = {60},
          doi = {10.1146/annurev-astro-052920-104505},
archivePrefix = {arXiv},
       eprint = {2110.05615},
 primaryClass = {astro-ph.GA},
       adsurl = {https://ui.adsabs.harvard.edu/abs/2022ARA&A..60...73S}
}

@ARTICLE{2011MNRAS.410..166L,
       author = {{Lintott}, Chris and {Schawinski}, Kevin and {Bamford}, Steven and {Slosar}, An{\r{a}}{\textthreequarters}e and {Land}, Kate and {Thomas}, Daniel and {Edmondson}, Edd and {Masters}, Karen and {Nichol}, Robert C. and {Raddick}, M. Jordan and et al.},
        title = "{Galaxy Zoo 1: data release of morphological classifications for nearly 900 000 galaxies}",
      journal = {\mnras},
         year = 2011,
        month = jan,
       volume = {410},
       number = {1},
        pages = {166-178},
          doi = {10.1111/j.1365-2966.2010.17432.x},
archivePrefix = {arXiv},
       eprint = {1007.3265},
 primaryClass = {astro-ph.GA},
       adsurl = {https://ui.adsabs.harvard.edu/abs/2011MNRAS.410..166L}
}

@ARTICLE{2000AJ....120.1579Y,
       author = {{York}, Donald G. and {Adelman}, J. and {Anderson}, Jr., John E. and {Anderson}, Scott F. and {Annis}, James and {Bahcall}, Neta A. and {Bakken}, J.~A. and {Barkhouser}, Robert and {Bastian}, Steven and {Berman}, Eileen and et al.},
        title = "{The Sloan Digital Sky Survey: Technical Summary}",
      journal = {\aj},
         year = 2000,
        month = sep,
       volume = {120},
       number = {3},
        pages = {1579-1587},
          doi = {10.1086/301513},
archivePrefix = {arXiv},
       eprint = {astro-ph/0006396},
 primaryClass = {astro-ph},
       adsurl = {https://ui.adsabs.harvard.edu/abs/2000AJ....120.1579Y}
}

@ARTICLE{2019NatAs...3..178P,
       author = {{Peterken}, Thomas G. and {Merrifield}, Michael R. and {Arag{\'o}n-Salamanca}, Alfonso and {Drory}, Niv and {Krawczyk}, Coleman M. and {Masters}, Karen L. and {Weijmans}, Anne-Marie and {Westfall}, Kyle B.},
        title = "{A direct test of density wave theory in a grand-design spiral galaxy}",
      journal = {Nature Astronomy},
         year = 2019,
        month = feb,
       volume = {3},
        pages = {178-182},
          doi = {10.1038/s41550-018-0627-5},
archivePrefix = {arXiv},
       eprint = {1809.08048},
 primaryClass = {astro-ph.GA},
       adsurl = {https://ui.adsabs.harvard.edu/abs/2019NatAs...3..178P}
}

@ARTICLE{2025A&A...701A.183Q,
       author = {{Querejeta}, Miguel and {Meidt}, Sharon E. and {Cao}, Yixian and {Colombo}, Dario and {Emsellem}, Eric and {Garc{\'\i}a-Burillo}, Santiago and {Klessen}, Ralf S. and {Koch}, Eric W. and {Leroy}, Adam K. and {Ruiz-Garc{\'\i}a}, Marina and et al.},
        title = "{Azimuthal offsets in spiral arms of nearby galaxies}",
      journal = {\aap},
         year = 2025,
        month = sep,
       volume = {701},
          eid = {A183},
        pages = {A183},
          doi = {10.1051/0004-6361/202556175},
archivePrefix = {arXiv},
       eprint = {2509.01668},
 primaryClass = {astro-ph.GA},
       adsurl = {https://ui.adsabs.harvard.edu/abs/2025A&A...701A.183Q}
}

@ARTICLE{2025ApJ...981..115S,
       author = {{Speights}, Jason C. and {Aust}, Virginia and {Lu}, Qinyan},
        title = "{Spiral Density Waves in the Multiple-armed Galaxy NGC 628}",
      journal = {\apj},
         year = 2025,
        month = mar,
       volume = {981},
       number = {2},
          eid = {115},
        pages = {115},
          doi = {10.3847/1538-4357/adb1b3},
       adsurl = {https://ui.adsabs.harvard.edu/abs/2025ApJ...981..115S}
}

@ARTICLE{2022MNRAS.512..366A,
       author = {{Abdeen}, Shameer and {Davis}, Benjamin L. and {Eufrasio}, Rafael and {Kennefick}, Daniel and {Kennefick}, Julia and {Miller}, Ryan and {Shields}, Deanna and {Monson}, Erik B. and {Bassett}, Calla and {O'Mara}, Harry},
        title = "{Evidence in favour of density wave theory through age gradients observed in star formation history maps and spatially resolved stellar clusters}",
      journal = {\mnras},
         year = 2022,
        month = may,
       volume = {512},
       number = {1},
        pages = {366-377},
          doi = {10.1093/mnras/stac459},
archivePrefix = {arXiv},
       eprint = {2010.14540},
 primaryClass = {astro-ph.GA},
       adsurl = {https://ui.adsabs.harvard.edu/abs/2022MNRAS.512..366A}
}

@ARTICLE{2019MNRAS.487.1808M,
       author = {{Masters}, Karen L. and {Lintott}, Chris J. and {Hart}, Ross E. and {Kruk}, Sandor J. and {Smethurst}, Rebecca J. and {Casteels}, Kevin V. and {Keel}, William C. and {Simmons}, Brooke D. and {Stanescu}, Dennis O. and {Tate}, Jean and et al.},
        title = "{Galaxy Zoo: unwinding the winding problem - observations of spiral bulge prominence and arm pitch angles suggest local spiral galaxies are winding}",
      journal = {\mnras},
         year = 2019,
        month = aug,
       volume = {487},
       number = {2},
        pages = {1808-1820},
          doi = {10.1093/mnras/stz1153},
archivePrefix = {arXiv},
       eprint = {1904.11436},
 primaryClass = {astro-ph.GA},
       adsurl = {https://ui.adsabs.harvard.edu/abs/2019MNRAS.487.1808M}
}

@ARTICLE{2019MNRAS.490.1470P,
       author = {{Pringle}, J.~E. and {Dobbs}, C.~L.},
        title = "{The evolution of pitch angles of spiral arms}",
      journal = {\mnras},
         year = 2019,
        month = nov,
       volume = {490},
       number = {1},
        pages = {1470-1473},
          doi = {10.1093/mnras/stz2694},
archivePrefix = {arXiv},
       eprint = {1909.10291},
 primaryClass = {astro-ph.GA},
       adsurl = {https://ui.adsabs.harvard.edu/abs/2019MNRAS.490.1470P}
}

@ARTICLE{2020MNRAS.493.2111G,
       author = {{Griv}, Evgeny and {Gedalin}, Michael and {Shih}, I.-Chun and {Hou}, Li-Gang and {Jiang}, Ing-Guey},
        title = "{The nearby spiral density-wave structure of the Galaxy: line-of-sight velocities of the Gaia DR2 main-sequence A, F, G, and K stars}",
      journal = {\mnras},
         year = 2020,
        month = apr,
       volume = {493},
       number = {2},
        pages = {2111-2126},
          doi = {10.1093/mnras/staa357},
       adsurl = {https://ui.adsabs.harvard.edu/abs/2020MNRAS.493.2111G}
}

@ARTICLE{2021RAA....21....9H,
       author = {{He}, Zhi-Hong and {Xu}, Ye and {Hou}, Li-Gang},
        title = "{Search for age pattern across spiral arms of the Milky Way}",
      journal = {Research in Astronomy and Astrophysics},
         year = 2021,
        month = jan,
       volume = {21},
       number = {1},
          eid = {009},
        pages = {009},
          doi = {10.1088/1674-4527/21/1/9},
       adsurl = {https://ui.adsabs.harvard.edu/abs/2021RAA....21....9H}
}

@ARTICLE{2026MNRAS.548ag774G,
       author = {{Grand}, Robert J.~J. and {Fragkoudi}, Francesca and {Pakmor}, R{\"u}diger and {G{\'o}mez}, Facundo A. and {van de Voort}, F. Freeke and {Bieri}, Rebekka and {Townson}, Sophie},
        title = "{The diverse nature of spiral arms in the AURIGA SUPERSTARS cosmological hydrodynamic simulations}",
      journal = {\mnras},
         year = 2026,
        month = jun,
       volume = {548},
       number = {4},
          eid = {stag774},
        pages = {stag774},
          doi = {10.1093/mnras/stag774},
archivePrefix = {arXiv},
       eprint = {2602.15108},
 primaryClass = {astro-ph.GA},
       adsurl = {https://ui.adsabs.harvard.edu/abs/2026MNRAS.548ag774G}
}

@ARTICLE{2012MNRAS.426L..46A,
       author = {{Athanassoula}, E.},
        title = "{Manifold-driven spirals in N-body barred galaxy simulations}",
      journal = {\mnras},
         year = 2012,
        month = oct,
       volume = {426},
       number = {1},
        pages = {L46-L50},
          doi = {10.1111/j.1745-3933.2012.01320.x},
archivePrefix = {arXiv},
       eprint = {1207.4590},
 primaryClass = {astro-ph.GA},
       adsurl = {https://ui.adsabs.harvard.edu/abs/2012MNRAS.426L..46A}
}

@BOOK{1996ssgd.book.....B,
       author = {{Bertin}, G. and {Lin}, C.~C.},
       title = "{Spiral structure in galaxies a density wave theory}",
       year = 1996,
       publisher = {MIT Press},
       adsurl = {https://ui.adsabs.harvard.edu/abs/1996ssgd.book.....B}
}

@ARTICLE{2013A&A...553A..77G,
       author = {{Grand}, R.~J.~J. and {Kawata}, D. and {Cropper}, M.},
        title = "{Spiral arm pitch angle and galactic shear rate in N-body simulations of disc galaxies}",
      journal = {\aap},
         year = 2013,
        month = may,
       volume = {553},
          eid = {A77},
        pages = {A77},
          doi = {10.1051/0004-6361/201321308},
archivePrefix = {arXiv},
       eprint = {1209.4083},
 primaryClass = {astro-ph.GA},
       adsurl = {https://ui.adsabs.harvard.edu/abs/2013A&A...553A..77G}
}

@ARTICLE{2017MNRAS.467..179G,
       author = {{Grand}, Robert J.~J. and {G{\'o}mez}, Facundo A. and {Marinacci}, Federico and {Pakmor}, R{\"u}diger and {Springel}, Volker and {Campbell}, David J.~R. and {Frenk}, Carlos S. and {Jenkins}, Adrian and {White}, Simon D.~M.},
        title = "{The Auriga Project: the properties and formation mechanisms of disc galaxies across cosmic time}",
      journal = {\mnras},
         year = 2017,
        month = may,
       volume = {467},
       number = {1},
        pages = {179-207},
          doi = {10.1093/mnras/stx071},
archivePrefix = {arXiv},
       eprint = {1610.01159},
 primaryClass = {astro-ph.GA},
       adsurl = {https://ui.adsabs.harvard.edu/abs/2017MNRAS.467..179G}
}

@ARTICLE{2018MNRAS.479L.108K,
       author = {{Kawata}, Daisuke and {Baba}, Junichi and {Ciuc{\v{a}}}, Ioana and {Cropper}, Mark and {Grand}, Robert J.~J. and {Hunt}, Jason A.~S. and {Seabroke}, George},
        title = "{Radial distribution of stellar motions in Gaia DR2}",
      journal = {\mnras},
         year = 2018,
        month = sep,
       volume = {479},
       number = {1},
        pages = {L108-L112},
          doi = {10.1093/mnrasl/sly107},
archivePrefix = {arXiv},
       eprint = {1804.10175},
 primaryClass = {astro-ph.GA},
       adsurl = {https://ui.adsabs.harvard.edu/abs/2018MNRAS.479L.108K}
}

@ARTICLE{2017MNRAS.470.3819B,
       author = {{Buta}, Ronald J.},
        title = "{Galactic rings revisited. II. Dark gaps and the locations of resonances in early-to-intermediate-type disc galaxies}",
      journal = {\mnras},
         year = 2017,
        month = oct,
       volume = {470},
       number = {4},
        pages = {3819-3849},
          doi = {10.1093/mnras/stx1392},
       adsurl = {https://ui.adsabs.harvard.edu/abs/2017MNRAS.470.3819B}
}

@ARTICLE{2025NewAR.10001721H,
       author = {{Hunt}, Jason A.~S. and {Vasiliev}, Eugene},
        title = "{Milky Way dynamics in light of Gaia}",
      journal = {\nar},
         year = 2025,
        month = jun,
       volume = {100},
          eid = {101721},
        pages = {101721},
          doi = {10.1016/j.newar.2024.101721},
archivePrefix = {arXiv},
       eprint = {2501.04075},
 primaryClass = {astro-ph.GA},
       adsurl = {https://ui.adsabs.harvard.edu/abs/2025NewAR.10001721H}
}

@ARTICLE{2019MNRAS.484.3291T,
       author = {{Trick}, Wilma H. and {Coronado}, Johanna and {Rix}, Hans-Walter},
        title = "{The Galactic disc in action space as seen by Gaia DR2}",
      journal = {\mnras},
         year = 2019,
        month = apr,
       volume = {484},
       number = {3},
        pages = {3291-3306},
          doi = {10.1093/mnras/stz209},
archivePrefix = {arXiv},
       eprint = {1805.03653},
 primaryClass = {astro-ph.GA},
       adsurl = {https://ui.adsabs.harvard.edu/abs/2019MNRAS.484.3291T}
}

@ARTICLE{2019MNRAS.484.3154S,
       author = {{Sellwood}, J.~A. and {Trick}, Wilma H. and {Carlberg}, R.~G. and {Coronado}, Johanna and {Rix}, Hans-Walter},
        title = "{Discriminating among theories of spiral structure using Gaia DR2}",
      journal = {\mnras},
         year = 2019,
        month = apr,
       volume = {484},
       number = {3},
        pages = {3154-3167},
          doi = {10.1093/mnras/stz140},
archivePrefix = {arXiv},
       eprint = {1810.03325},
 primaryClass = {astro-ph.GA},
       adsurl = {https://ui.adsabs.harvard.edu/abs/2019MNRAS.484.3154S}
}

@ARTICLE{2000AJ....119..800D,
       author = {{Dehnen}, Walter},
        title = "{The Effect of the Outer Lindblad Resonance of the Galactic Bar on the Local Stellar Velocity Distribution}",
      journal = {\aj},
         year = 2000,
        month = feb,
       volume = {119},
       number = {2},
        pages = {800-812},
          doi = {10.1086/301226},
archivePrefix = {arXiv},
       eprint = {astro-ph/9911161},
 primaryClass = {astro-ph},
       adsurl = {https://ui.adsabs.harvard.edu/abs/2000AJ....119..800D}
}

@ARTICLE{2008A&A...490..135A,
       author = {{Antoja}, T. and {Figueras}, F. and {Fern{\'a}ndez}, D. and {Torra}, J.},
        title = "{Origin and evolution of moving groups. I. Characterization in the observational kinematic-age-metallicity space}",
      journal = {\aap},
         year = 2008,
        month = oct,
       volume = {490},
       number = {1},
        pages = {135-150},
          doi = {10.1051/0004-6361:200809519},
archivePrefix = {arXiv},
       eprint = {0809.0511},
 primaryClass = {astro-ph},
       adsurl = {https://ui.adsabs.harvard.edu/abs/2008A&A...490..135A}
}

@ARTICLE{2020ApJ...890..117D,
       author = {{D'Onghia}, Elena and {L. Aguerri}, J. Alfonso},
        title = "{Trojans in the Solar Neighborhood}",
      journal = {\apj},
         year = 2020,
        month = feb,
       volume = {890},
       number = {2},
          eid = {117},
        pages = {117},
          doi = {10.3847/1538-4357/ab6bd6},
archivePrefix = {arXiv},
       eprint = {1907.08484},
 primaryClass = {astro-ph.GA},
       adsurl = {https://ui.adsabs.harvard.edu/abs/2020ApJ...890..117D}
}

@ARTICLE{2019MNRAS.490.1026H,
       author = {{Hunt}, Jason A.~S. and {Bub}, Mathew W. and {Bovy}, Jo and {Mackereth}, J. Ted and {Trick}, Wilma H. and {Kawata}, Daisuke},
        title = "{Signatures of resonance and phase mixing in the Galactic disc}",
      journal = {\mnras},
         year = 2019,
        month = nov,
       volume = {490},
       number = {1},
        pages = {1026-1043},
          doi = {10.1093/mnras/stz2667},
archivePrefix = {arXiv},
       eprint = {1904.10968},
 primaryClass = {astro-ph.GA},
       adsurl = {https://ui.adsabs.harvard.edu/abs/2019MNRAS.490.1026H}
}

@INPROCEEDINGS{1981seng.proc..111T,
       author = {{Toomre}, A.},
        title = "{What amplifies the spirals}",
    booktitle = {Structure and Evolution of Normal Galaxies},
         year = 1981,
       editor = {{Fall}, S.~M. and {Lynden-Bell}, D.},
        month = jan,
        pages = {111-136},
       adsurl = {https://ui.adsabs.harvard.edu/abs/1981seng.proc..111T}
}

@ARTICLE{2013ApJ...766...34D,
       author = {{D'Onghia}, Elena and {Vogelsberger}, Mark and {Hernquist}, Lars},
        title = "{Self-perpetuating Spiral Arms in Disk Galaxies}",
      journal = {\apj},
         year = 2013,
        month = mar,
       volume = {766},
       number = {1},
          eid = {34},
        pages = {34},
          doi = {10.1088/0004-637X/766/1/34},
archivePrefix = {arXiv},
       eprint = {1204.0513},
 primaryClass = {astro-ph.GA},
       adsurl = {https://ui.adsabs.harvard.edu/abs/2013ApJ...766...34D}
}

@ARTICLE{2024A&A...690A.147H,
       author = {{Haywood}, Misha and {Khoperskov}, Sergey and {Cerqui}, Valeria and {Di Matteo}, Paola and {Katz}, David and {Snaith}, Owain},
        title = "{Timing the Milky Way bar formation and the accompanying radial migration episode}",
      journal = {\aap},
         year = 2024,
        month = oct,
       volume = {690},
          eid = {A147},
        pages = {A147},
          doi = {10.1051/0004-6361/202348767},
archivePrefix = {arXiv},
       eprint = {2403.08963},
 primaryClass = {astro-ph.GA},
       adsurl = {https://ui.adsabs.harvard.edu/abs/2024A&A...690A.147H}
}

@ARTICLE{2025ApJ...983L..10Z_AMR,
       author = {{Zhang}, HanYuan and {Belokurov}, Vasily and {Evans}, N. Wyn and {Sanders}, Jason L. and {Lu}, Yuxi(Lucy) and {Cao}, Chengye and {Myeong}, GyuChul and {Dillamore}, Adam M. and {Kane}, Sarah G. and {Li}, Zhao-Yu},
        title = "{Observational Constraints of Radial Migration in the Galactic Disk Driven by the Slowing Bar}",
      journal = {\apjl},
         year = 2025,
        month = apr,
       volume = {983},
       number = {1},
          eid = {L10},
        pages = {L10},
          doi = {10.3847/2041-8213/adc261},
archivePrefix = {arXiv},
       eprint = {2502.02642},
 primaryClass = {astro-ph.GA},
       adsurl = {https://ui.adsabs.harvard.edu/abs/2025ApJ...983L..10Z}
}

@ARTICLE{2024arXiv241108944H,
       author = {{Hamilton}, Chris and {Modak}, Shaunak and {Tremaine}, Scott},
        title = "{Why is the Galactic disk so cool?}",
      journal = {arXiv e-prints},
         year = 2024,
        month = nov,
          eid = {arXiv:2411.08944},
        pages = {arXiv:2411.08944},
          doi = {10.48550/arXiv.2411.08944},
archivePrefix = {arXiv},
       eprint = {2411.08944},
 primaryClass = {astro-ph.GA},
       adsurl = {https://ui.adsabs.harvard.edu/abs/2024arXiv241108944H}
}

@ARTICLE{2010ApJ...721.1878S,
       author = {{Saha}, Kanak and {Tseng}, Yao-Huan and {Taam}, Ronald E.},
        title = "{The Effect of Bars and Transient Spirals on the Vertical Heating in Disk Galaxies}",
      journal = {\apj},
         year = 2010,
        month = oct,
       volume = {721},
       number = {2},
        pages = {1878-1890},
          doi = {10.1088/0004-637X/721/2/1878},
archivePrefix = {arXiv},
       eprint = {1008.0787},
 primaryClass = {astro-ph.CO},
       adsurl = {https://ui.adsabs.harvard.edu/abs/2010ApJ...721.1878S}
}

@ARTICLE{2016MNRAS.460L..94G,
       author = {{Grand}, Robert J.~J. and {Springel}, Volker and {Kawata}, Daisuke and {Minchev}, Ivan and {S{\'a}nchez-Bl{\'a}zquez}, Patricia and {G{\'o}mez}, Facundo A. and {Marinacci}, Federico and {Pakmor}, R{\"u}diger and {Campbell}, David J.~R.},
        title = "{Spiral-induced velocity and metallicity patterns in a cosmological zoom simulation of a Milky Way-sized galaxy}",
      journal = {\mnras},
         year = 2016,
        month = jul,
       volume = {460},
       number = {1},
        pages = {L94-L98},
          doi = {10.1093/mnrasl/slw086},
archivePrefix = {arXiv},
       eprint = {1604.01027},
 primaryClass = {astro-ph.GA},
       adsurl = {https://ui.adsabs.harvard.edu/abs/2016MNRAS.460L..94G}
}

@ARTICLE{2009MNRAS.397.1599Q,
       author = {{Quillen}, A.~C. and {Minchev}, Ivan and {Bland-Hawthorn}, Joss and {Haywood}, Misha},
        title = "{Radial mixing in the outer Milky Way disc caused by an orbiting satellite}",
      journal = {\mnras},
         year = 2009,
        month = aug,
       volume = {397},
       number = {3},
        pages = {1599-1606},
          doi = {10.1111/j.1365-2966.2009.15054.x},
archivePrefix = {arXiv},
       eprint = {0903.1851},
 primaryClass = {astro-ph.GA},
       adsurl = {https://ui.adsabs.harvard.edu/abs/2009MNRAS.397.1599Q}
}

@ARTICLE{2022MNRAS.516.5067C,
       author = {{Carr}, Christopher and {Johnston}, Kathryn V. and {Laporte}, Chervin F.~P. and {Ness}, Melissa K.},
        title = "{Migration and heating in the galactic disc from encounters between Sagittarius and the Milky Way}",
      journal = {\mnras},
         year = 2022,
        month = nov,
       volume = {516},
       number = {4},
        pages = {5067-5083},
          doi = {10.1093/mnras/stac2403},
archivePrefix = {arXiv},
       eprint = {2201.04133},
 primaryClass = {astro-ph.GA},
       adsurl = {https://ui.adsabs.harvard.edu/abs/2022MNRAS.516.5067C}
}

@ARTICLE{2025arXiv251209987Z,
       author = {{Zhang}, HanYuan and {Belokurov}, Vasily and {Sanders}, Jason L. and {Evans}, N. Wyn and {Chemaly}, David and {Kawata}, Daisuke and {Funakoshi}, Natsuki and {Frankel}, Neige and {Kane}, Sarah G. and {Koposov}, Sergey E.},
        title = "{Orbital migration and heating history of the Galactic disc: a transition between the bimodal discs}",
      journal = {arXiv e-prints},
         year = 2025,
        month = dec,
          eid = {arXiv:2512.09987},
        pages = {arXiv:2512.09987},
          doi = {10.48550/arXiv.2512.09987},
archivePrefix = {arXiv},
       eprint = {2512.09987},
 primaryClass = {astro-ph.GA},
       adsurl = {https://ui.adsabs.harvard.edu/abs/2025arXiv251209987Z}
}

@ARTICLE{2025ApJ...991..139G,
       author = {{Graf}, Russell L. and {Wetzel}, Andrew and {Bailin}, Jeremy and {Orr}, Matthew E.},
        title = "{Inside-out versus Upside-down: The Origin and Evolution of Metallicity Radial Gradients in FIRE Simulations of Milky Way-mass Galaxies and the Essential Role of Gas Mixing}",
      journal = {\apj},
         year = 2025,
        month = oct,
       volume = {991},
       number = {2},
          eid = {139},
        pages = {139},
          doi = {10.3847/1538-4357/adfa07},
archivePrefix = {arXiv},
       eprint = {2410.21377},
 primaryClass = {astro-ph.GA},
       adsurl = {https://ui.adsabs.harvard.edu/abs/2025ApJ...991..139G}
}

@ARTICLE{2021MNRAS.505.4586B,
       author = {{Bellardini}, Matthew A. and {Wetzel}, Andrew and {Loebman}, Sarah R. and {Faucher-Gigu{\`e}re}, Claude-Andr{\'e} and {Ma}, Xiangcheng and {Feldmann}, Robert},
        title = "{3D gas-phase elemental abundances across the formation histories of Milky Way-mass galaxies in the FIRE simulations: initial conditions for chemical tagging}",
      journal = {\mnras},
         year = 2021,
        month = aug,
       volume = {505},
       number = {3},
        pages = {4586-4607},
          doi = {10.1093/mnras/stab1606},
archivePrefix = {arXiv},
       eprint = {2102.06220},
 primaryClass = {astro-ph.GA},
       adsurl = {https://ui.adsabs.harvard.edu/abs/2021MNRAS.505.4586B}
}

@ARTICLE{2026arXiv260407076K,
       author = {{Koller}, Maria and {Maiolino}, Roberto and {{\"U}bler}, Hannah and {Duan}, Qiao and {Scholtz}, Jan and {Arribas}, Santiago and {Baker}, William M. and {Carniani}, Stefano and {Charlot}, Stephane and {Curti}, Mirko and et al.},
        title = "{Metal Mayhem at $\rm z \sim 7-10$: Diversity and Evolution of Gas-Phase Metallicity Gradients}",
      journal = {arXiv e-prints},
         year = 2026,
        month = apr,
          eid = {arXiv:2604.07076},
        pages = {arXiv:2604.07076},
          doi = {10.48550/arXiv.2604.07076},
archivePrefix = {arXiv},
       eprint = {2604.07076},
 primaryClass = {astro-ph.GA},
       adsurl = {https://ui.adsabs.harvard.edu/abs/2026arXiv260407076K}
}

@ARTICLE{2017MNRAS.467.5022H,
       author = {{Herpich}, Jakob and {Tremaine}, Scott and {Rix}, Hans-Walter},
        title = "{Galactic disc profiles and a universal angular momentum distribution from statistical physics}",
      journal = {\mnras},
         year = 2017,
        month = jun,
       volume = {467},
       number = {4},
        pages = {5022-5032},
          doi = {10.1093/mnras/stx352},
archivePrefix = {arXiv},
       eprint = {1612.03171},
 primaryClass = {astro-ph.GA},
       adsurl = {https://ui.adsabs.harvard.edu/abs/2017MNRAS.467.5022H}
}

@ARTICLE{2026A&A...708A.252F,
       author = {{Fiteni}, Karl and {Anderson}, Stuart Robert and {Debattista}, Victor. P. and {Caruana}, Joseph and {Amarante}, Jo{\~a}o A.~S. and {Gough-Kelly}, Steven and {Eyer}, Laurent and {Beraldo e Silva}, Leandro and {Khachaturyants}, Tigran and {Cuomo}, Virginia},
        title = "{The edge of the Milky Way's star-forming disc: Evidence from a 'U-shaped' stellar age profile}",
      journal = {\aap},
         year = 2026,
        month = apr,
       volume = {708},
          eid = {A252},
        pages = {A252},
          doi = {10.1051/0004-6361/202558144},
archivePrefix = {arXiv},
       eprint = {2603.18737},
 primaryClass = {astro-ph.GA},
       adsurl = {https://ui.adsabs.harvard.edu/abs/2026A&A...708A.252F}
}

@ARTICLE{2023MNRAS.525.3318H,
       author = {{Hawkins}, Keith},
        title = "{Chemical cartography with LAMOST and Gaia reveal azimuthal and spiral structure in the Galactic disc}",
      journal = {\mnras},
         year = 2023,
        month = nov,
       volume = {525},
       number = {3},
        pages = {3318-3329},
          doi = {10.1093/mnras/stad1244},
archivePrefix = {arXiv},
       eprint = {2207.04542},
 primaryClass = {astro-ph.GA},
       adsurl = {https://ui.adsabs.harvard.edu/abs/2023MNRAS.525.3318H}
}

@ARTICLE{2013A&A...553A.102D,
       author = {{Di Matteo}, P. and {Haywood}, M. and {Combes}, F. and {Semelin}, B. and {Snaith}, O.~N.},
        title = "{Signatures of radial migration in barred galaxies: Azimuthal variations in the metallicity distribution of old stars}",
      journal = {\aap},
         year = 2013,
        month = may,
       volume = {553},
          eid = {A102},
        pages = {A102},
          doi = {10.1051/0004-6361/201220539},
archivePrefix = {arXiv},
       eprint = {1301.2545},
 primaryClass = {astro-ph.GA},
       adsurl = {https://ui.adsabs.harvard.edu/abs/2013A&A...553A.102D}
}

@ARTICLE{2018A&A...611L...2K,
       author = {{Khoperskov}, S. and {Di Matteo}, P. and {Haywood}, M. and {Combes}, F.},
        title = "{Stellar metallicity variations across spiral arms in disk galaxies with multiple populations}",
      journal = {\aap},
         year = 2018,
        month = mar,
       volume = {611},
          eid = {L2},
        pages = {L2},
          doi = {10.1051/0004-6361/201732521},
archivePrefix = {arXiv},
       eprint = {1801.08711},
 primaryClass = {astro-ph.GA},
       adsurl = {https://ui.adsabs.harvard.edu/abs/2018A&A...611L...2K}
}

@ARTICLE{2016AN....337..949F,
       author = {{Famaey}, B. and {Monari}, G. and {Siebert}, A.},
        title = "{The Milky Way disk non-axisymmetries and galactoseismology}",
      journal = {Astronomische Nachrichten},
         year = 2016,
        month = sep,
       volume = {337},
       number = {8-9},
        pages = {949},
          doi = {10.1002/asna.201612405},
       adsurl = {https://ui.adsabs.harvard.edu/abs/2016AN....337..949F}
}

@ARTICLE{2025MNRAS.537.1620D,
       author = {{Debattista}, Victor P. and {Khachaturyants}, Tigran and {Amarante}, Jo{\~a}o A.~S. and {Carr}, Christopher and {Beraldo e Silva}, Leandro and {Laporte}, Chervin F.~P.},
        title = "{Azimuthal metallicity variations, spiral structure, and the failure of radial actions based on assuming axisymmetry}",
      journal = {\mnras},
         year = 2025,
        month = feb,
       volume = {537},
       number = {2},
        pages = {1620-1645},
          doi = {10.1093/mnras/staf035},
archivePrefix = {arXiv},
       eprint = {2402.08356},
 primaryClass = {astro-ph.GA},
       adsurl = {https://ui.adsabs.harvard.edu/abs/2025MNRAS.537.1620D}
}

@ARTICLE{1997AJ....114..376C,
       author = {{Castro}, Sandra and {Rich}, R. Michael and {Grenon}, Michel and {Barbuy}, Beatriz and {McCarthy}, James K.},
        title = "{High-Resolution Abundance Analysis of Very Metal-rich Stars in the Solar Neighborhood}",
      journal = {\aj},
         year = 1997,
        month = jul,
       volume = {114},
        pages = {376-387},
          doi = {10.1086/118481},
archivePrefix = {arXiv},
       eprint = {astro-ph/9704220},
 primaryClass = {astro-ph},
       adsurl = {https://ui.adsabs.harvard.edu/abs/1997AJ....114..376C}
}

@ARTICLE{2024MNRAS.535..392L,
       author = {{Lu}, Yuxi (Lucy) and {Minchev}, Ivan and {Buck}, Tobias and {Khoperskov}, Sergey and {Steinmetz}, Matthias and {Libeskind}, Noam and {Cescutti}, Gabriele and {Freeman}, Ken C. and {Ratcliffe}, Bridget},
        title = "{There is no place like home - finding birth radii of stars in the Milky Way}",
      journal = {\mnras},
         year = 2024,
        month = nov,
       volume = {535},
       number = {1},
        pages = {392-405},
          doi = {10.1093/mnras/stae2364},
archivePrefix = {arXiv},
       eprint = {2212.04515},
 primaryClass = {astro-ph.GA},
       adsurl = {https://ui.adsabs.harvard.edu/abs/2024MNRAS.535..392L}
}

@ARTICLE{2022AJ....164...85M,
       author = {{Myers}, Natalie and {Donor}, John and {Spoo}, Taylor and {Frinchaboy}, Peter M. and {Cunha}, Katia and {Price-Whelan}, Adrian M. and {Majewski}, Steven R. and {Beaton}, Rachael L. and {Zasowski}, Gail and {O'Connell}, Julia and et al.},
        title = "{The Open Cluster Chemical Abundances and Mapping Survey. VI. Galactic Chemical Gradient Analysis from APOGEE DR17}",
      journal = {\aj},
         year = 2022,
        month = sep,
       volume = {164},
       number = {3},
          eid = {85},
        pages = {85},
          doi = {10.3847/1538-3881/ac7ce5},
archivePrefix = {arXiv},
       eprint = {2206.13650},
 primaryClass = {astro-ph.GA},
       adsurl = {https://ui.adsabs.harvard.edu/abs/2022AJ....164...85M}
}

@ARTICLE{Lingard2021,
       author = {{Lingard}, Timothy and {Masters}, Karen L. and {Krawczyk}, Coleman and {Lintott}, Chris and {Kruk}, Sandor and {Simmons}, Brooke and {Keel}, William and {Nichol}, Robert C. and {Baeten}, Elisabeth},
        title = "{Galaxy zoo builder: Morphological dependence of spiral galaxy pitch angle}",
      journal = {\mnras},
         year = 2021,
        month = jul,
       volume = {504},
       number = {3},
        pages = {3364-3374},
          doi = {10.1093/mnras/stab1072},
archivePrefix = {arXiv},
       eprint = {2105.04500},
 primaryClass = {astro-ph.GA},
       adsurl = {https://ui.adsabs.harvard.edu/abs/2021MNRAS.504.3364L}
}

@ARTICLE{Kennicutt1981,
       author = {{Kennicutt}, R.~C., Jr.},
        title = "{The shapes of spiral arms along the Hubble sequence.}",
      journal = {\aj},
         year = 1981,
        month = dec,
       volume = {86},
        pages = {1847-1858},
          doi = {10.1086/113064},
       adsurl = {https://ui.adsabs.harvard.edu/abs/1981AJ.....86.1847K}
}

@ARTICLE{Eilers2020,
       author = {{Eilers}, Anna-Christina and {Hogg}, David W. and {Rix}, Hans-Walter and {Frankel}, Neige and {Hunt}, Jason A.~S. and {Fouvry}, Jean-Baptiste and {Buck}, Tobias},
        title = "{The Strength of the Dynamical Spiral Perturbation in the Galactic Disk}",
      journal = {\apj},
         year = 2020,
        month = sep,
       volume = {900},
       number = {2},
          eid = {186},
        pages = {186},
          doi = {10.3847/1538-4357/abac0b},
archivePrefix = {arXiv},
       eprint = {2003.01132},
 primaryClass = {astro-ph.GA},
       adsurl = {https://ui.adsabs.harvard.edu/abs/2020ApJ...900..186E}
}

@ARTICLE{Reid2019,
       author = {{Reid}, M.~J. and {Menten}, K.~M. and {Brunthaler}, A. and {Zheng}, X.~W. and {Dame}, T.~M. and {Xu}, Y. and {Li}, J. and {Sakai}, N. and {Wu}, Y. and {Immer}, K. and {Zhang}, B. and {Sanna}, A. and {Moscadelli}, L. and {Rygl}, K.~L.~J. and {Bartkiewicz}, A. and {Hu}, B. and {Quiroga-Nu{\~n}ez}, L.~H. and {van Langevelde}, H.~J.},
        title = "{Trigonometric Parallaxes of High-mass Star-forming Regions: Our View of the Milky Way}",
      journal = {\apj},
         year = 2019,
        month = nov,
       volume = {885},
       number = {2},
          eid = {131},
        pages = {131},
          doi = {10.3847/1538-4357/ab4a11},
archivePrefix = {arXiv},
       eprint = {1910.03357},
 primaryClass = {astro-ph.GA},
       adsurl = {https://ui.adsabs.harvard.edu/abs/2019ApJ...885..131R}
}

@ARTICLE{Rubin83,
       author = {{Rubin}, V.~C.},
        title = "{The Rotation of Spiral Galaxies}",
      journal = {Science},
         year = "1983",
        month = "Jun",
       volume = {220},
       number = {4604},
        pages = {1339-1344},
          doi = {10.1126/science.220.4604.1339},
       adsurl = {https://ui.adsabs.harvard.edu/abs/1983Sci...220.1339R}
}

@ARTICLE{Sofue09,
       author = {{Sofue}, Yoshiaki and {Honma}, Mareki and {Omodaka}, Toshihiro},
        title = "{Unified Rotation Curve of the Galaxy -- Decomposition into de Vaucouleurs Bulge, Disk, Dark Halo, and the 9-kpc Rotation Dip --}",
      journal = {Publications of the Astronomical Society of Japan},
         year = "2009",
        month = "Feb",
       volume = {61},
        pages = {227},
          doi = {10.1093/pasj/61.2.227},
archivePrefix = {arXiv},
       eprint = {0811.0859},
 primaryClass = {astro-ph},
       adsurl = {https://ui.adsabs.harvard.edu/abs/2009PASJ...61..227S}
}

@ARTICLE{NFW97,
       author = {{Navarro}, Julio F. and {Frenk}, Carlos S. and {White}, Simon D.~M.},
        title = "{A Universal Density Profile from Hierarchical Clustering}",
      journal = {\apj},
         year = 1997,
        month = dec,
       volume = {490},
       number = {2},
        pages = {493-508},
          doi = {10.1086/304888},
archivePrefix = {arXiv},
       eprint = {astro-ph/9611107},
 primaryClass = {astro-ph},
       adsurl = {https://ui.adsabs.harvard.edu/abs/1997ApJ...490..493N}
}

@ARTICLE{BR13,
       author = {{Bovy}, Jo and {Rix}, Hans-Walter},
        title = "{A Direct Dynamical Measurement of the Milky Way's Disk Surface Density Profile, Disk Scale Length, and Dark Matter Profile at 4 kpc <\raisebox{-0.5ex}\textasciitilde R <\raisebox{-0.5ex}\textasciitilde 9 kpc}",
      journal = {\apj},
         year = 2013,
        month = dec,
       volume = {779},
       number = {2},
          eid = {115},
        pages = {115},
          doi = {10.1088/0004-637X/779/2/115},
archivePrefix = {arXiv},
       eprint = {1309.0809},
 primaryClass = {astro-ph.GA},
       adsurl = {https://ui.adsabs.harvard.edu/abs/2013ApJ...779..115B}
}

@ARTICLE{CG02,
       author = {{Cox}, Donald P. and {G{\'o}mez}, Gilberto C.},
        title = "{Analytical Expressions for Spiral Arm Gravitational Potential and Density}",
      journal = {\apjs},
         year = 2002,
        month = oct,
       volume = {142},
       number = {2},
        pages = {261-267},
          doi = {10.1086/341946},
archivePrefix = {arXiv},
       eprint = {astro-ph/0207635},
 primaryClass = {astro-ph},
       adsurl = {https://ui.adsabs.harvard.edu/abs/2002ApJS..142..261C}
}

@ARTICLE{wang_2020,
       author = {{Wang}, Long and {Iwasawa}, Masaki and {Nitadori}, Keigo and {Makino}, Junichiro},
        title = "{PETAR: a high-performance N-body code for modelling massive collisional stellar systems}",
      journal = {\mnras},
         year = 2020,
        month = sep,
       volume = {497},
       number = {1},
        pages = {536-555},
          doi = {10.1093/mnras/staa1915},
archivePrefix = {arXiv},
       eprint = {2006.16560},
 primaryClass = {astro-ph.IM},
       adsurl = {https://ui.adsabs.harvard.edu/abs/2020MNRAS.497..536W}
}

@ARTICLE{Wetzel16,
       author = {{Wetzel}, Andrew R. and {Hopkins}, Philip F. and {Kim}, Ji-hoon and {Faucher-Gigu{\`e}re}, Claude-Andr{\'e} and {Kere{\v{s}}}, Du{\v{s}}an and {Quataert}, Eliot},
        title = "{Reconciling Dwarf Galaxies with {\ensuremath{\Lambda}}CDM Cosmology: Simulating a Realistic Population of Satellites around a Milky Way-mass Galaxy}",
      journal = {\apjl},
         year = 2016,
        month = aug,
       volume = {827},
       number = {2},
          eid = {L23},
        pages = {L23},
          doi = {10.3847/2041-8205/827/2/L23},
archivePrefix = {arXiv},
       eprint = {1602.05957},
 primaryClass = {astro-ph.GA},
       adsurl = {https://ui.adsabs.harvard.edu/abs/2016ApJ...827L..23W}
}

@ARTICLE{Hopkins18,
       author = {{Hopkins}, Philip F. and {Wetzel}, Andrew and {Kere{\v{s}}}, Du{\v{s}}an and {Faucher-Gigu{\`e}re}, Claude-Andr{\'e} and {Quataert}, Eliot and {Boylan-Kolchin}, Michael and {Murray}, Norman and {Hayward}, Christopher C. and {Garrison-Kimmel}, Shea and {Hummels}, Cameron and {Feldmann}, Robert and {Torrey}, Paul and {Ma}, Xiangcheng and {Angl{\'e}s-Alc{\'a}zar}, Daniel and {Su}, Kung-Yi and {Orr}, Matthew and {Schmitz}, Denise and {Escala}, Ivanna and {Sanderson}, Robyn and {Grudi{\'c}}, Michael Y. and {Hafen}, Zachary and {Kim}, Ji-Hoon and {Fitts}, Alex and {Bullock}, James S. and {Wheeler}, Coral and {Chan}, T.~K. and {Elbert}, Oliver D. and {Narayanan}, Desika},
        title = "{FIRE-2 simulations: physics versus numerics in galaxy formation}",
      journal = {\mnras},
         year = 2018,
        month = oct,
       volume = {480},
       number = {1},
        pages = {800-863},
          doi = {10.1093/mnras/sty1690},
archivePrefix = {arXiv},
       eprint = {1702.06148},
 primaryClass = {astro-ph.GA},
       adsurl = {https://ui.adsabs.harvard.edu/abs/2018MNRAS.480..800H}
}

@ARTICLE{GT81,
       author = {{Goldreich}, P. and {Tremaine}, S.},
        title = "{The origin of the eccentricities of the rings of Uranus}",
      journal = {\apj},
         year = 1981,
        month = feb,
       volume = {243},
        pages = {1062-1075},
          doi = {10.1086/158671},
       adsurl = {https://ui.adsabs.harvard.edu/abs/1981ApJ...243.1062G}
}

@ARTICLE{2021ApJ...910...17P,
       author = {{Price-Whelan}, Adrian M. and {Hogg}, David W. and {Johnston}, Kathryn V. and {Ness}, Melissa K. and {Rix}, Hans-Walter and {Beaton}, Rachael L. and {Brownstein}, Joel R. and {Garc{\'\i}a-Hern{\'a}ndez}, D.~A. and {Hasselquist}, Sten and {Hayes}, Christian R. and et al.},
        title = "{Orbital Torus Imaging: Using Element Abundances to Map Orbits and Mass in the Milky Way}",
      journal = {\apj},
         year = 2021,
        month = mar,
       volume = {910},
       number = {1},
          eid = {17},
        pages = {17},
          doi = {10.3847/1538-4357/abe1b7},
archivePrefix = {arXiv},
       eprint = {2012.00015},
 primaryClass = {astro-ph.GA},
       adsurl = {https://ui.adsabs.harvard.edu/abs/2021ApJ...910...17P}
}

@ARTICLE{BSW13,
       author = {{Baba}, Junichi and {Saitoh}, Takayuki R. and {Wada}, Keiichi},
        title = "{Dynamics of Non-steady Spiral Arms in Disk Galaxies}",
      journal = {\apj},
         year = "2013",
        month = "Jan",
       volume = {763},
       number = {1},
          eid = {46},
        pages = {46},
          doi = {10.1088/0004-637X/763/1/46},
archivePrefix = {arXiv},
       eprint = {1211.5401},
 primaryClass = {astro-ph.GA},
       adsurl = {https://ui.adsabs.harvard.edu/abs/2013ApJ...763...46B}
}

@ARTICLE{BW67,
   author = {{Barbanis}, B. and {Woltjer}, L.},
    title = "{Orbits in Spiral Galaxies and the Velocity Dispersion of Population i Stars}",
  journal = {\apj},
     year = 1967,
    month = nov,
   volume = 150,
    pages = {461-+},
      doi = {10.1086/149349},
   adsurl = {http://adsabs.harvard.edu/abs/1967ApJ...150..461B}
}

@BOOK{BT87,
   author = {{Binney}, J. and {Tremaine}, S.},
    title = "{Galactic dynamics}",
booktitle = {Princeton, NJ, Princeton University Press, 1987, 747 p.},
     year = 1987,
     publisher = {Princeton University Press},
   adsurl = {http://adsabs.harvard.edu/abs/1987gady.book.....B}
}

@BOOK{BT08,
  author = {{Binney}, J. and {Tremaine}, S.},
    title = "{Galactic Dynamics: Second Edition}",
booktitle = {Galactic Dynamics: Second Edition, by James Binney and Scott Tremaine.~ISBN 978-0-691-13026-2 (HB).~Published by Princeton University Press, Princeton, NJ USA, 2008.},
     year = 2008,
publisher = {Princeton University Press},
   adsurl = {http://adsabs.harvard.edu/abs/2008gady.book.....B}
}

@ARTICLE{BKW12,
   author = {{Bird}, J.~C. and {Kazantzidis}, S. and {Weinberg}, D.~H.},
    title = "{Radial mixing in galactic discs: the effects of disc structure and satellite bombardment}",
  journal = {\mnras},
archivePrefix = "arXiv",
   eprint = {1104.0933},
 primaryClass = "astro-ph.GA",
     year = 2012,
    month = feb,
   volume = 420,
    pages = {913-925},
      doi = {10.1111/j.1365-2966.2011.19728.x},
   adsurl = {http://adsabs.harvard.edu/abs/2012MNRAS.420..913B}
}

@ARTICLE{BP99,
   author = {{Block}, D.~L. and {Puerari}, I.},
    title = "{Toward a dust penetrated classification of the evolved stellar Population II disks of galaxies}",
  journal = {\aap},
   eprint = {arXiv:astro-ph/9811060},
     year = 1999,
    month = feb,
   volume = 342,
    pages = {627-642},
   adsurl = {http://adsabs.harvard.edu/abs/1999A%26A...342..627B}
}

@ARTICLE{Bovy15,
       author = {{Bovy}, Jo},
        title = "{galpy: A python Library for Galactic Dynamics}",
      journal = {ApJS},
         year = 2015,
        month = Feb,
       volume = {216},
          eid = {29},
        pages = {29},
          doi = {10.1088/0067-0049/216/2/29},
       adsurl = {https://ui.adsabs.harvard.edu/#abs/2015ApJS..216...29B}
}

@ARTICLE{CS85,
   author = {{Carlberg}, R.~G. and {Sellwood}, J.~A.},
    title = "{Dynamical evolution in galactic disks}",
  journal = {\apj},
     year = 1985,
    month = may,
   volume = 292,
    pages = {79-89},
      doi = {10.1086/163134},
   adsurl = {http://adsabs.harvard.edu/abs/1985ApJ...292...79C}
}

@ARTICLE{Casagrande11,
   author = {{Casagrande}, L. and {Sch{\"o}nrich}, R. and {Asplund}, M. and 
	{Cassisi}, S. and {Ram{\'{\i}}rez}, I. and {Mel{\'e}ndez}, J. and 
	{Bensby}, T. and {Feltzing}, S.},
    title = "{New constraints on the chemical evolution of the solar neighbourhood and Galactic disc(s). Improved astrophysical parameters for the Geneva-Copenhagen Survey}",
  journal = {\aap},
archivePrefix = "arXiv",
   eprint = {1103.4651},
 primaryClass = "astro-ph.GA",
     year = 2011,
    month = jun,
   volume = 530,
    pages = {A138+},
      doi = {10.1051/0004-6361/201016276},
   adsurl = {http://adsabs.harvard.edu/abs/2011A%26A...530A.138C}
}

@ARTICLE{Contopoulos78,
   author = {{Contopoulos}, G.},
    title = "{Periodic orbits near the particle resonance in galaxies}",
  journal = {\aap},
     year = 1978,
    month = mar,
   volume = 64,
    pages = {323-332},
   adsurl = {http://adsabs.harvard.edu/abs/1978A%26A....64..323C}
}

@INPROCEEDINGS{DRL17,
       author = {{Debattista}, Victor P. and {Ro{\v{s}}kar}, Rok and {Loebman}, Sarah R.},
        title = "{The Impact of Stellar Migration on Disk Outskirts}",
    booktitle = {Outskirts of Galaxies},
         year = 2017,
       editor = {{Knapen}, Johan H. and {Lee}, Janice C. and {Gil de Paz}, Armando},
       volume = {434},
        month = Jan,
        series = {},
        pages = {77},
          doi = {10.1007/978-3-319-56570-5_3},
 primaryClass = {astro-ph.GA},
       adsurl = {https://ui.adsabs.harvard.edu/#abs/2017ASSL..434...77D}
}

@ARTICLE{Dehnen99,
   author = {{Dehnen}, W.},
    title = "{Approximating Stellar Orbits: Improving on Epicycle Theory}",
  journal = {\aj},
   eprint = {arXiv:astro-ph/9906081},
     year = 1999,
    month = sep,
   volume = 118,
    pages = {1190-1200},
      doi = {10.1086/301009},
   adsurl = {http://adsabs.harvard.edu/abs/1999AJ....118.1190D}
}

@ARTICLE{deJong10,
   author = {{de Jong}, J.~T.~A. and {Yanny}, B. and {Rix}, H.-W. and {Dolphin}, A.~E. and 
	{Martin}, N.~F. and {Beers}, T.~C.},
    title = "{Mapping the Stellar Structure of the Milky Way Thick Disk and Halo Using SEGUE Photometry}",
  journal = {\apj},
archivePrefix = "arXiv",
   eprint = {0911.3900},
 primaryClass = "astro-ph.GA",
     year = 2010,
    month = may,
   volume = 714,
    pages = {663-674},
      doi = {10.1088/0004-637X/714/1/663},
   adsurl = {http://adsabs.harvard.edu/abs/2010ApJ...714..663D}
}

@ARTICLE{DL05,
   author = {{Dias}, W.~S. and {L{\'e}pine}, J.~R.~D.},
    title = "{Direct Determination of the Spiral Pattern Rotation Speed of the Galaxy}",
  journal = {\apj},
   eprint = {arXiv:astro-ph/0503083},
     year = 2005,
    month = aug,
   volume = 629,
    pages = {825-831},
      doi = {10.1086/431456},
   adsurl = {http://adsabs.harvard.edu/abs/2005ApJ...629..825D}
}

@ARTICLE{Drimmel00,
   author = {{Drimmel}, R.},
    title = "{Evidence for a two-armed spiral in the Milky Way}",
  journal = {\aap},
   eprint = {arXiv:astro-ph/0005241},
     year = 2000,
    month = jun,
   volume = 358,
    pages = {L13-L16},
   adsurl = {http://adsabs.harvard.edu/abs/2000A%26A...358L..13D}
}

@ARTICLE{Edvardsson93,
   author = {{Edvardsson}, B. and {Andersen}, J. and {Gustafsson}, B. and 
	{Lambert}, D.~L. and {Nissen}, P.~E. and {Tomkin}, J.},
    title = "{The Chemical Evolution of the Galactic Disk - Part One - Analysis and Results}",
  journal = {\aap},
     year = 1993,
    month = aug,
   volume = 275,
    pages = {101-+},
   adsurl = {http://adsabs.harvard.edu/abs/1993A%26A...275..101E}
}

@ARTICLE{Elmegreen11,
   author = {{Elmegreen}, D.~M. and {Elmegreen}, B.~G. and {Yau}, A. and 
	{Athanassoula}, E. and {Bosma}, A. and {Buta}, R.~J. and {Helou}, G. and 
	{Ho}, L.~C. and {Gadotti}, D.~A. and {Knapen}, J.~H. and {Laurikainen}, E. and 
	{Madore}, B.~F. and {Masters}, K.~L. and {Meidt}, S.~E. and 
	{Men{\'e}ndez-Delmestre}, K. and {Regan}, M.~W. and {Salo}, H. and 
	{Sheth}, K. and {Zaritsky}, D. and {Aravena}, M. and {Skibba}, R. and 
	{Hinz}, J.~L. and {Laine}, J. and {Gil de Paz}, A. and {Mu{\~n}oz-Mateos}, J.-C. and 
	{Seibert}, M. and {Mizusawa}, T. and {Kim}, T. and {Erroz Ferrer}, S.
	},
    title = "{Grand Design and Flocculent Spirals in the Spitzer Survey of Stellar Structure in Galaxies (S$^{4}$G)}",
  journal = {\apj},
archivePrefix = "arXiv",
   eprint = {1106.4840},
 primaryClass = "astro-ph.GA",
     year = 2011,
    month = aug,
   volume = 737,
      eid = {32},
    pages = {32},
      doi = {10.1088/0004-637X/737/1/32},
   adsurl = {http://adsabs.harvard.edu/abs/2011ApJ...737...32E}
}

@ARTICLE{EEL03,
   author = {{Elmegreen}, B.~G. and {Elmegreen}, D.~M. and {Leitner}, S.~N.
	},
    title = "{A Turbulent Origin for Flocculent Spiral Structure in Galaxies}",
  journal = {\apj},
   eprint = {arXiv:astro-ph/0305049},
     year = 2003,
    month = jun,
   volume = 590,
    pages = {271-283},
      doi = {10.1086/374860},
   adsurl = {http://adsabs.harvard.edu/abs/2003ApJ...590..271E}
}

@BOOK{Elmegreen98,
    author = {{Elmegreen}, D.~M.},
    title = "{Galaxies and galactic structure}",
	booktitle = {Galaxies and galactic structure},
    year = 1998,
    publisher = {Prentice Hall},
    adsurl = {http://adsabs.harvard.edu/abs/1998ggs..book.....E},
    notes = {QB 857 E455 1998}
}

@ARTICLE{Elmegreen99,
   author = {{Elmegreen}, D.~M. and {Chromey}, F.~R. and {Bissell}, B.~A. and 
	{Corrado}, K.},
    title = "{K'-Band Observations of Underlying Symmetric Structure in Flocculent Galaxies}",
  journal = {\aj},
     year = 1999,
    month = dec,
   volume = 118,
    pages = {2618-2624},
      doi = {10.1086/301127},
   adsurl = {http://adsabs.harvard.edu/abs/1999AJ....118.2618E}
}

@ARTICLE{EE84,
   author = {{Elmegreen}, D.~M. and {Elmegreen}, B.~G.},
    title = "{Blue and near-infrared surface photometry of spiral structure in 34 nonbarred grand design and flocculent galaxies}",
  journal = {\apjs},
     year = 1984,
    month = jan,
   volume = 54,
    pages = {127-149},
      doi = {10.1086/190922},
   adsurl = {http://adsabs.harvard.edu/abs/1984ApJS...54..127E}
}

@ARTICLE{EE83,
   author = {{Elmegreen}, B.~G. and {Elmegreen}, D.~M.},
    title = "{Flocculent and grand design spiral galaxies in groups - Time scales for the persistence of grand design spiral structures}",
  journal = {\apj},
     year = 1983,
    month = apr,
   volume = 267,
    pages = {31-34},
      doi = {10.1086/160842},
   adsurl = {http://adsabs.harvard.edu/abs/1983ApJ...267...31E}
}

@ARTICLE{FFT01,
   author = {{Fern{\'a}ndez}, D. and {Figueras}, F. and {Torra}, J.},
    title = "{Kinematics of young stars. II. Galactic spiral structure}",
  journal = {\aap},
   eprint = {arXiv:astro-ph/0103212},
     year = 2001,
    month = jun,
   volume = 372,
    pages = {833-850},
      doi = {10.1051/0004-6361:20010366},
   adsurl = {http://adsabs.harvard.edu/abs/2001A%26A...372..833F}
}

@ARTICLE{Frankel18,
       author = {{Frankel}, Neige and {Rix}, Hans-Walter and {Ting}, Yuan-Sen and {Ness},
        Melissa and {Hogg}, David W.},
        title = "{Measuring Radial Orbit Migration in the Galactic Disk}",
      journal = {\apj},
         year = 2018,
        month = Oct,
       volume = {865},
          eid = {96},
        pages = {96},
          doi = {10.3847/1538-4357/aadba5},
 primaryClass = {astro-ph.GA},
       adsurl = {https://ui.adsabs.harvard.edu/#abs/2018ApJ...865...96F}
}

@ARTICLE{Freeman70,
   author = {{Freeman}, K.~C.},
    title = "{On the Disks of Spiral and S0 Galaxies}",
  journal = {\apj},
     year = 1970,
    month = jun,
   volume = 160,
    pages = {811},
      doi = {10.1086/150474},
   adsurl = {http://adsabs.harvard.edu/abs/1970ApJ...160..811F}
}

@ARTICLE{GL65,
   author = {{Goldreich}, P. and {Lynden-Bell}, D.},
    title = "{II. Spiral arms as sheared gravitational instabilities}",
  journal = {\mnras},
     year = 1965,
   volume = 130,
    pages = {125},
   adsurl = {http://adsabs.harvard.edu/abs/1965MNRAS.130..125G}
}

@ARTICLE{GT82,
   author = {{Goldreich}, P. and {Tremaine}, S.},
    title = "{The dynamics of planetary rings}",
  journal = {\araa},
     year = 1982,
   volume = 20,
    pages = {249-283},
      doi = {10.1146/annurev.aa.20.090182.001341},
   adsurl = {http://adsabs.harvard.edu/abs/1982ARA%26A..20..249G}
}

@ARTICLE{Grand16,
   author = {{Grand}, R.~J.~J. and {Springel}, V. and {G{\'o}mez}, F.~A. and 
	{Marinacci}, F. and {Pakmor}, R. and {Campbell}, D.~J.~R. and 
	{Jenkins}, A.},
    title = "{Vertical disc heating in Milky Way-sized galaxies in a cosmological context}",
  journal = {\mnras},
archivePrefix = "arXiv",
   eprint = {1512.02219},
     year = 2016,
    month = jun,
   volume = 459,
    pages = {199-219},
      doi = {10.1093/mnras/stw601},
   adsurl = {http://adsabs.harvard.edu/abs/2016MNRAS.459..199G}
}

@ARTICLE{GKC12,
   author = {{Grand}, R.~J.~J. and {Kawata}, D. and {Cropper}, M.},
    title = "{The dynamics of stars around spiral arms}",
  journal = {\mnras},
archivePrefix = "arXiv",
   eprint = {1112.0019},
 primaryClass = "astro-ph.GA",
     year = 2012,
    month = apr,
   volume = 421,
    pages = {1529-1538},
      doi = {10.1111/j.1365-2966.2012.20411.x},
   adsurl = {http://adsabs.harvard.edu/abs/2012MNRAS.421.1529G}
}

@ARTICLE{Grenon87,
   author = {{Grenon}, M.},
    title = "{Past and present metal abundance gradient in the Galactic disc}",
  journal = {JApA},
     year = 1987,
    month = jun,
   volume = 8,
    pages = {123-139},
      doi = {10.1007/BF02714310},
   adsurl = {http://adsabs.harvard.edu/abs/1987JApA....8..123G}
}

@ARTICLE{Hayden15,
   author = {{Hayden}, M.~R. and {Bovy}, J. and {Holtzman}, J.~A. and {Nidever}, D.~L. and 
	{Bird}, J.~C. and {Weinberg}, D.~H. and {Andrews}, B.~H. and 
	{Majewski}, S.~R. and {Allende Prieto}, C. and {Anders}, F. and 
	{Beers}, T.~C. and {Bizyaev}, D. and {Chiappini}, C. and {Cunha}, K. and 
	{Frinchaboy}, P. and {Garc{\'{\i}}a-Her{\'n}andez}, D.~A. and 
	{Garc{\'{\i}}a P{\'e}rez}, A.~E. and {Girardi}, L. and {Harding}, P. and 
	{Hearty}, F.~R. and {Johnson}, J.~A. and {M{\'e}sz{\'a}ros}, S. and 
	{Minchev}, I. and {O'Connell}, R. and {Pan}, K. and {Robin}, A.~C. and 
	{Schiavon}, R.~P. and {Schneider}, D.~P. and {Schultheis}, M. and 
	{Shetrone}, M. and {Skrutskie}, M. and {Steinmetz}, M. and {Smith}, V. and 
	{Wilson}, J.~C. and {Zamora}, O. and {Zasowski}, G.},
    title = "{Chemical Cartography with APOGEE: Metallicity Distribution Functions and the Chemical Structure of the Milky Way Disk}",
  journal = {\apj},
archivePrefix = "arXiv",
   eprint = {1503.02110},
     year = 2015,
    month = aug,
   volume = 808,
      eid = {132},
    pages = {132},
      doi = {10.1088/0004-637X/808/2/132},
   adsurl = {http://adsabs.harvard.edu/abs/2015ApJ...808..132H}
}

@ARTICLE{Juric08,
   author = {{Juri{\'c}}, M. and {Ivezi{\'c}}, {\v Z}. and {Brooks}, A. and 
	{Lupton}, R.~H. and {Schlegel}, D. and {Finkbeiner}, D. and 
	{Padmanabhan}, N. and {Bond}, N. and {Sesar}, B. and {Rockosi}, C.~M. and 
	{Knapp}, G.~R. and {Gunn}, J.~E. and {Sumi}, T. and {Schneider}, D.~P. and 
	{Barentine}, J.~C. and {Brewington}, H.~J. and {Brinkmann}, J. and 
	{Fukugita}, M. and {Harvanek}, M. and {Kleinman}, S.~J. and 
	{Krzesinski}, J. and {Long}, D. and {Neilsen}, Jr., E.~H. and 
	{Nitta}, A. and {Snedden}, S.~A. and {York}, D.~G.},
    title = "{The Milky Way Tomography with SDSS. I. Stellar Number Density Distribution}",
  journal = {\apj},
   eprint = {arXiv:astro-ph/0510520},
     year = 2008,
    month = feb,
   volume = 673,
    pages = {864-914},
      doi = {10.1086/523619},
   adsurl = {http://adsabs.harvard.edu/abs/2008ApJ...673..864J}
}

@ARTICLE{Kordopatis15,
   author = {{Kordopatis}, G. and {Binney}, J. and {Gilmore}, G. and {Wyse}, R.~F.~G. and 
	{Belokurov}, V. and {McMillan}, P.~J. and {Hatfield}, P. and 
	{Grebel}, E.~K. and {Steinmetz}, M. and {Navarro}, J.~F. and 
	{Seabroke}, G. and {Minchev}, I. and {Chiappini}, C. and {Bienaym{\'e}}, O. and 
	{Bland-Hawthorn}, J. and {Freeman}, K.~C. and {Gibson}, B.~K. and 
	{Helmi}, A. and {Munari}, U. and {Parker}, Q. and {Reid}, W.~A. and 
	{Siebert}, A. and {Siviero}, A. and {Zwitter}, T.},
    title = "{The rich are different: evidence from the RAVE survey for stellar radial migration}",
  journal = {\mnras},
archivePrefix = "arXiv",
   eprint = {1412.5649},
     year = 2015,
    month = mar,
   volume = 447,
    pages = {3526-3535},
      doi = {10.1093/mnras/stu2726},
   adsurl = {http://adsabs.harvard.edu/abs/2015MNRAS.447.3526K}
}

@ARTICLE{Lacey84,
   author = {{Lacey}, C.~G.},
    title = "{The influence of massive gas clouds on stellar velocity dispersions in galactic discs}",
  journal = {\mnras},
     year = 1984,
    month = jun,
   volume = 208,
    pages = {687-707},
   adsurl = {http://adsabs.harvard.edu/abs/1984MNRAS.208..687L}
}

@ARTICLE{LMD01,
   author = {{L{\'e}pine}, J.~R.~D. and {Mishurov}, Y.~N. and {Dedikov}, S.~Y.
	},
    title = "{A New Model for the Spiral Structure of the Galaxy: Superposition of 2- and 4-armed Patterns}",
  journal = {\apj},
   eprint = {arXiv:astro-ph/0001216},
     year = 2001,
    month = jan,
   volume = 546,
    pages = {234-247},
      doi = {10.1086/318225},
   adsurl = {http://adsabs.harvard.edu/abs/2001ApJ...546..234L}
}

@ARTICLE{LS64,
   author = {{Lin}, C.~C. and {Shu}, F.~H.},
    title = "{On the Spiral Structure of Disk Galaxies.}",
  journal = {\apj},
     year = 1964,
    month = aug,
   volume = 140,
    pages = {646},
      doi = {10.1086/147955},
   adsurl = {http://adsabs.harvard.edu/abs/1964ApJ...140..646L}
}

@ARTICLE{LS66,
   author = {{Lin}, C.~C. and {Shu}, F.~H.},
    title = "{On the Spiral Structure of Disk Galaxies, II. Outline of a Theory of Density Waves}",
  journal = {Proceedings of the National Academy of Science},
     year = 1966,
    month = feb,
   volume = 55,
    pages = {229-234},
      doi = {10.1073/pnas.55.2.229},
   adsurl = {http://adsabs.harvard.edu/abs/1966PNAS...55..229L}
}

@ARTICLE{LYS69,
   author = {{Lin}, C.~C. and {Yuan}, C. and {Shu}, F.~H.},
    title = "{On the Spiral Structure of Disk Galaxies. III. Comparison with Observations}",
  journal = {\apj},
     year = 1969,
    month = mar,
   volume = 155,
    pages = {721-+},
      doi = {10.1086/149907},
   adsurl = {http://adsabs.harvard.edu/abs/1969ApJ...155..721L}
}

@ARTICLE{Loebman11,
   author = {{Loebman}, S.~R. and {Ro{\v s}kar}, R. and {Debattista}, V.~P. and 
	{Ivezi{\'c}}, {\v Z}. and {Quinn}, T.~R. and {Wadsley}, J.},
    title = "{The Genesis of the Milky Way's Thick Disk Via Stellar Migration}",
  journal = {\apj},
archivePrefix = "arXiv",
   eprint = {1009.5997},
 primaryClass = "astro-ph.GA",
     year = 2011,
    month = aug,
   volume = 737,
    pages = {8-+},
      doi = {10.1088/0004-637X/737/1/8},
   adsurl = {http://adsabs.harvard.edu/abs/2011ApJ...737....8L}
}

@ARTICLE{Loebman16,
   author = {{Loebman}, S.~R. and {Debattista}, V.~P. and {Nidever}, D.~L. and 
	{Hayden}, M.~R. and {Holtzman}, J.~A. and {Clarke}, A.~J. and 
	{Ro{\v s}kar}, R. and {Valluri}, M.},
    title = "{Imprints of Radial Migration on the Milky Way{\rsquo}s Metallicity Distribution Functions}",
  journal = {\apjl},
archivePrefix = "arXiv",
   eprint = {1511.06369},
     year = 2016,
    month = feb,
   volume = 818,
      eid = {L6},
    pages = {L6},
      doi = {10.3847/2041-8205/818/1/L6},
   adsurl = {http://adsabs.harvard.edu/abs/2016ApJ...818L...6L}
}

@ARTICLE{LBK72,
   author = {{Lynden-Bell}, D. and {Kalnajs}, A.~J.},
    title = "{On the generating mechanism of spiral structure}",
  journal = {\mnras},
     year = 1972,
   volume = 157,
    pages = {1},
   adsurl = {http://adsabs.harvard.edu/abs/1972MNRAS.157....1L}
}

@ARTICLE{Ma02,
   author = {{Ma}, J.},
    title = "{Properties of disks and spiral arms along the Hubble sequence}",
  journal = {\aap},
   eprint = {astro-ph/0203220},
     year = 2002,
    month = jun,
   volume = 388,
    pages = {389-395},
      doi = {10.1051/0004-6361:20020414},
   adsurl = {http://adsabs.harvard.edu/abs/2002A%26A...388..389M}
}

@ARTICLE{Ma99,
   author = {{Ma}, J. and {Zhao}, J.~L. and {Shu}, C.~G. and {Peng}, Q.~H.
	},
    title = "{Some statistical properties of spiral galaxies}",
  journal = {\aap},
   eprint = {arXiv:astro-ph/0002476},
     year = 1999,
    month = oct,
   volume = 350,
    pages = {31-37},
   adsurl = {http://adsabs.harvard.edu/abs/1999A%26A...350...31M}
}

@ARTICLE{MPQ70,
   author = {{Miller}, R.~H. and {Prendergast}, K.~H. and {Quirk}, W.~J.},
    title = "{Numerical Experiments on Spiral Structure}",
  journal = {\apj},
     year = 1970,
    month = sep,
   volume = 161,
    pages = {903},
      doi = {10.1086/150593},
   adsurl = {http://adsabs.harvard.edu/abs/1970ApJ...161..903M}
}

@ARTICLE{Minchev12,
   author = {{Minchev}, I. and {Famaey}, B. and {Quillen}, A.~C. and {Di Matteo}, P. and 
	{Combes}, F. and {Vlaji{\'c}}, M. and {Erwin}, P. and {Bland-Hawthorn}, J.
	},
    title = "{Evolution of galactic discs: multiple patterns, radial migration, and disc outskirts}",
  journal = {\aap},
archivePrefix = "arXiv",
   eprint = {1203.2621},
 primaryClass = "astro-ph.GA",
     year = 2012,
    month = dec,
   volume = 548,
      eid = {A126},
    pages = {A126},
      doi = {10.1051/0004-6361/201219198},
   adsurl = {http://adsabs.harvard.edu/abs/2012A%26A...548A.126M}
}

@ARTICLE{Minchev11,
   author = {{Minchev}, I. and {Famaey}, B. and {Combes}, F. and {Di Matteo}, P. and 
	{Mouhcine}, M. and {Wozniak}, H.},
    title = "{Radial migration in galactic disks caused by resonance overlap of multiple patterns: Self-consistent simulations}",
  journal = {\aap},
archivePrefix = "arXiv",
   eprint = {1006.0484},
 primaryClass = "astro-ph.GA",
     year = 2011,
    month = mar,
   volume = 527,
    pages = {A147+},
      doi = {10.1051/0004-6361/201015139},
   adsurl = {http://adsabs.harvard.edu/abs/2011A%26A...527A.147M}
}

@ARTICLE{MF10,
   author = {{Minchev}, I. and {Famaey}, B.},
    title = "{A New Mechanism for Radial Migration in Galactic Disks: Spiral-Bar Resonance Overlap}",
  journal = {ApJ},
archivePrefix = "arXiv",
   eprint = {0911.1794},
 primaryClass = "astro-ph.GA",
     year = 2010,
    month = oct,
   volume = 722,
    pages = {112-121},
      doi = {10.1088/0004-637X/722/1/112},
   adsurl = {http://adsabs.harvard.edu/abs/2010ApJ...722..112M}
}

@ARTICLE{MQ06,
   author = {{Minchev}, I. and {Quillen}, A.~C.},
    title = "{Radial heating of a galactic disc by multiple spiral density waves}",
  journal = {MNRAS},
   eprint = {arXiv:astro-ph/0511037},
     year = 2006,
    month = may,
   volume = 368,
    pages = {623-636},
      doi = {10.1111/j.1365-2966.2006.10129.x},
   adsurl = {http://adsabs.harvard.edu/abs/2006MNRAS.368..623M}
}

@BOOK{Hagihara72,
    author = {{Hagihara}, Y.},
    title = "{Celestial mechanics.  pt.1: Perturbation theory}",
    year = 1972,
    volume = 2,
    publisher = {MIT Press},
    adsurl = {https://ui.adsabs.harvard.edu/abs/1972ceme.book.....H}
}

@ARTICLE{BM11,
       author = {{Binney}, James and {McMillan}, Paul},
        title = "{Models of our Galaxy - II}",
      journal = {\mnras},
         year = 2011,
        month = may,
       volume = {413},
       number = {3},
        pages = {1889-1898},
          doi = {10.1111/j.1365-2966.2011.18268.x},
archivePrefix = {arXiv},
       eprint = {1101.0747},
 primaryClass = {astro-ph.GA},
       adsurl = {https://ui.adsabs.harvard.edu/abs/2011MNRAS.413.1889B}
}

@ARTICLE{Binney10,
       author = {{Binney}, James},
        title = "{Distribution functions for the Milky Way}",
      journal = {\mnras},
         year = 2010,
        month = feb,
       volume = {401},
       number = {4},
        pages = {2318-2330},
          doi = {10.1111/j.1365-2966.2009.15845.x},
archivePrefix = {arXiv},
       eprint = {0910.1512},
 primaryClass = {astro-ph.GA},
       adsurl = {https://ui.adsabs.harvard.edu/abs/2010MNRAS.401.2318B}
}

@ARTICLE{Papayannopoulos79a,
   author = {{Papayannopoulos}, T.},
    title = "{Orbits near the particle resonance of a galaxy. I - Numerical study}",
  journal = {\aap},
     year = 1979,
    month = aug,
   volume = 77,
    pages = {75-85},
   adsurl = {http://adsabs.harvard.edu/abs/1979A%26A....77...75P}
}

@ARTICLE{Papayannopoulos79b,
   author = {{Papayannopoulos}, T.},
    title = "{Orbits near the particle resonance of a galaxy. II - Theoretical study}",
  journal = {\aap},
     year = 1979,
    month = oct,
   volume = 79,
    pages = {197-203},
   adsurl = {http://adsabs.harvard.edu/abs/1979A%26A....79..197P}
}

@ARTICLE{RS12,
   author = {{Radburn-Smith}, D.~J. and {Ro{\v s}kar}, R. and {Debattista}, V.~P. and 
	{Dalcanton}, J.~J. and {Streich}, D. and {de Jong}, R.~S. and 
	{Vlaji{\'c}}, M. and {Holwerda}, B.~W. and {Purcell}, C.~W. and 
	{Dolphin}, A.~E. and {Zucker}, D.~B.},
    title = "{Outer-disk Populations in NGC 7793: Evidence for Stellar Radial Migration}",
  journal = {\apj},
archivePrefix = "arXiv",
   eprint = {1206.1057},
 primaryClass = "astro-ph.CO",
     year = 2012,
    month = jul,
   volume = 753,
      eid = {138},
    pages = {138},
      doi = {10.1088/0004-637X/753/2/138},
   adsurl = {http://adsabs.harvard.edu/abs/2012ApJ...753..138R}
}

@ARTICLE{RZ95,
   author = {{Rix}, H.-W. and {Zaritsky}, D.},
    title = "{Nonaxisymmetric Structures in the Stellar Disks of Galaxies}",
  journal = {\apj},
   eprint = {arXiv:astro-ph/9505111},
     year = 1995,
    month = jul,
   volume = 447,
    pages = {82},
      doi = {10.1086/175858},
   adsurl = {http://adsabs.harvard.edu/abs/1995ApJ...447...82R}
}

@ARTICLE{Roskar12,
   author = {{Ro{\v s}kar}, R. and {Debattista}, V.~P. and {Quinn}, T.~R. and 
	{Wadsley}, J.},
    title = "{Radial migration in disc galaxies - I. Transient spiral structure and dynamics}",
  journal = {\mnras},
archivePrefix = "arXiv",
   eprint = {1110.4413},
 primaryClass = "astro-ph.GA",
     year = 2012,
    month = nov,
   volume = 426,
    pages = {2089-2106},
      doi = {10.1111/j.1365-2966.2012.21860.x},
   adsurl = {http://adsabs.harvard.edu/abs/2012MNRAS.426.2089R}
}

@ARTICLE{Roskar08a,
   author = {{Ro{\v s}kar}, R. and {Debattista}, V.~P. and {Stinson}, G.~S. and 
	{Quinn}, T.~R. and {Kaufmann}, T. and {Wadsley}, J.},
    title = "{Beyond Inside-Out Growth: Formation and Evolution of Disk Outskirts}",
  journal = {ApJL},
archivePrefix = "arXiv",
   eprint = {0710.5523},
     year = 2008,
    month = mar,
   volume = 675,
    pages = {L65-L68},
      doi = {10.1086/586734},
   adsurl = {http://adsabs.harvard.edu/abs/2008ApJ...675L..65R}
}

@ARTICLE{SB09b,
   author = {{Sch{\"o}nrich}, R. and {Binney}, J.},
    title = "{Origin and structure of the Galactic disc(s)}",
  journal = {MNRAS},
archivePrefix = "arXiv",
   eprint = {0907.1899},
 primaryClass = "astro-ph.GA",
     year = 2009,
    month = nov,
   volume = 399,
    pages = {1145-1156},
      doi = {10.1111/j.1365-2966.2009.15365.x},
   adsurl = {http://adsabs.harvard.edu/abs/2009MNRAS.399.1145S}
}

@ARTICLE{SJ98,
   author = {{Seigar}, M.~S. and {James}, P.~A.},
    title = "{The structure of spiral galaxies - II. Near-infrared properties of spiral arms}",
  journal = {\mnras},
   eprint = {arXiv:astro-ph/9803254},
     year = 1998,
    month = sep,
   volume = 299,
    pages = {685-698},
      doi = {10.1046/j.1365-8711.1998.01779.x},
   adsurl = {http://adsabs.harvard.edu/abs/1998MNRAS.299..685S}
}

@ARTICLE{SC14,
   author = {{Sellwood}, J.~A. and {Carlberg}, R.~G},
    title = "{Transient Spirals As Superposed Instabilities}",
  journal = {\apj},
 primaryClass = "astro-ph.GA",
     year = 2014,
    month = apr,
   volume = 785,
    pages = {137-+},
}

@ARTICLE{Sellwood11,
   author = {{Sellwood}, J.~A.},
    title = "{The lifetimes of spiral patterns in disc galaxies}",
  journal = {\mnras},
archivePrefix = "arXiv",
   eprint = {1008.2737},
 primaryClass = "astro-ph.CO",
     year = 2011,
    month = jan,
   volume = 410,
    pages = {1637-1646},
      doi = {10.1111/j.1365-2966.2010.17545.x},
   adsurl = {http://adsabs.harvard.edu/abs/2011MNRAS.410.1637S}
}

@ARTICLE{SB02,
   author = {{Sellwood}, J.~A. and {Binney}, J.~J.},
    title = "{Radial mixing in galactic discs}",
  journal = {MNRAS},
   eprint = {arXiv:astro-ph/0203510},
     year = 2002,
    month = nov,
   volume = 336,
    pages = {785-796},
      doi = {10.1046/j.1365-8711.2002.05806.x},
   adsurl = {http://adsabs.harvard.edu/abs/2002MNRAS.336..785S}
}

@ARTICLE{SSS12,
   author = {{Solway}, M. and {Sellwood}, J.~A. and {Sch{\"o}nrich}, R.},
    title = "{Radial migration in galactic thick discs}",
  journal = {\mnras},
archivePrefix = "arXiv",
   eprint = {1202.1418},
 primaryClass = "astro-ph.GA",
     year = 2012,
    month = may,
   volume = 422,
    pages = {1363-1383},
      doi = {10.1111/j.1365-2966.2012.20712.x},
   adsurl = {http://adsabs.harvard.edu/abs/2012MNRAS.422.1363S}
}

@ARTICLE{SS53,
   author = {{Spitzer}, Jr., L. and {Schwarzschild}, M.},
    title = "{The Possible Influence of Interstellar Clouds on Stellar Velocities. II.}",
  journal = {\apj},
     year = 1953,
    month = jul,
   volume = 118,
    pages = {106-+},
      doi = {10.1086/145730},
   adsurl = {http://adsabs.harvard.edu/abs/1953ApJ...118..106S}
}

@ARTICLE{Struck11,
   author = {{Struck}, C. and {Dobbs}, C.~L. and {Hwang}, J.-S.},
    title = "{Slowly breaking waves: the longevity of tidally induced spiral structure}",
  journal = {\mnras},
archivePrefix = "arXiv",
   eprint = {1102.4817},
 primaryClass = "astro-ph.CO",
     year = 2011,
    month = jul,
   volume = 414,
    pages = {2498-2510},
      doi = {10.1111/j.1365-2966.2011.18568.x},
   adsurl = {http://adsabs.harvard.edu/abs/2011MNRAS.414.2498S}
}

@ARTICLE{2016A&A...595A...1G,
       author = {{Gaia Collaboration} and {Prusti}, T. and {de Bruijne}, J.~H.~J. and {Brown}, A.~G.~A. and {Vallenari}, A. and {Babusiaux}, C. and {Bailer-Jones}, C.~A.~L. and {Bastian}, U. and {Biermann}, M. and {Evans}, D.~W. and et al.},
        title = "{The Gaia mission}",
      journal = {\aap},
         year = 2016,
        month = nov,
       volume = {595},
          eid = {A1},
        pages = {A1},
          doi = {10.1051/0004-6361/201629272},
archivePrefix = {arXiv},
       eprint = {1609.04153},
 primaryClass = {astro-ph.IM},
       adsurl = {https://ui.adsabs.harvard.edu/abs/2016A&A...595A...1G}
}

@ARTICLE{2022AJ....164..207D,
       author = {{DESI Collaboration} and {Abareshi}, B. and {Aguilar}, J. and {Ahlen}, S. and {Alam}, Shadab and {Alexander}, David M. and {Alfarsy}, R. and {Allen}, L. and {Allende Prieto}, C. and {Alves}, O. and {Ameel}, J. and {Armengaud}, E. and {Asorey}, J. and {Aviles}, Alejandro and {Bailey}, S. and {Balaguera-Antol{\'\i}nez}, A. and {Ballester}, O. and {Baltay}, C. and {Bault}, A. and {Beltran}, S.~F. and {Benavides}, B. and {BenZvi}, S. and {Berti}, A. and {Besuner}, R. and {Beutler}, Florian and {Bianchi}, D. and {Blake}, C. and {Blanc}, P. and {Blum}, R. and {Bolton}, A. and {Bose}, S. and {Bramall}, D. and {Brieden}, S. and {Brodzeller}, A. and {Brooks}, D. and {Brownewell}, C. and {Buckley-Geer}, E. and {Cahn}, R.~N. and {Cai}, Z. and {Canning}, R. and {Capasso}, R. and {Carnero Rosell}, A. and {Carton}, P. and {Casas}, R. and {Castander}, F.~J. and {Cervantes-Cota}, J.~L. and {Chabanier}, S. and {Chaussidon}, E. and {Chuang}, C. and {Circosta}, C. and {Cole}, S. and {Cooper}, A.~P. and {da Costa}, L. and {Cousinou}, M.-C. and {Cuceu}, A. and {Davis}, T.~M. and {Dawson}, K. and {de la Cruz-Noriega}, R. and {de la Macorra}, A. and {de Mattia}, A. and {Della Costa}, J. and {Demmer}, P. and {Derwent}, M. and {Dey}, A. and {Dey}, B. and {Dhungana}, G. and {Ding}, Z. and {Dobson}, C. and {Doel}, P. and {Donald-McCann}, J. and {Donaldson}, J. and {Douglass}, K. and {Duan}, Y. and {Dunlop}, P. and {Edelstein}, J. and {Eftekharzadeh}, S. and {Eisenstein}, D.~J. and {Enriquez-Vargas}, M. and {Escoffier}, S. and {Evatt}, M. and {Fagrelius}, P. and {Fan}, X. and {Fanning}, K. and {Fawcett}, V.~A. and {Ferraro}, S. and {Ereza}, J. and {Flaugher}, B. and {Font-Ribera}, A. and {Forero-Romero}, J.~E. and {Frenk}, C.~S. and {Fromenteau}, S. and {G{\"a}nsicke}, B.~T. and {Garcia-Quintero}, C. and {Garrison}, L. and {Gazta{\~n}aga}, E. and {Gerardi}, F. and {Gil-Mar{\'\i}n}, H. and {Gontcho A Gontcho}, S. and {Gonzalez-Morales}, Alma X. and {Gonzalez-de-Rivera}, G. and {Gonzalez-Perez}, V. and {Gordon}, C. and {Graur}, O. and {Green}, D. and {Grove}, C. and {Gruen}, D. and {Gutierrez}, G. and {Guy}, J. and {Hahn}, C. and {Harris}, S. and {Herrera}, D. and {Herrera-Alcantar}, Hiram K. and {Honscheid}, K. and {Howlett}, C. and {Huterer}, D. and {Ir{\v{s}}i{\v{c}}}, V. and {Ishak}, M. and {Jelinsky}, P. and {Jiang}, L. and {Jimenez}, J. and {Jing}, Y.~P. and {Joyce}, R. and {Jullo}, E. and {Juneau}, S. and {Kara{\c{c}}ayl{\i}}, N.~G. and {Karamanis}, M. and {Karcher}, A. and {Karim}, T. and {Kehoe}, R. and {Kent}, S. and {Kirkby}, D. and {Kisner}, T. and {Kitaura}, F. and {Koposov}, S.~E. and {Kov{\'a}cs}, A. and {Kremin}, A. and {Krolewski}, Alex and {L'Huillier}, B. and {Lahav}, O. and {Lambert}, A. and {Lamman}, C. and {Lan}, Ting-Wen and {Landriau}, M. and {Lane}, S. and {Lang}, D. and {Lange}, J.~U. and {Lasker}, J. and {Le Guillou}, L. and {Leauthaud}, A. and {Le Van Suu}, A. and {Levi}, Michael E. and {Li}, T.~S. and {Magneville}, C. and {Manera}, M. and {Manser}, Christopher J. and {Marshall}, B. and {Martini}, Paul and {McCollam}, W. and {McDonald}, P. and {Meisner}, Aaron M. and {Mena-Fern{\'a}ndez}, J. and {Meneses-Rizo}, J. and {Mezcua}, M. and {Miller}, T. and {Miquel}, R. and {Montero-Camacho}, P. and {Moon}, J. and {Moustakas}, J. and {Mueller}, E. and {Mu{\~n}oz-Guti{\'e}rrez}, Andrea and {Myers}, Adam D. and {Nadathur}, S. and {Najita}, J. and {Napolitano}, L. and {Neilsen}, E. and {Newman}, Jeffrey A. and {Nie}, J.~D. and {Ning}, Y. and {Niz}, G. and {Norberg}, P. and {Noriega}, Hern{\'a}n E. and {O'Brien}, T. and {Obuljen}, A. and {Palanque-Delabrouille}, N. and {Palmese}, A. and {Zhiwei}, P. and {Pappalardo}, D. and {PENG}, X. and {Percival}, W.~J. and {Perruchot}, S. and {Pogge}, R. and {Poppett}, C. and {Porredon}, A. and {Prada}, F. and {Prochaska}, J. and {Pucha}, R. and {P{\'e}rez-Fern{\'a}ndez}, A. and {P{\'e}rez-R{\`a}fols}, I. and {Rabinowitz}, D. and {Raichoor}, A.},
        title = "{Overview of the Instrumentation for the Dark Energy Spectroscopic Instrument}",
      journal = {\aj},
         year = 2022,
        month = nov,
       volume = {164},
       number = {5},
          eid = {207},
        pages = {207},
          doi = {10.3847/1538-3881/ac882b},
archivePrefix = {arXiv},
       eprint = {2205.10939},
 primaryClass = {astro-ph.IM},
       adsurl = {https://ui.adsabs.harvard.edu/abs/2022AJ....164..207D}
}

@ARTICLE{Toomre64,
   author = {{Toomre}, A.},
    title = "{On the gravitational stability of a disk of stars}",
  journal = {\apj},
     year = 1964,
    month = may,
   volume = 139,
    pages = {1217-1238},
      doi = {10.1086/147861},
   adsurl = {http://adsabs.harvard.edu/abs/1964ApJ...139.1217T}
}

@ARTICLE{vanderKruit87,
   author = {{van der Kruit}, P.~C.},
    title = "{The radial distribution of surface brightness in galactic disks}",
  journal = {\aap},
     year = 1987,
    month = feb,
   volume = 173,
    pages = {59-80},
   adsurl = {http://adsabs.harvard.edu/abs/1987A%26A...173...59V}
}

@ARTICLE{VCdON16,
   author = {{Vera-Ciro}, C. and {D'Onghia}, E. and {Navarro}, J.~F.},
    title = "{The Imprint of Radial Migration on the Vertical Structure of Galaxy Disks}",
  journal = {\apj},
archivePrefix = "arXiv",
   eprint = {1605.03575},
     year = 2016,
    month = dec,
   volume = 833,
      eid = {42},
    pages = {42},
      doi = {10.3847/1538-4357/833/1/42},
   adsurl = {http://adsabs.harvard.edu/abs/2016ApJ...833...42V}
}

@ARTICLE{VC14,
   author = {{Vera-Ciro}, C. and {D'Onghia}, E. and {Navarro}, J. and {Abadi}, M.
	},
    title = "{The Effect of Radial Migration on Galactic Disks}",
  journal = {\apj},
archivePrefix = "arXiv",
   eprint = {1405.3317},
     year = 2014,
    month = oct,
   volume = 794,
      eid = {173},
    pages = {173},
      doi = {10.1088/0004-637X/794/2/173},
   adsurl = {http://adsabs.harvard.edu/abs/2014ApJ...794..173V}
}

@ARTICLE{WBS11,
   author = {{Wada}, K. and {Baba}, J. and {Saitoh}, T.~R.},
    title = "{Interplay between Stellar Spirals and the Interstellar Medium in Galactic Disks}",
  journal = {\apj},
archivePrefix = "arXiv",
   eprint = {1104.1287},
     year = 2011,
    month = jul,
   volume = 735,
      eid = {1},
    pages = {1},
      doi = {10.1088/0004-637X/735/1/1},
   adsurl = {http://adsabs.harvard.edu/abs/2011ApJ...735....1W}
}

@ARTICLE{Wielen77,
   author = {{Wielen}, R.},
    title = "{The diffusion of stellar orbits derived from the observed age-dependence of the velocity dispersion}",
  journal = {\aap},
     year = 1977,
    month = sep,
   volume = 60,
    pages = {263-275},
   adsurl = {http://adsabs.harvard.edu/abs/1977A%26A....60..263W}
}

@ARTICLE{WFD96,
   author = {{Wielen}, R. and {Fuchs}, B. and {Dettbarn}, C.},
    title = "{On the birth-place of the Sun and the places of formation of other nearby stars}",
  journal = {\aap},
     year = 1996,
    month = oct,
   volume = 314,
    pages = {438-+},
   adsurl = {http://adsabs.harvard.edu/abs/1996A%26A...314..438W}
}

@ARTICLE{WS89,
   author = {{Wyse}, R.~F.~G. and {Silk}, J.},
    title = "{Star formation rates and abundance gradients in disk galaxies}",
  journal = {ApJ},
     year = 1989,
    month = apr,
   volume = 339,
    pages = {700-711},
      doi = {10.1086/167329},
   adsurl = {http://adsabs.harvard.edu/abs/1989ApJ...339..700W}
}

@ARTICLE{2026arXiv260700077S,
       author = {{Steel}, Cecilia and {Wetzel}, Andrew and {Kang}, Rori and {McCluskey}, Fiona and {Loebman}, Sarah and {Daniel}, Kathryne J.},
        title = "{Hot or Cold? Radial Redistribution of Stars in FIRE Simulations of Milky Way-Mass Galaxies and the Asymmetry of Inward versus Outward Migrators}",
      journal = {arXiv e-prints},
         year = {\the\year},
        month = jun,
          eid = {arXiv:2607.00077},
        pages = {arXiv:2607.00077},
archivePrefix = {arXiv},
       eprint = {2607.00077},
 primaryClass = {astro-ph.GA},
       adsurl = {https://ui.adsabs.harvard.edu/abs/2026arXiv260700077S}
}

@article{Smock26,
  author       = {{Smock}, Amy and {Daniel}, Kathryne J. and {Rampalli}, Rayna and Newton, Elisabeth R. and McAuley, Olivia and Hyman, S{\'o}ley and Chatur, Lipika},
  title        = {Wrinkles in Time. II. Stellar Age Trends in Kinematic Signatures from Transient Spiral Structure},
  journal      = {\apj},
  year         = {2026},
  doi          = {10.3847/1538-4357/ae7367},
        month = jul,
       volume = {1005},
       number = {2},
        pages = {151},  
        note  = {Accepted},
  entrysubtype = {refereed}
}

@ARTICLE{Wisz25,
       author = {{Wisz}, M.~E. and {Masters}, Karen L. and {Daniel}, Kathryne J. and {Stark}, David V. and {Belfiore}, Francesco},
        title = "{The Impacts of Bars, Spirals, and Bulge Size on Gas-phase Metallicity Gradients in MaNGA Galaxies}",
      journal = {\apj},
         year = 2025,
        month = apr,
       volume = {983},
       number = {1},
          eid = {57},
        pages = {57},
          doi = {10.3847/1538-4357/adbb6f},
archivePrefix = {arXiv},
       eprint = {2502.10922},
 primaryClass = {astro-ph.GA},
       adsurl = {https://ui.adsabs.harvard.edu/abs/2025ApJ...983...57W}
}

@ARTICLE{BeS23,
       author = {{Beraldo e Silva}, Leandro and {Debattista}, Victor P. and {Anderson}, Stuart Robert and {Valluri}, Monica and {Erwin}, Peter and {Daniel}, Kathryne J. and {Deg}, Nathan},
        title = "{Orbital Support and Evolution of Flat Profiles of Bars (Shoulders)}",
      journal = {\apj},
         year = 2023,
        month = sep,
       volume = {955},
       number = {1},
          eid = {38},
        pages = {38},
          doi = {10.3847/1538-4357/ace976},
archivePrefix = {arXiv},
       eprint = {2303.04828},
 primaryClass = {astro-ph.GA},
       adsurl = {https://ui.adsabs.harvard.edu/abs/2023ApJ...955...38B}
}

@ARTICLE{Quinn26,
       author = {{Quinn}, J.~R. and {Loebman}, S.~R. and {Daniel}, K.~J. and {Beraldo e Silva}, L. and {Wetzel}, A. and {Debattista}, V.~P. and {Arora}, A. and {Ansar}, S. and {McCluskey}, F. and {Masoumi}, D.},
        title = "{Spiral Structure Properties, Dynamics, and Evolution in Milky Way-mass Galaxy Simulations}",
      journal = {\apj},
         year = 2026,
        month = feb,
       volume = {997},
       number = {2},
          eid = {363},
        pages = {363},
          doi = {10.3847/1538-4357/ae2be1},
archivePrefix = {arXiv},
       eprint = {2507.22793},
 primaryClass = {astro-ph.GA},
       adsurl = {https://ui.adsabs.harvard.edu/abs/2026ApJ...997..363Q}
}

@ARTICLE{McClure25,
       author = {{McClure}, Rachel Lee and {Beane}, Angus and {D'Onghia}, Elena and {Filion}, Carrie and {Daniel}, Kathryne J.},
        title = "{The impact of classical bulges on stellar bars and boxy-peanut-X features in disc galaxies}",
      journal = {\mnras},
         year = 2025,
        month = feb,
       volume = {537},
       number = {2},
        pages = {1475-1488},
          doi = {10.1093/mnras/staf107},
archivePrefix = {arXiv},
       eprint = {2410.08277},
 primaryClass = {astro-ph.GA},
       adsurl = {https://ui.adsabs.harvard.edu/abs/2025MNRAS.537.1475M}
}

@ARTICLE{Wiggins25,
       author = {{Wiggins}, Alessa I. and {Quinn}, Jamie R. and {Oeur}, Micah and {Loebman}, Sarah R. and {Frinchaboy}, Peter M. and {Daniel}, Kathryne J. and {McCluskey}, Fiona and {Otto}, Jonah M. and {Woodward}, Hannah R. and {D'Onghia}, Elena and et al.},
        title = "{Understanding the Origin and Dynamical Evolution of the Unique Open Star Cluster Berkeley 20 Using FIRE Simulations}",
      journal = {\apjl},
         year = 2025,
        month = dec,
       volume = {995},
       number = {1},
          eid = {L25},
        pages = {L25},
          doi = {10.3847/2041-8213/ae21bf},
archivePrefix = {arXiv},
       eprint = {2511.14958},
 primaryClass = {astro-ph.GA},
       adsurl = {https://ui.adsabs.harvard.edu/abs/2025ApJ...995L..25W}
}

@ARTICLE{DW15,
       author = {{Daniel}, Kathryne J. and {Wyse}, Rosemary F.~G.},
        title = "{Constraints on radial migration in spiral galaxies - I. Analytic criterion for capture at corotation}",
      journal = {\mnras},
         year = 2015,
        month = mar,
       volume = {447},
       number = {4},
        pages = {3576-3592},
          doi = {10.1093/mnras/stu2683},
archivePrefix = {arXiv},
       eprint = {1412.6110},
 primaryClass = {astro-ph.GA},
       adsurl = {https://ui.adsabs.harvard.edu/abs/2015MNRAS.447.3576D}
}

@ARTICLE{Khachaturyants22a,
       author = {{Khachaturyants}, Tigran and {Beraldo e Silva}, Leandro and {Debattista}, Victor P. and {Daniel}, Kathryne J.},
        title = "{Bending waves excited by irregular gas inflow along warps}",
      journal = {\mnras},
         year = 2022,
        month = may,
       volume = {512},
       number = {3},
        pages = {3500-3519},
          doi = {10.1093/mnras/stac606},
archivePrefix = {arXiv},
       eprint = {2203.03741},
 primaryClass = {astro-ph.GA},
       adsurl = {https://ui.adsabs.harvard.edu/abs/2022MNRAS.512.3500K}
}
\appendix

\section{Derivation of radial excursions around the \Lag}\label{s:appendix}

Equation~\ref{eqn:intmotionLspiralphi1} is,
\begin{equation}\label{eqn:AintmotionLspiralphi1}
    \dot{\phi_1} + 2\Omega_0 \dfrac{R_1}{R_0} = - \Phi_s\dfrac{m}{R_0^2} \int dt \sin[-\phi_1(t)].
\end{equation}
By adopting the form for its \azi\ position given in \eq~\ref{eqn:phi1max},
\begin{equation}
    \phi_1(t) = |\phi_1| \cos(\omega t +\delta),
\end{equation}
\eq~\ref{eqn:intmotionLspiralphi1} (now \eq~\ref{eqn:AintmotionLspiralphi1}) can be expressed in the form,
\begin{equation}\label{eqn:intmotionLspiralphi12}
    \dot{\phi_1} + 2\Omega_0 \dfrac{R_1}{R_0} = -\Phi_s \dfrac{m}{R_0^2 \omega} \int d\tau \sin(- |\phi_1| \cos \tau) 
\end{equation}
where $\omega$ is the frequency of oscillation about the \Lag\ and $\tau = \omega t$.  This can be explicitly solved by expanding $\sin x$ near \Lag\ where $x=0$.  The general form for the Maclaurin series is,
\begin{equation}
    \sin x = \sum_{k=0}^\infty \dfrac{(-1)^k \, x^{1+2k}}{(1+2k)!}.
\end{equation}
From \eq~\ref{eqn:intmotionLspiralphi12}, $x=-\phi_1(t) =-|\phi_1| \cos(\omega t)$ and so,
\begin{equation}\label{eq:sinxexpansion}
    \sin(-|\phi_1| \cos \tau) = \sum_{k=0}^\infty \dfrac{(-1)^{1+k} \,|\phi_1|^{1+2k} \,\cos ^{1+2k} \tau}{(1+2k)!}.
\end{equation}
The integral over the period of oscillation $\tau$ in \eq~\ref{eq:sinxexpansion} is separable from the other terms such that,
\begin{equation}
    \int d\tau \sin(-|\phi_1| \cos \tau) = \sum_{k=0}^\infty \dfrac{(-1)^{1+k} \, |\phi_1|^{1+2k}}{(1+2k)! }\int d\tau  \cos ^{1+2k} \tau.
\end{equation}
Keeping the first two terms of order $\sin\tau$ of the integral series we find eqn. (\ref{eqn:intmotionLspiralphi12}) becomes,
\begin{equation}
\begin{array}{cl}
\dot{\phi_1} + 2\Omega_0 \dfrac{R_1}{R_0} & \approx \Phi_s \dfrac{m}{R_0^2 \omega}  \sin \tau \left(|\phi_1|-\dfrac{|\phi_1|^3}{3!}+...\right)\\
&\\
& \approx \Phi_s \dfrac{m |\phi_1|}{R_0^2 \omega}  \sin \tau .
\end{array}
\end{equation}
We now have the following expression for the angular velocity in the rotating frame for an orbit oscillating about an \Lag,
\begin{equation}
    \dot{\phi_1} = \Phi_s \dfrac{m |\phi_1|}{R_0^2 \omega}  \sin \tau - 2\Omega_0 \dfrac{R_1}{R_0} .
\end{equation}
Plugging this result into eqn.(\ref{eqn:motionLspiralR1}), we find,
\begin{equation}\label{eq:R1EoM}
    \ddot{R_1} +\kappa^2 R_1 = \dfrac{\Phi_s}{R_0} \left(\alpha \sin(|\phi_1|\cos\tau) +2m\dfrac{\Omega_0}{\omega} |\phi_1|\sin\tau  \right).
\end{equation}
This is a non-homogeneous second order differential equation. The first term on the right hand side arises from \eq~\ref{eqn:dPdR} and corresponds to the force associated with the radial gradient in the potential.  It can be interpreted as a distortion from an elliptical orbit about the \Lag in the rotating frame.  The second term comes from the time integral of the torque, which is the angular impulse or change in angular momentum, in \eq~\ref{eqn:dPdp}.

\end{document}